%% file: susymelons-arxiv-v3.tex
\documentclass[11pt]{article}
\pdfoutput=1
\usepackage{jheppub}
\usepackage{tabu}
\usepackage[vcentermath]{youngtab}
\usepackage[usenames,dvipsnames,table]{xcolor}
\usepackage{graphicx,subfig}
\usepackage{amsmath,amssymb,amsthm,amsfonts,multirow,array,bm,bbm,esint}
\usepackage[mathscr]{eucal}
\usepackage[bbgreekl]{mathbbol}
\usepackage{epsf,grffile}
\usepackage{slashed}
\usepackage[numbers,sort&compress]{natbib}
\usepackage{array,tikz-cd}
\usepackage{float}

\usepackage{rotfloat}
\usepackage{rotating}
\usepackage{pdflscape}
\usepackage{array}
\newcolumntype{H}{>{\setbox0=\hbox\bgroup}c<{\egroup}@{}}
\usepackage{afterpage}

\usetikzlibrary{snakes}
\usetikzlibrary{shapes.misc}

\include{susymelons-macros}

\title{On Melonic Supertensor Models}

\author{Chi-Ming Chang, Sean Colin-Ellerin,  Mukund Rangamani}

\affiliation[]{
Center for Quantum Mathematics and Physics (QMAP),  \\
Department of Physics, University of California, Davis, CA 95616 USA.}

\emailAdd{wychang@ucdavis.edu, scolinellerin@ucdavis.edu, mukund@physics.ucdavis.edu}

\abstract{ 
We investigate a class of supersymmetric quantum mechanical theories (with two supercharges) having tensor-valued degrees of freedom which are dominated by melon diagrams in  the large $N$ limit. One motivation was to examine the interplay between supersymmetry and melonic dominance and  potential implications for building toy models of holography. We find a definite tension between supersymmetry (with dynamical bosons) and melonic dominance in this class of systems. More specifically, our theories attain a low energy non-supersymmetric conformal fixed point. The origin of supersymmetry breaking lies in the need to regularize bosonic and fermionic degrees of freedom independently. We investigate various aspects of the low energy spectrum  and also comment on related examples with different numbers of supercharges. Along the way we also derive some 
technical results for $SL(2,{\mathbb R})$ wavefunctions for fermionic excitations.
}

\begin{document}
\maketitle


\section{Introduction}
\label{sec:intro}

Despite the holographic AdS/CFT correspondence having been discovered more than two decades ago, 
the \emph{raison d'\^etre} for planar field theories to have classical gravitational duals has as yet proven elusive to formulate. While we have various necessary conditions such as the existence of a sparse spectrum of light states in the planar limit the full set of sufficient conditions are yet to be discovered. Part of the issue is that while planar field theories are easy to attain by taking suitable 't Hooft-like large $N$ limits, canonical representatives are either too simple (e.g., planar vector models)  or too difficult to solve analytically (e.g., planar matrix models). The simplicity/complexity in the field theory analysis translates into the dual picture a correspondence notion of complexity/simplicity, preserving the overall intransigence of the system from revealing the rationale for the duality. One might hope that identifying theories which lie in some intermediate domain between the aforementioned would potentially aid in our attempts to understand the origins of geometry from field theory.

A promising arena for such explorations which has attracted lots of recent attention is the family of large $N$ melonic models. Interest in these theories stems from the success of the quantum mechanical  model, the Sachdev-Ye-Kitaev (SYK) model, described by Kitaev \cite{Kitaev:2015aa} building on an earlier construction of Sachdev and Ye  \cite{Sachdev:1992fk}. The model consists of  $N$ fermions with a random (disordered) multi-fermion interaction. The free fermion system in the UV flows to an IR fixed point with emergent conformal symmetry in the strongly coupled planar limit \cite{Kitaev:2015aa,Maldacena:2016hyu}. While the conformal symmetry is, strictly speaking, broken away from the IR limit, it turns out that the gapless modes capture some of the essential physics, which furthermore, bears close resemblance  to that of black holes in holographic systems. The sub-sector of the theory (essentially a single mode, the Schwarzian field) controlling the emergent conformal symmetry and its breaking is dual to a two dimensional dilaton gravity theory, the Jackiw-Teitelboim (JT) theory \cite{Maldacena:2016hyu,Maldacena:2016upp}.  A key intriguing feature is that the system saturates the chaos bound \cite{Maldacena:2015waa}, which indicates that it is maximally scrambling just as black holes in situations with dynamical gravity.  All told, the relative simplicity coupled with intricate dynamical behaviour with features that resemble more conventional gauge/gravity duals, makes the model a compelling study. For a selection of literature, see
\cite{Sachdev:1992fk,Parcollet:1997ysb,Parcollet_1999} for early works on disordered systems which led up the SYK model, 
 \cite{Sachdev:2015efa,Gross:2016kjj,Davison:2016ngz,Bulycheva:2017uqj,Yoon:2017nig,Bhattacharya:2017vaz} for generalizations to models with global (flavor) symmetries, and \cite{Anninos:2016szt,Fu:2016vas,Murugan:2017eto,Yoon:2017gut,Peng:2017spg,Narayan:2017hvh,Bulycheva:2018qcp} for supersymmetric generalizations. The spectrum and higher point-couplings are analyzed in \cite{Polchinski:2016xgd,Gross:2017hcz,Gross:2017aos}. The bulk duals of these are further explored in \cite{Maldacena:2016upp,Jensen:2016pah,Mandal:2017thl,Das:2017pif,Das:2017hrt,Das:2017wae,Gaikwad:2018dfc,Nayak:2018qej,Forste:2017apw} and a detailed discussion of the Schwarzian theory and near AdS$_2$ dynamics can be found in \cite{Stanford:2017thb,Kitaev:2017awl,Qi:2018rqm}.

It is interesting to examine if the SYK model is unique in its ability to capture features of holographic dualities. One reason for seeking generalizations is to ascertain if we can find a genuine quantum system sans disorder.\footnote{ Disordered systems are classical superpositions of different realizations of a  quantum system and therefore preclude a  well-defined Hilbert space in the theory (after disorder averaging).} Consequently, other models have been constructed with similar physics in the large $N$ limit without any disorder. These constructions take inspiration from models examined in the context of triangulations of manifolds in higher dimensions \cite{Bonzom:2011zz,Carrozza:2015adg}, and broadly fall  into one of two classes: the class of colored tensor models exemplified by the Gurau-Witten (GW) model \cite{Witten:2016iux,Bonzom:2011zz}, and the class of uncolored models exemplified by the  Carrozza-Tanasa-Klebanov-Tarnopolsky  model \cite{Klebanov:2016xxf,Carrozza:2015adg}. hese models are further explored in \cite{Nishinaka:2016nxg,Gurau:2016lzk,Peng:2016mxj,Krishnan:2016bvg,Ferrari:2017ryl,Narayan:2017qtw,Chaudhuri:2017vrv,deMelloKoch:2017bvv,Azeyanagi:2017drg,Giombi:2017dtl,Bulycheva:2017ilt,Choudhury:2017tax,Krishnan:2017lra,Krishnan:2017txw,Prakash:2017hwq,Benedetti:2018goh,Krishnan:2018hhu,Klebanov:2018nfp,Gubser:2018yec}; a recent review of the subject is \cite{Delporte:2018iyf}. We will collectively refer to these as \emph{melonic tensor models}. 

 In their simplest incarnations, these models comprise of  $O(N)$ tensor-valued fermionic fields with a particular class of multi-fermion vertices that ensure melonic dominance in the large $N$ limit. This ensures that the leading behaviour of the theory shares features such as the emergent near-conformal symmetry at low energies, and the saturation of the chaos bound. However, thanks to the large symmetry group\footnote{ The symmetry group is roughly $O(N)^M$ for some $M$ depending on the specifics of the model  (one may consider gauging it or part thereof).}  the low energy theory also comprises of other light degrees of freedom and peculiar thermodynamics  \cite{Bulycheva:2017ilt,Choudhury:2017tax}.

From a holographic perspective though a curious feature is that these quantum mechanical systems are devoid of supersymmetry. Let us first note that it is a debatable proposition as to whether supersymmetry is necessary for field theory to have classical gravity holographic duals.  While non-supersymmetric AdS vacua with low curvature on the string or Planck scale, $\ell_{AdS} \gg \ell_s, \ell_{_P}$, suffer from pathologies prompting conjectures that they are perhaps forbidden \cite{Ooguri:2016pdq}, there is no a-priori argument precluding theories with classical higher spin or stringy duals.\footnote{ Several examples of non-supersymmetric large $N$ field theories with classical master fields involving some form of gravitational interactions exist: eg., the classical higher spin theories dual to vector models, or stringy duals of  the symmetric orbifold CFT in two dimensions. We should also note that a non-supersymmetric theory could potentially capture some features of the supersymmetric model, say the high temperature thermodynamics, as exemplified by the ungauged D0-brane quantum mechanics theory, cf., \cite{Maldacena:2018vsr,Berkowitz:2018qhn}.} Indeed, the SYK model beyond the Schwarzian mode dynamics would be expected to be dual to a stringy bulk theory. However, the simplest quantum mechanical system that one hopes would capture gravitational dynamics of string/M-theory is the D0-brane quantum mechanics with sixteen supercharges \cite{Banks:1996vh}. It is therefore intriguing to ask if inclusion of supersymmetry reveals some further simplification to the analysis of melonic quantum mechanical models.  Various groups have addressed aspects of this question earlier: for instance a supersymmetric version of SYK model was analyzed first in \cite{Anninos:2016szt,Fu:2016vas} (with four, one and two supercharges). This was extended to two dimensions in search of melonic 2d CFTs in \cite{Murugan:2017eto}. Analysis of correlation functions in the model with two supercharges was carried out in 
\cite{Peng:2017spg,Bulycheva:2018qcp}. Supersymmetric tensor models were proposed in  \cite{Peng:2016mxj} -- these involve some additional augmentation involving  `mesonic'  operators in the theory. In the SYK case the essential features are preserved with the inclusion of supersymmetry (though there is signal of supersymmetry breaking in the one supercharge theory \cite{Fu:2016vas}).

We undertake an analysis of supersymmetric tensor models with the aim of ascertaining whether any simplification may be attained. Philosophically our models are different from the aforementioned (see below) and involve a simple generalization involving tensor-valued superfields with suitable superpotential couplings. We find several peculiarities with the main result being that supersymmetry does not appear to aid in the very least! While we will explicitly analyze the theory with two supercharges $\mathcal{N}=2$ we do find that increasing the amount of supersymmetry does not materially affect the story (if anything it makes it worse). The issue will turn out to be the dynamical bosonic fields that are present in the multiplets, which induce an effective UV divergence in the theory (despite it being quantum mechanics). The essential feature may already be seen in a simple bosonic tensor model which we revisit to provide some intuition \cite{Klebanov:2016xxf,Murugan:2017eto,Azeyanagi:2017drg,Azeyanagi:2017mre}.\footnote{ A theory of bosonic tensors with melonic vertices has a Hamiltonian that is unbounded from below. This feature while problematic will not affect the analysis we will undertake. Of course, this issue is mitigated in the supersymmetric context since the Hamiltonian being built from the supercharges will be bounded. \label{fn:bospot}}

Let us summarize some of the salient features of our analysis: we start with a quantum mechanical theory with tensor-valued $\mathcal{N}=2$ real superfields $\Psi^{a_1\, \cdots \,a_{q-1}}$ transforming in the fundamental representation of $O(N)^{q-1}$ (with $q\geq 4$). Apart from a canonical kinetic term we will include a single $q$-body superpotential term given by the melonic contraction, uplifting the fermion model of \cite{Klebanov:2016xxf} (who already mention our model as  a potential generalization). While the fermionic theory has a $q$-fermion vertex, our model has a melonic Yukawa term with fermions appearing at most bilinearly (and coupled thence to $q-2$ bosons).  Despite this change, we find that the system admits a (suitably regulated) RG flow that ends up at a non-trivial IR fixed point with emergent conformal invariance. The IR fixed point that we find however breaks supersymmetry -- the spectrum of singlet excitations does not fit into a supermultiplet. This is in contrast to the finite $N$ theory where we have unbroken supersymmetry (the theory has a non-trivial Witten index). 

Supersymmetry breaking at large $N$ is  of course possible as first illustrated in \cite{Affleck:1983pn}. One potential rationale has to be the emergence of a continuum in the spectrum owing to $N\to \infty$. A plausible mechanism may be attributed to the presence of $\mathcal{O}(N^2)$ light excitations in the theory arising from the global  $O(N)^{q-1}$ rotations of the tensor indices.\footnote{ We thank Steve Shenker for this suggestion.} This feature was illustrated explicitly for the fermionic uncolored tensor model in \cite{Choudhury:2017tax} with the light-modes being described by a non-linear sigma model  with target space being the group manifold for $O(N)^{q-1}$. It seems natural to conjecture that the supersymmetric theory will lead to a similar situation.

In our discussion however, it appears that there is an inherent tension between melonic dominance and supersymmetry.  We will see that the origins of supersymmetry breaking lie in having to explicitly regularize bosonic and fermionic degrees of freedom independently, lending credence to the idea that supersymmetry is broken explicitly along the RG flow rather than dynamically in the IR. This appears to be consistent with our analysis of the low energy spectrum which does not reveal the presence of a goldstino as would be the case with spontaneous breaking \cite{Witten:1981nf}.

We undertake a careful analysis of the model arguing for a particular regularization scheme that attains the IR fixed point identified from a naive solution of the truncated Schwinger-Dyson equations. Having established the existence of a non-supersymmetric fixed point, we turn to the spectrum of composite operators in the theory focusing on the singlet sector. In contrast to earlier studies of related systems we have both bosonic and fermionic composite operators. We work out the spectrum of excitations for both kinds of operators; doing so requires some new technical machinery to analyze fermionic excitations. Representing the four-point function in the bose-fermi OPE channel  involves a new set of conformal eigenfunctions. They can be viewed as $SL(2,{\mathbb R})$ wavefunctions with twisted boundary conditions or equivalently wavefunctions that are Hermitian with respect to a modified norm (we are not aware of this having been discussed in the literature before).

\paragraph{Outline of the paper:} The paper is organized as follows. We begin in  \S\ref{sec:BosonicKT} by  reconsidering  the bosonic tensor model. While this is not a viable quantum system as the potential has negative directions of $N>2$ (footnote \ref{fn:bospot}), it serves to illustrate the issues with the RG flow. We use it to argue for  our regularization scheme of the UV divergences  (present all along the flow) that are present for melonic tensor models with dynamical bosons. We regulate the UV divergences by fine-tuning a bare mass in the UV theory. This also serves to address  issues discussed in \cite{Azeyanagi:2017drg,Azeyanagi:2017mre,Murugan:2017eto} and noted in \cite{Klebanov:2016xxf} for such theories.

In  \S\ref{sec:SUSYKT}, we turn to our primary exhibit:  the $\mathcal{N}=2$ supersymmetric tensor model. We demonstrate that supersymmetry is unbroken for finite $N$ and then turn to the RG flow. We compute in \S\ref{sec:melons} the renormalized self-energy of the theory at large $N$ using the regularization scheme from  \S\ref{sec:BosonicKT} and exhibit a strong coupling IR fixed point where supersymmetry is broken. We also compute a set of 4-point functions for theory in \S\ref{sec:4pt}, taking the opportunity to generalize some results relating to generic external states. In particular, as we have both fermionic and bosonic fields, we will need $SL(2,{\mathbb R})$ wavefunctions with twisted boundary conditions; we derive these explicitly in the course of our analysis.

The appendices contain some additional observations about supersymmetric SYK and tensor models. In Appendix \ref{sec:N=1a4KT} we explore tensor models with different supersymmetries and in each case we find some tension with melonic dominance. 
Appendices~\ref{sec:sl2Rwf} and \ref{sec:useint} collect technical details relevant for the 4-point function computations. The former details the $SL(2,{\mathbb R})$ wavefunctions that we require for our analysis, while the latter summarizes a useful basis of integrals that enter into our computations.

\section{Bosonic tensor model revisited}
\label{sec:BosonicKT}

Let us consider bosonic tensors $\phi^{a_1 a_2\ldots a_{q-1}}$ with distinguishable indices $a_i=1,\cdots,N$ and the (Euclidean) action\footnote{ We will denote Euclidean time by $\tau$ and refer to real-time by $t$.}
\begin{equation}\label{eqn:bosonAction}
S=\int d\tau \left(\frac{1}{2}\, \partial_{\tau} \phi^{a_1\ldots a_{q-1}}\partial_{\tau} \phi^{a_1\ldots a_{q-1}}+ 
\frac{1}{q}\,
g\, [\phi^q] \right),
\end{equation}
where $[\phi^q]$ denotes the special type of index contraction, where each pair of fields has exactly one index contracted between them.
\footnote{ For $q > 6$, this choice of index contraction structure is not unique (see \cite{Gubser:2018yec} for a detailed analysis). However, every interaction of this type has the same large $N$ limit so we choose one such interaction for our model. We thank Grigory Tarnopolsky for discussions on this point.} 
For $q=4$ we have the tetrahedral index contraction:
\begin{equation}
[\phi^4]=\phi^{a_1a_2a_3} \phi^{a_1b_2c_3} \phi^{d_1a_2c_3} \phi^{d_1 b_2a_3}.
\end{equation}
As  noted earlier the vertex $[\phi^q]$ results in a Hamiltonian that is not bounded  from below. We will proceed for now ignoring this issue. It will be helpful to often simplify notation and suppress the tensor indices except when we need to illustrate particular contractions. To this end, let us collectively denote the tenor indices by an index $A_q$ and write $\phi^{A_q}$  for our basic field.\footnote{ We hope it is not overly confusing to keep track of the fact that $\phi^{A_q}$ only has $(q-1)$ tensor indices.} 
\begin{equation}
\phi^{A_q} \equiv \phi^{a_1\ldots a_{q-1}}
\label{eq:compactnot}
\end{equation}	
\begin{figure}[h]
\centering
\begin{tikzpicture}
\draw [thick](-2,0)--(2,0);
\draw [thick, fill=gray] (0,0) circle (0.5cm);
\draw (0,0) node{$G$};

\draw (2.5,0) node{$=$};

\draw [thick](3,0)--(6,0);
\draw (6.5,0) node{$+$};

\draw [thick](7,0)--(8,0);
\draw[thick] (8,0) to[out=60,in=210] (9,1);
\draw[thick, fill=gray] (9.5,1) circle (0.5cm);
\draw (9.5,1) node{$G$};
\draw[thick]  (10,1) to[out=-30,in=120] (11,0);
\draw (9.5,0) node{$\vdots$};

\draw[thick] (8,0) to[out=-60,in=150] (9,-1);
\draw[thick, fill=gray] (9.5,-1) circle (0.5cm);
\draw (9.5,-1) node{$G$};
\draw[thick]  (10,-1) to[out=30,in=-120] (11,0);

\draw [thick](11,0)--(12,0);
\draw[thick, fill=gray] (12.5,0) circle (0.5cm);
\draw (12.5,0) node{$G$};
\draw [thick](13,0)--(14,0);

\end{tikzpicture}
\caption{ The leading order large $N$ contribution to the boson propagator which leads to the Schwinger-Dyson equation \eqref{eqn:SDeqn}.}
\label{Fig:bosSD}
\end{figure}
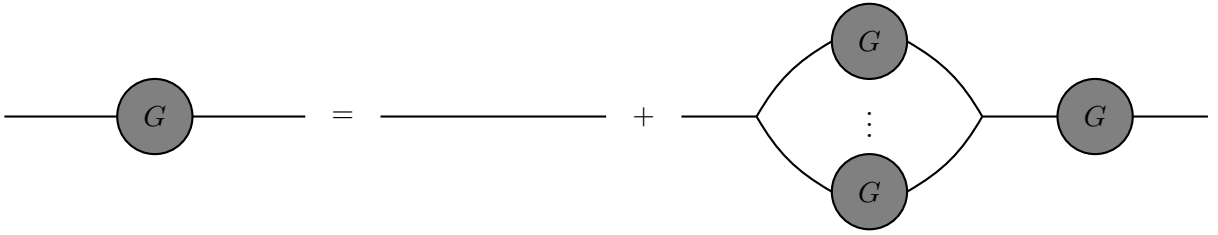

In the large $N$ limit, the theory is dominated by melon diagrams (see Fig.~\ref{Fig:bosSD}) with the dimension one effective coupling 
\begin{equation}
J\equiv g^\frac{2}{q+2} \,  N^{\frac{(q-1)(q-2)}{ 2(q+2)}}\,.
\label{eqn:Jbos}
\end{equation}	
Consider the two-point function
\begin{equation}
\begin{split}
\vev{\mathcal{T} \left(\phi^{A_q} (\tau_1)\, \phi^{B_q} (\tau_2)\right)} 
&=
	G(\tau_1-\tau_2)\, \delta^{A_q\, B_q} \\
&\equiv
G(\tau_1-\tau_2)\,\delta^{a_1 b_1}\,\cdots\,\delta^{a_{q-1} b_{q-1}}.
\end{split}
\end{equation}
The Green's function $G(\tau)$ can be solved by aid of the Schwinger-Dyson equation
\begin{equation}\label{eqn:SDeqn}
\widetilde G(\omega)= \frac{1}{ \omega^2-\widetilde\Sigma(\omega)},\qquad
\Sigma(\tau)=J^{q+2} G(\tau)^{q-1},
\end{equation}
where $\widetilde G(\omega)=\int d\tau\, e^{i\omega \tau}G(\tau)$ is the Fourier transform of $G(\tau)$ and similarly for $\widetilde \Sigma(\omega)$.

In the strong coupling limit or equivalently the low energy limit, the Schwinger-Dyson equation reduces to 
\begin{equation}\label{eqn:SDconformal}
\widetilde G_c(\omega)\widetilde\Sigma_c(\omega)=-1,\qquad\Sigma_c(\tau)=J^{q+2} G_c(\tau)^{q-1},
\end{equation}
which exhibits the reparametrization symmetry
\begin{equation}
\begin{split}
G_c(\tau_1-\tau_2)&\to\left[f'(\tau_1)f'(\tau_2)\right]^\frac{1}{q} G_c(f(\tau_1)-f(\tau_2)),
\\
\Sigma_c(\tau_1-\tau_2)&\to\left[f'(\tau_1)f'(\tau_2)\right]^\frac{q-1}{ q} \Sigma_c(f(\tau_1)-f(\tau_2)).
\end{split}
\label{eqn:reparametrization}
\end{equation}
Consider the conformal ansatz
\begin{equation}\label{eqn:ConformalAnsatz}
G_c(\tau)= \frac{b}{ |\tau|^{2\Delta}}.
\end{equation}
The equations \eqref{eqn:SDconformal} are solved by (we used \eqref{eqn:inta1})
\begin{equation}\label{eqn:bostensorsol}
\Delta={1\over q},\quad b^q J^{q+2}\pi=\left(\frac{1}{2} -{1\over q}\right) \cot \left(\frac{\pi }{q}\right).
\end{equation}

There is an apparent contradiction of this simple solution \cite{Azeyanagi:2017mre}. The conformal ansatz \eqref{eqn:ConformalAnsatz} is manifestly  positive everywhere, so the Fourier transforms $\widetilde G_c(\omega)$ and $\widetilde\Sigma_c(\omega)$ should both be positive functions. However, this contradicts the first equation in \eqref{eqn:SDconformal}.

The contradiction is due to the divergences in the Fourier integral of the conformal Green's function $G_c(\tau)$ and self-energy $\Sigma_c(\tau)$. The Fourier transform of $G_c(\tau)$ suffers from a long distance divergence, while the Fourier transform of $\Sigma_c(\tau)$ suffers from a short distance divergence. The long distance divergence can be easily regularized with an IR cut-off, e.g., by turning on a non-zero temperature.

 The conformal ansatz has a thermal regulator given by a reparametrization \eqref{eqn:reparametrization} which compactifies the real Euclidean time line to a circle. Using  $f(\tau)=\tan{\pi\tau\over \beta}$,
\begin{equation}\label{eqn:conformalG}
G_c(\tau)=b\left|\pi\over \beta\sin{\pi\tau\over \beta}\right|^{2\Delta}.
\end{equation}
The Fourier integral of $\widetilde G_c(\omega)$ is over a finite range $\tau\in [-{\beta\over 2},{\beta\over 2}]$ and therefore converges now. However, the  Fourier transform of $\Sigma_c(\tau)$ at finite temperature
\begin{equation}
\widetilde\Sigma_c(\omega_n)=J^{q+2}b^{q-1}\int^{\beta\over 2}_{-{\beta\over 2}}d\tau\cos(\omega_n\tau)\left|\pi\over \beta\sin{\pi\tau\over \beta}\right|^{2(q-1)\Delta},\quad \omega_n={2\pi n\over \beta},
\end{equation}
still suffers the short distance divergence at $\tau=0$. One can regularize the integral by first performing the integral for $2(q-1)\Delta<1$, and then analytic continuing the result to $\Delta={1\over q}$. In this regularization scheme, the function $\widetilde\Sigma_c(\omega_n)$ is everywhere negative, and the Schwinger-Dyson equations in the conformal limit \eqref{eqn:SDconformal} are satisfied.  

The solution we find at strong coupling has the following salient features. First, the self-energy at zero frequency gives an IR effective mass; 
using \eqref{eqn:inta2},
\begin{equation}
m^2_{\rm eff}=-\widetilde\Sigma_c(0)=J^{q+2}b^{q-1}\frac{\pi ^{2 \Delta  (q-1)-\frac{1}{2}}\Gamma \left((1-q)
   \Delta +\frac{1}{2}\right)}{ \beta ^{2 \Delta  (q-1)-1} \Gamma ((1-q) \Delta +1)}={(\beta J)^{1+{2\over q}}\over \beta^2}\left[\frac{2q \tan \frac{\pi
   }{q}}{\pi(q-2)}\right]^{\frac{1}{q}}\frac{ \pi ^{\frac{1}{2}}  \Gamma \left(\frac{q-1}{q}\right)}{  \Gamma \left(\frac{1}{2}-\frac{1}{q}\right)} .
\end{equation}
This self-energy correction vanishes in the zero temperature limit $\beta\to \infty$. Second, since the classical potential in the action \eqref{eqn:bosonAction} is not bounded from below the classical vacuum $\phi^{A_q}=0$ is an unstable critical point of the classical potential. The induced IR effective mass converts the point at $\phi^{A_q}=0$ to a metastable vacuum of the theory.

As pointed out by other authors \cite{Azeyanagi:2017drg,Azeyanagi:2017mre}, the Schwinger-Dyson equation \eqref{eqn:SDeqn} is still problematic away from the strong coupling limit. By unitarity, $\widetilde G(\omega)$ should be real and strictly positive. By the second equation in \eqref{eqn:SDeqn}, $\widetilde \Sigma(0)$ should also be positive.  However, the first equation in \eqref{eqn:SDeqn} at $\omega=0$ requires $\widetilde G(0)\widetilde \Sigma(0)=-1$. Relatedly, earlier attempts to solve the Schwinger-Dyson equation \eqref{eqn:SDeqn} by numerical iteration also consequently fail \cite{Murugan:2017eto}.

One can take inspiration from the strong coupling IR limit and enquire if  one can continue to attribute this tension to a divergent self-energy $\Sigma(\tau)$ even away from the conformal limit. We however need a different regularization scheme, for the analytic continuation of the conformal dimension $\Delta$ is only defined in the conformal limit. 

To resolve the contradiction, we need to fine tune the UV action \eqref{eqn:bosonAction}. Since the bosonic tensor field $\phi^{A_q}$ 
has mass dimension $-\frac{1}{2} $, the action \eqref{eqn:bosonAction} admits a relevant mass deformation
\begin{equation}
S_{\rm mass}=\int d\tau\,\frac{1}{2} m_{\rm bare}^2\, \phi^{A_q} \, \phi^{B_q} \, \delta_{A_q B_q} \,.
\end{equation}
Under the renormalization group flow, the bare mass $m_{\rm bare}$ would be renormalized. For the RG flow to end on a conformal fixed point, we would like to fine tune the bare mass such that in the low energy (strong coupling) limit the renormalized mass approaches the IR effective mass $m_{\rm eff}$,
\begin{equation}\label{eqn:renorCond}
\lim_{\beta J\to\infty}(\beta J)^{-1-{2\over q}}\left[m_{\rm bare}^2-\widetilde\Sigma(0)\right]={m^2_{\rm eff}\over(\beta J)^{1+{2\over q}}}={1\over \beta^2}\left[\frac{2q \tan \frac{\pi
   }{q}}{\pi(q-2)}\right]^{\frac{1}{q}}\frac{ \pi ^{\frac{1}{2}}  \Gamma \left(\frac{q-1}{q}\right)}{  \Gamma \left(\frac{1}{2}-\frac{1}{q}\right)}.
\end{equation}
There are many choices of the bare mass $m_{\rm bare}$ as a function of the dimensionless coupling $\beta J$ such that the renormalization condition \eqref{eqn:renorCond} is satisfied.  Different choices correspond to different UV theories which all flow to the same IR fixed point with the conformal two-point function \eqref{eqn:conformalG}. 

We pick the simplest possibility for the bare mass
\begin{equation}\label{eqn:renorChoice}
m_{\rm bare}^2=\widetilde \Sigma(0)+m^2_{\rm eff},
\end{equation}
which gives the renormalized Schwinger-Dyson equation
\begin{equation}\label{eqn:massiveSDeqn}
\widetilde G(\omega)={1\over \omega^2+m^2_{\rm eff}-\left[\widetilde\Sigma(\omega)-\widetilde\Sigma(0)\right]},\quad\Sigma(\tau)=J^{q+2} G(\tau)^{q-1}.
\end{equation}
Since only the difference of the self-energy $\widetilde\Sigma(\omega)-\widetilde\Sigma(0)$ appears in the equation, the Schwinger-Dyson equation is free from the short distance divergences in the Fourier integral. 

\begin{figure}[htbp]
\centering
\includegraphics[width=.5\textwidth]{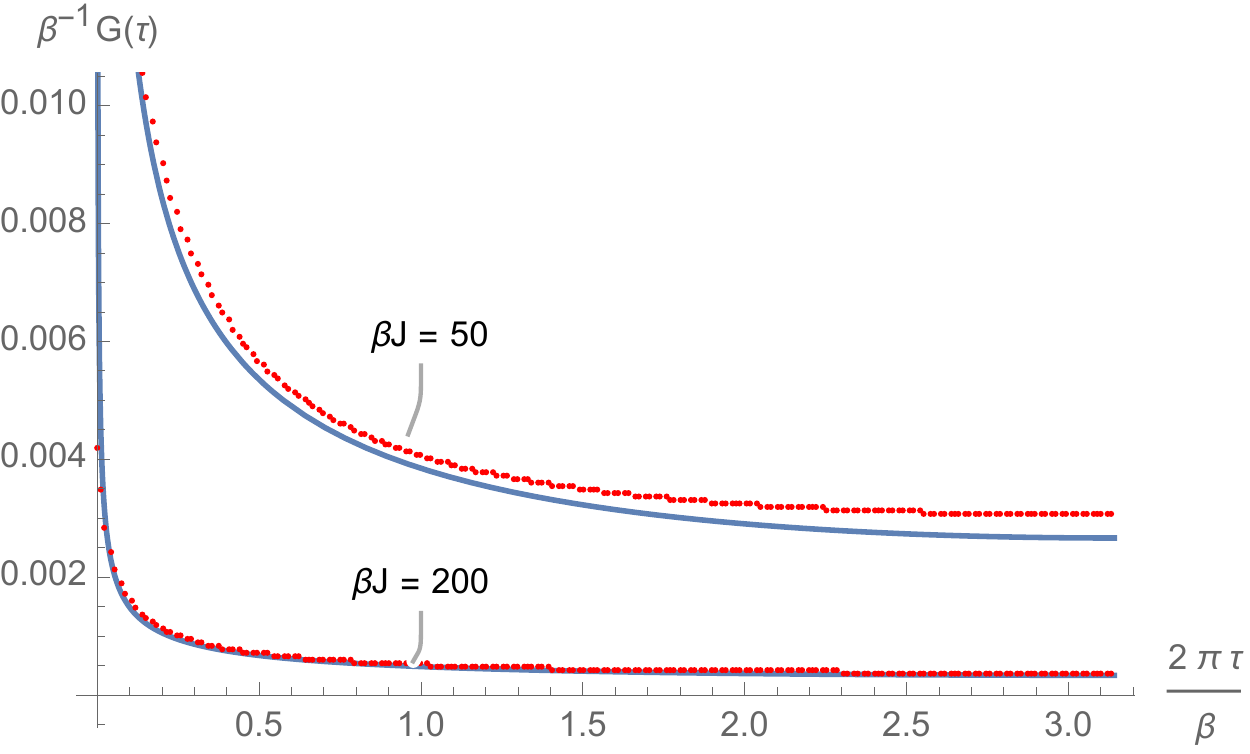}
\caption{Comparison of numerical (red) and analytic (blue) solutions of the regularized bosonic Schwinger-Dyson equations \eqref{eqn:massiveSDeqn} for two different values of $\beta J$ as indicated. The numerical simulation is carried out with the imaginary time circle discretized by a lattice with $200$ points (see footnote \ref{fn:resolution}). }
\label{Fig:bosG}
\end{figure}

To validate our renormalization condition \eqref{eqn:renorCond} (or equivalently \eqref{eqn:renorChoice}), we numerically solve the renormalized Schwinger-Dyson equation \eqref{eqn:massiveSDeqn}, and compare the numerical solution of large $\beta J$ with the analytic solution in the conformal limit \eqref{eqn:conformalG}. The result is shown in Fig.~\ref{Fig:bosG}.\footnote{ We have checked that increased resolution by working with say $\mathcal{O}(10^4)$ grid points as opposed to 200 grid points in Fig.~\ref{Fig:bosG} does not show any discernible  qualitative difference. We thank Douglas Stanford for raising this issue. \label{fn:resolution}}
As is clear from the plot the regulated Schwinger-Dyson equation converges clearly onto the anticipated IR fixed point, lending support for our procedure. 

Note that the problem is unique to bosonic degrees of freedom. Fermionic tensor models are much better behaved; indeed, the self-energy integral suffers from no UV divergence issues either in the conformal limit or along the flow. The reason can be traced to the Fermi statistics which in the IR limit give rise to a conformal propagator $G_c(\tau) = \frac{b}{|\tau|^{2\Delta}}\, \text{sgn}(\tau)$ at zero temperature. The sign function ends up ensuring the self-energy is free of divergences. 
We will take inspiration from this analysis for the case of the supersymmetric tensor model we introduce shortly.

\section{The $\mathcal{N}=2$ supersymmetric tensor model}
\label{sec:SUSYKT}

We now turn to the main model we wish to analyze, a quantum mechanical  supertensor model with $\mathcal{N}=2$ supersymmetry.  This amount of supersymmetry turns out to provide an interesting interaction term. Similar attempts to construct a theory with one supercharge lead to an interaction involving an odd number of fermions, while higher number of supercharges lead to derivative couplings between the component fields 
(see Appendix~\ref{sec:N=1a4KT}).

We will start by introducing the model. It will be convenient to start out in superspace ${\mathbb R}^{1|2}$ with  coordinates $t,\theta, \bar{\theta}$ ($t$ is the real time coordinate). The basic superfield $\Phi^{A_q}$ will be tensor-valued as in the bosonic model, so much of the structure is actually quite straightforward to intuit if we stick to superspace.

\subsection{The model}
\label{sec:n2}

We consider an $\mathcal{N}=2$ supersymmetric model in $(0+1)$-dimensions with superfields $\Phi^{A_q}$ transforming  in the $(q-1)$-fundamental representation 
of $O(N)^{q-1}$ for $q \geq 4$ even. These superfields can be written in terms of component fields on superspace as
\begin{equation}\label{eqn:superfield}
\Phi^{A_q}(t,\theta,\bar{\theta}) = \phi^{A_q}(t)+
i\,\theta\, \bar{\psi}^{A_q}(t)+i\,\bar{\theta}\, \psi^{A_q}(t)+ \theta\bar{\theta}\, F^{A_q}(t),
\end{equation}
where $\phi^{A_q} ,F^{A_q}$ are bosonic and $\psi^{A_q},\bar{\psi}^{A_q}$ are fermionic.

The action will be given as a superspace integral with canonical kinetic terms along with a  superpotential
 $W(\Phi^{A_q})$. Taking inspiration from the non-supersymmetric tensor models, the superpotential will be taken to be  the $q$-point interaction $[\Phi^{q}]$, the index contraction for the tensors being identical to the case of the bosonic model discussed in \S\ref{sec:BosonicKT}. Such a model was first proposed in \cite{Klebanov:2016xxf}. The action then is given by
\begin{equation}\label{eqn:susyAction}
S = \int dt\,d\bar{\theta}\,d\theta\, \left( 
	\frac{1}{2}D_\theta\Phi^{A_q}\;D_{\bar{\theta}}\Phi^{A_q}+\frac{1}{q}\, g\, [\Phi^{q}] 
	\right),
\end{equation}
where $D_\theta = \partial_{\theta}-i\, \bar{\theta}\, \partial_{t}$ and $D_{\bar{\theta}} = \partial_{\bar{\theta}}-i\, \theta\, \partial_{t}$ are the superderivations.  In terms of component fields, the action can be evaluated to be  
\begin{equation}
\begin{split}
S &= \int dt\,\frac{1}{2}\bigg(i\,\bar{\psi}^{A_q} \; \partial_{t}\psi^{A_q}   
		 -i \, \partial_{t}\bar{\psi}^{A_q}\; \psi^{A_q}  
		 + (\partial_{t}\phi^{A_q})^2 + (F^{A_q})^2  \bigg)
\\ & 
\hspace{2cm} +\; \frac{1}{q}\, g\, \bigg(\sum_{\mathrm{perms.}}[\phi^{q-2}\bar{\psi}\psi]+\sum_{\mathrm{perms.}}[\phi^{q-1}F]\bigg),
\end{split}
\label{eqn:susyActioncompfields}
\end{equation}
where the sums run over all possible rearrangements of the $\bar{\psi},\psi$ fields and the $F$ field, respectively, within the special contraction structure indicated by the square brackets. At this point, we could integrate out the auxiliary field. This will however induce scalar interaction terms with tensor contraction structure differing from the chosen one to ensure melonic dominance. While the end result will be equivalent, we prefer to leave the auxiliary field in place to make the melonic dominance manifest in the analysis to follow.

The supersymmetry generators are
\begin{equation}\label{eqn:susyGenerators}
Q = \partial_{\theta} + i\, \bar{\theta}\, \partial_{t} \qquad
\bar{Q} = \partial_{\bar{\theta}} + i\,\theta\, \partial_{t}
\end{equation}
with corresponding supersymmetry transformations
\begin{equation}
\label{eqn:susyTransformations}
\begin{split}
\delta \phi^{A_q} &=
	 i\, (\bar{\epsilon} \, \bar{\psi}^{A_q}+\epsilon \, \psi^{A_q})
\,,
\qquad \,
  \delta F^{A_q} = \bar{\epsilon}\, \partial_{t}\bar{\psi}^{A_q} 
   -\epsilon\,\partial_{t}\psi^{A_q}\,,
   	  \,\\
 \delta \psi^{A_q} &= 
 	\bar{\epsilon} \, (i\, F^{A_q}-\partial_{t}\phi^{A_q}) \,, 
 	\qquad
 \delta \bar{\psi}^{A_q} = 
 	\epsilon\, (-i\, F^{A_q}-\partial_{t}\phi^{A_q} ) \,. \\
 \end{split}
\end{equation}
Using the Noether procedure, we obtain the corresponding conserved supercharges,
\begin{equation}
\label{eqn:susyCharges}
\begin{split}
Q &= 
	\partial_{t}\phi^{A_q} \bar{\psi}^{A_q}+\frac{ig}{q}\sum_{\mathrm{perms.}}[\bar{\psi}\phi^{q-1}]
\\
\bar{Q} &= 
	\partial_{t}\phi^{A_q} \psi^{A_q}-\frac{ig}{q}\sum_{\mathrm{perms.}}[\psi\phi^{q-1}].
\end{split}
\end{equation}
The Hamiltonian is, of course, $H = \frac{1}{2}\{Q,\bar{Q}\}$ and it now has a bounded spectrum (unlike the bosonic tensor model considered in \S\ref{sec:BosonicKT}).

 The theory has a $U(1)$ R-symmetry under which $\psi^{A_q}$ has charge $+1$ and $\bar{\psi}^{A_q}$ has charge $-1$, while $\phi^{A_q}$ and $F^{A_q}$ are uncharged. The supercharges are normalized such that $Q$ and $\bar{Q}$ have $R$-charge $-1$ and $+1$, respectively.

\subsection{Hilbert space at finite $N$}
\label{sec:}

We first turn to an examination of the theory at finite $N$ where we expect usual behaviour as a supersymmetric quantum mechanical theory. First, let us examine the Witten index  to determine whether or not supersymmetry is broken and get a sense for the BPS sector of the theory. Since the Witten index is invariant under deformations of the theory, we can compute it in the free theory where $g=0$ \cite{Witten:1982df}. This is easy to do for we can evaluate the free partition function on a Euclidean circle with periodic fermions, which ensures that we are evaluating $\Tr{\left((-1)^F \, e^{-\beta H}\right)}$, with the Boltzmann factor providing a suitable regulator. 
Integrating out the auxiliary field, we obtain (suppressing tensor indices for convenience)
\begin{equation}
\label{eqn:WittenIndex}
\begin{split}
\Tr(-1)^{F} 
&= 
	\int\mathcal{D}\phi\,\mathcal{D}\psi\,\mathcal{D}\bar{\psi}|_{P}\; e^{-S_{E}[\phi,\psi,\bar{\psi}]} \\
&= 
	\bigg[\int\mathcal{D}\phi\,e^{-\frac{1}{2}
	\int_{0}^{\beta} dt\,\phi(-\partial_{t}^2)\phi}\int\mathcal{D}\psi\,\mathcal{D}\bar{\psi}|_{P}e^{-\int_{0}^{\beta}dt\,\bar{\psi}(\partial_{t}\psi)}\bigg]^{N^{q-1}}
		 \\	&= (-1)^{N} \neq 0\,.
\end{split}
\end{equation}
In the final expression we have used the fact that the parity of $N^{q-1}$ equals the parity of $N$ for any even $q$. From the non-vanishing Witten index, we can conclude that supersymmetry is not broken in the theory at finite $N$.

One can check this computation by explicitly constructing the BPS sector. From the canonical quantization of 
the fermions $\psi^{A_q},\bar{\psi}^{A_q}$, we have the Hilbert space for a given $\psi^{A_q}$ (i.e., with fixed tensor components): 
\begin{equation}
\mathcal{H}^{A_q} = 
	L^{2}(\mathbb{R},\mathbb{C})\, \ket{0}_{A_q} \oplus 
	L^{2}(\mathbb{R},\mathbb{C})\, \bar{\psi}^{A_q}\, \ket{0}_{A_q}, \qquad \psi^{A_q}\ket{0}_{A_q} = 0.
\label{eq:}
\end{equation}	
 Thus, the full Hilbert space of the theory is obtained by summing over all possible tensors 
\begin{equation}
\mathcal{H} = 
	\bigotimes_{a_{1}, \ldots, a_{q-1}=1}^{N}\bigg(L^{2}(\mathbb{R},\mathbb{C})\, \ket{0}_{A_q} 
		\oplus L^{2}(\mathbb{R},\mathbb{C})\,\bar{\psi}^{A_q}\, \ket{0}_{A_q}\bigg).
\label{eq:}
\end{equation}	

 To determine the $Q$-cohomology, we seek states $\ket{\chi}$ such that $Q\ket{\chi}=\bar{Q}\ket{\chi} = 0$. 
 One can show that there exists only one such state 
\begin{equation}
\ket{\chi} = 
	\exp \left(
	-\sum_{i=1}^{q-1}\, \sum_{a_i=1}^{N}\, \frac{g}{q} \,
	\int d\phi^{A_q}  \frac{d}{dF^{A_{q}}}[F\, \phi^{q-1}] \right) 
	\displaystyle \bigotimes_{a_i=1}^{N}\;  \bar{\psi}^{A_q}\, \ket{0}_{A_q} ,
\label{eq:susyvac}
\end{equation}	
The statistics of the state is determined by the parity of $N^{q-1}$ as can be see from the fermion creation operator count.  This agrees with the Witten index computation and we conclude that, at finite $N$, there exists one supersymmetric ground state whose parity depends on the parity of $N$.

However, the arguments used above in the computation of the Witten index and $Q$-cohomology can potentially  break down as $N \to \infty$. Usually this is associated with the appearance of a new continuum in the spectrum
or the vacuum running away to infinity, as is well documented in large $N$ quantum mechanical models \cite{Affleck:1983pn}. To understand potential issues arising in the large $N$ limit, it will suffice to examine the spectrum of the theory as carried out in \cite{Bulycheva:2017ilt,Choudhury:2017tax,Klebanov:2018nfp}. These authors find that the theory admits $\mathcal{O}(N^2)$ light modes in the spectrum generically. One way to intuit their presence is to realize that the theory in the absence of the kinetic term actually admits a large global symmetry group $O(N)^{q-1}$. Away from the IR limit, the irrelevant kinetic term breaks this explicitly and leaves behind a set of Goldstone fields which may be associated with  time-dependent $O(N)^{q-1}$ rotations. The presence of these modes has the potential to open up a continuum in the large $N$ limit, spoiling our analysis of the Witten index (by invalidating the localization argument used to set $g=0$ in the Witten index computation \cite{Witten:1982df}). We also see another sign of trouble in the norm of the supersymmetric ground state \eqref{eq:susyvac} vanishing in the large $N$ limit. 
In fact, soon we will find that the  low energy fixed point obtained by assuming melonic dominance  prefers to be non-supersymmetric.

\section{Melonic dominance and low energy conformal symmetry}
\label{sec:melons}

We now have all the ingredients at hand to analyze the dynamical behaviour of the model \eqref{eqn:susyAction} as a function of the coupling $g$. To this end we will first  compute the  two-point functions for the fundamental fields 
$\phi,\psi$ and $F$ for general $q$. Since it will be helpful to work in superspace directly, let us denote by $X$  the supercoordinate $X \equiv(\tau,\theta,\bar\theta)$. 

Consider then the  two-point function of the superfield $\Phi^{A_q}(X)\equiv \Phi^{a_1\ldots a_{q-1}}(X)$
\begin{equation}
\begin{split}
\mathcal{G}(X_{1},X_{2})& ={1\over N^{q-1}} \sum_{a_i=1}^N\langle \mathcal{T}\,(\Phi^{A_q}(X_{1})\Phi^{A_q}(X_{2})) \rangle,
\end{split}
\end{equation}
which can be expanded in terms of two-point functions of the component fields as
\begin{equation}
\mathcal{G}(X_{1},X_{2})=G^{\phi\phi}(\tau_{12})+\bar\theta_1\theta_2G^{\psi\bar\psi}(\tau_{12})+\theta_1\bar\theta_2 G^{\bar\psi\psi}(\tau_{12})+\theta_1\bar\theta_1\theta_2\bar\theta_2 G^{FF}(\tau_{12}),
\end{equation}
where $G^{\phi\phi}(\tau_{12})$, $G^{\psi\bar\psi}(\tau_{12})$, and $G^{FF}(\tau_{12})$ are
\begin{equation}
\begin{split}
&G^{\phi\phi}(\tau_{12})={1\over N^{q-1}} \sum_{a_i=1}^N\langle \mathcal{T}\,(\phi^{A_q} (\tau_{1})\phi^{A_q} (\tau_{2})) \rangle,
\\
&G^{\psi\bar\psi}(\tau_{12})={1\over N^{q-1}} \sum_{a_i=1}^N\langle \mathcal{T}\,(\psi^{A_q}(\tau_{1})\bar\psi^{A_q}(\tau_{2})) \rangle,
\\
&G^{\bar\psi\psi}(\tau_{12})={1\over N^{q-1}} \sum_{a_i=1}^N\langle \mathcal{T}\,(\bar\psi^{A_q}(\tau_{1})\psi^{A_q}(\tau_{2})) \rangle,
\\
&G^{FF}(\tau_{12})={1\over N^{q-1}} \sum_{a_i=1}^N\langle \mathcal{T}\,(F^{A_q}(\tau_{1})F^{A_q}(\tau_{2})) \rangle.
\end{split}
\end{equation}

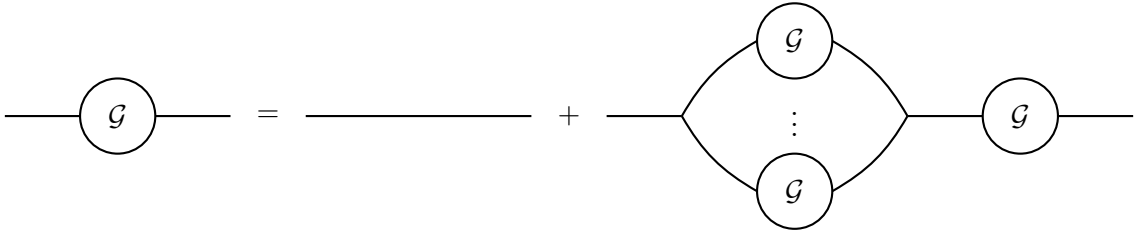
\begin{figure}[h]
\centering
\begin{tikzpicture}
\draw [thick](0,0)--(1,0);
\draw [thick](3,0)--(2,0);
\draw[thick] (1.5,0) circle (.5cm);
\draw (1.5,0) node{${\cal G}$};

\draw (3.5,0) node{$=$};

\draw [thick](4,0)--(7,0);
\draw (7.5,0) node{$+$};

\draw [thick](8,0)--(9,0);

\draw[thick] (9,0) to[out=60,in=210] (10,1);
\draw[thick] (10.5,1) circle (.5cm);
\draw (10.5,1) node{${\cal G}$};
\draw[thick]  (11,1) to[out=-30,in=120] (12,0);
\draw (10.5,0) node{$\vdots$};

\draw[thick] (9,0) to[out=-60,in=150] (10,-1);
\draw[thick] (10.5,-1) circle (.5cm);
\draw (10.5,-1) node{${\cal G}$};
\draw[thick]  (11,-1) to[out=30,in=-120] (12,0);

\draw [thick](12,0)--(13,0);
\draw[thick] (13.5,0) circle (.5cm);
\draw (13.5,0) node{${\cal G}$};
\draw [thick](14,0)--(15,0);

\end{tikzpicture}
\caption{Supergraph representation of the super-Schwinger-Dyson equation.}
\label{Fig:SD}
\end{figure}

An advantage of working directly with the superfields is that it is obvious that the large $N$ counting works in a manner similar to the bosonic model discussed in \S\ref{sec:BosonicKT}. We can immediately write down the super-Schwinger-Dyson equations satisfied by the super-propagator:
\begin{equation}\label{eqn:superSDeqn}
\frac{1}{2}[D_{\theta_{1}},D_{\bar{\theta}_{1}}]\,\mathcal{G}(X_{1},X_{3}) - 
\int dX_{2}\,\Sigma(X_{1},X_{2})\,\mathcal{G}(X_{2},X_{3}) = \delta(\tau_{13})\delta^2(\theta_{1}-\theta_{3}) \,,
\end{equation}
which is given in the large $N$ limit by an iterated sum over melon diagrams Fig.~\ref{Fig:SD}, viz., 
\begin{equation}\label{eqn:superSDeqn}
\begin{split}
\mathcal{G}(X_{1},X_{2}) &= \mathcal{G}_0(X_{1},X_{2})+\int dX_{3}\,dX_{4}\,\mathcal{G}_0(X_{1},X_{3})\,
\Sigma(X_{3},X_{4}) \, \mathcal{G}(X_{4},X_{2}) \,, \\
\Sigma(X_{1},X_{2}) &= J^{q}\,\mathcal{G}(X_{1},X_{2})^{q-1}\,, \qquad 
J\equiv g^{2\over q} N^{(q-1)(q-2)\over 2q} \,.
\end{split}
\end{equation}
In the above the free super-propagator $\mathcal{G}_0(X_1,X_2)$ is given by 
\begin{equation}
\mathcal{G}_0(X_{1},X_{2})=-\frac{1}{2} |\tau_{12}-\bar{\theta}_1\theta_2-\theta_1\bar{\theta}_2|
\label{eq:}
\end{equation}	
and can be obtained from solving the  free equation of motion $\frac{1}{2} [\,D_{\theta_1},D_{\bar{\theta}_1}]\mathcal{G}_0(X_{1},X_{2})=\delta^2(\theta_{12})\delta(\tau_{12})$.
Expanding out the super-Schwinger-Dyson equations gives three coupled Schwinger-Dyson equations for the component fields: 
\begin{equation}\label{eqn:compSDeqn}
\begin{split}
		-\partial_{\tau_1}^2 G^{\phi\phi}(\tau_{12}) &= \delta(\tau_{12}) + \int d\tau_{3}\,\Sigma^{\phi\phi}(\tau_{13})G^{\phi\phi}(\tau_{32}),
		\\  \partial_{\tau_1}G^{\psi\bar\psi}(\tau_{12}) &= \delta(\tau_{12}) + \int d\tau_{3}\,\Sigma^{\psi\bar\psi}(\tau_{13})G^{\psi\bar\psi}(\tau_{32}),
		\\ -G^{FF}(\tau_{12}) &= \delta(\tau_{12}) + \int d\tau_{3}\,\Sigma^{FF}(\tau_{13})G^{FF}(\tau_{32}),
\end{split}
\end{equation}
or equivalently, 
\begin{equation}\label{eqn:compSDeqn}
\begin{split}
		G^{\phi\phi}(\tau_{1},\tau_{2}) &= G_{0}^{\phi\phi}(\tau_{1},\tau_{2}) + \int d\tau_{3}\,d\tau_{4}\,G_{0}^{\phi\phi}(\tau_{1},\tau_{3})\Sigma^{\phi\phi}(\tau_{3},\tau_{4})G^{\phi\phi}(\tau_{4},\tau_{2})
		\\ G^{\psi\bar\psi}(\tau_{1},\tau_{2}) &= G_{0}^{\psi\bar\psi}(\tau_{1},\tau_{2}) + \int d\tau_{3}\,d\tau_{4}\,G_{0}^{\psi\bar\psi}(\tau_{1},\tau_{3})\Sigma^{\psi\bar\psi}(\tau_{3},\tau_{4})G^{\psi\bar\psi}(\tau_{4},\tau_{2})
		\\ G^{FF}(\tau_{1},\tau_{2}) &= G_{0}^{FF}(\tau_{1},\tau_{2}) + \int d\tau_{3}\,d\tau_{4}\,G_{0}^{FF}(\tau_{1},\tau_{3})\Sigma^{FF}(\tau_{3},\tau_{4})G^{FF}(\tau_{4},\tau_{2}) \,.
\end{split}
\end{equation}
The explicit form for the self-energy functions is given by: 
\begin{equation}\label{eqn:selfEnergies}
\begin{split}
\Sigma^{\phi\phi}(\tau) &= J^{q}\Big((q-1)(q-2)G^{\phi\phi}(\tau)^{q-3}G^{\psi\bar\psi}(\tau)G^{\bar\psi\psi}(\tau)+(q-1)G^{\phi\phi}(\tau)^{q-2}G^{FF}(\tau)\Big),
\\ 
\Sigma^{\psi\bar\psi}(\tau) &= J^{q}(q-1)G^{\psi\bar\psi}(\tau)G^{\phi\phi}(\tau)^{q-2},
\\ 
\Sigma^{FF}(\tau) &= J^{q}G^{\phi\phi}(\tau)^{q-1}.
\end{split}
\end{equation}
%

\subsection{The conformal fixed point and IR symmetries}
\label{sec:cfix}

These equations can be solved by standard techniques. For start, we transform to frequency space and pass to the IR limit, or equivalently the strong coupling limit, to obtain the simplified Schwinger-Dyson equations
\begin{equation}\label{eqn:SDeqnsIRlimit}
\widetilde G_c^{\phi\phi}(\omega)\widetilde\Sigma_c^{\phi\phi}(\omega) = \widetilde G_c^{\psi\bar\psi}(\omega)\widetilde \Sigma_c^{\psi\bar\psi}(\omega) = \widetilde G_c^{FF}(\omega)\widetilde\Sigma_c^{FF}(\omega) = -1,
\end{equation}
where the overhead $\sim$ denotes the Fourier transform and the subscript $c$ denotes the conformal limit.

\paragraph{The fixed point solution:} We can now attempt to solve the truncated equations assuming a flow to a conformal fixed point by picking an  ans\"atze\footnote{ Note that the coefficients $b_\phi$, $b_\psi$ and $b_F$ are dimensionful, and the dimensionless combinations are $b_\phi J^{2\Delta_\phi+1}$, $b_\psi J^{2\Delta_\psi}$ and $b_F J^{2\Delta_F-1}$.} 
\begin{equation}\label{eqn:conformalAnsatz}
G_c^{\phi\phi}(\tau) = \frac{b_{\phi}}{ |\tau|^{2\Delta_{\phi}}}, \qquad 
G_c^{\psi\bar\psi}(\tau) = \frac{b_{\psi}\, \mathrm{sgn}(\tau)}{|\tau|^{2\Delta_{\psi}}}, \qquad 
G_c^{FF}(\tau) = \frac{b_{F}}{|\tau|^{2\Delta_{F}}}.
\end{equation}
This implies that $G^{\psi \bar{\psi}}_c(\tau) = G^{\bar{\psi}\psi}_c(\tau)$. Plugging (\ref{eqn:conformalAnsatz}) into (\ref{eqn:SDeqnsIRlimit}) gives the constraints on the conformal dimensions
\begin{equation}\label{eqn:confdimConstraints}
(q-2)\Delta_{\phi}+2\Delta_{\psi} = 1 \qquad \mathrm{and} \qquad (q-1)\Delta_{\phi} + \Delta_{F} = 1,
\end{equation}
along with an additional constraint
\begin{equation}\label{eqn:extraConstraint}
\begin{split}
1 - &(q-1)\frac{\sin^2(\pi\Delta_{\phi})\Gamma(1-2\Delta_{\phi})\Gamma(2\Delta_{\phi}-1)}{\sin^{2}(\pi(q-1)\Delta_{\phi})\Gamma(2(q-1)\Delta_{\phi}-1)\Gamma(1-2(q-1)\Delta_{\phi})} =
		\\	& \qquad \qquad \qquad \qquad \qquad \qquad (q-2)\frac{\Gamma(1-2\Delta_{\phi})\Gamma(2\Delta_{\phi}-1)\sin^2(\pi\Delta_{\phi})}{\Gamma((q-2)\Delta_{\phi})\Gamma((2-q)\Delta_{\phi})\sin^2(\frac{\pi}{2}(q-2)\Delta_{\phi})}.
\end{split}
\end{equation}
The solutions of (\ref{eqn:extraConstraint}) for various values of $q$ are given in Table~\ref{table:confDims}.
It is clear from  the relation between conformal dimensions \eqref{eqn:confdimConstraints}, even without inspecting the explicit solutions, that the low energy fixed point breaks supersymmetry. 

\begin{table}[h]
\begin{center}
	\begin{tabular}{| l | l | l | l |}
	\hline
	$q$ & $\Delta_{\phi}$ & $\Delta_{\psi}$ & $\Delta_{F}$ \\
	\hline
	$4$ & $\frac{1}{6}$ & $\frac{1}{3}$ & $\frac{1}{2}$ \\
	$6$ & $0.109$ & $0.282$ & $0.456$ \\
	$8$ & $0.081$ & $0.258$ & $0.436$ \\
	$100$ & $0.006$ & $0.196$ & $0.386$ \\
	$\infty$ & 0 & $0.191$ & $0.382$ \\
	\hline
	\end{tabular}
	\caption{Conformal dimensions of fields for various values of $q$.}
	\label{table:confDims}
\end{center}
\end{table}

The low energy equations \eqref{eqn:SDeqnsIRlimit} are invariant under a scaling symmetry:
\ie\label{eqn:globalScaling}
&G^{\phi \phi}(\tau_1,\tau_2)\to \lambda^4G^{\phi\phi}(\tau_1,\tau_2),& G^{FF}(\tau_1,\tau_2)\to \lambda^{4(1-q)}G^{FF}(\tau_1,\tau_2),
\\
&G^{\psi \bar\psi}(\tau_1,\tau_2)\to \lambda^{2(2-q)}G^{\psi \bar\psi}(\tau_1,\tau_2),& G^{\bar\psi \psi}(\tau_1,\tau_2)\to \lambda^{2(2-q)}G^{\bar\psi \psi}(\tau_1,\tau_2).
\fe	
Hence, the coefficients $b_\phi$, $b_\psi$, $b_F$ are not determined completely. Only the products $b_\phi^{q-2}b_\psi^2$ and $b_\phi^{q-1}b_F$ are fixed
\begin{equation}\label{eqn:bsol}
\begin{split}
 b_\phi^{q-2}b_\psi^2J^q&=\frac{(q-2) \left(\Delta _F-1\right) \cot \left(\frac{\pi  (q-2) \left(\Delta
   _F-1\right)}{2 (q-1)}\right)}{2 \pi(q-1)^2},
\\
 b_\phi^{q-1}b_FJ^q&=\frac{1}{2 \pi }\left(1-2 \Delta _F\right)
   \cot \left(\pi  \Delta _F\right).
 \end{split}
\end{equation}
Note that a similar statement holds for the supersymmetric SYK model discussed in \cite{Fu:2016vas}, though there one can further use supersymmetry to fix this additional parameter. We do not have this additional freedom.

\paragraph{Local symmetries in the IR:} In the low energy limit, the truncated Schwinger-Dyson equations have a large set of local symmetries. These are typically broken by the kinetic term which we ignore in the deep infrared. Let us record the symmetries that are visible in the truncated theory for future reference:
\begin{itemize}
\item The time reparametrization symmetry discussed in the bosonic model, \eqref{eqn:reparametrization} continues to apply in the supersymmetric Schwinger-Dyson equations. The breaking of this symmetry by the UV dynamics leads to the Schwarzian action \cite{Maldacena:2016hyu} for the lone reparametrization mode.  
\item In the deep infrared we have an $U(1)$ affine algebra arising as a low-energy version of the $U(1)_R$ symmetry. This acts on the bilinear-propagator fields as
\begin{equation}
G^{\psi \bar\psi}(\tau_1,\tau_2)\to e^{ia(\tau_1)-ia(\tau_2)}G^{\psi \bar\psi}(\tau_1,\tau_2),\quad G^{\bar\psi \psi}(\tau_1,\tau_2)\to e^{-ia(\tau_1)+ia(\tau_2)}G^{\bar\psi \psi}(\tau_1,\tau_2).
\label{eq:u1affine}
\end{equation}	
The reality condition $G^{\bar\psi \psi}(\tau_1,\tau_2)=G^{\psi \bar\psi}(\tau_1,\tau_2)^*$ implies that $a(\tau)$ is a real function. The effective action for $a(\tau)$ can be inferred from standard analysis and is similar to the discussions of the charged SYK model  \cite{Davison:2016ngz}. 

\item The theory has in addition a scaling symmetry identified in \eqref{eqn:globalScaling} which entails that we only have enough information to fix two of the three parameters in the Green's function. cf., \eqref{eqn:bsol}. This symmetry acts locally in the IR as:
\begin{equation}
\begin{split}
&G^{\phi \phi}(\tau_1,\tau_2)\to \left[\lambda(\tau_1)\lambda(\tau_2)\right]^{2}G^{\phi\phi}(\tau_1,\tau_2),\quad G^{FF}(\tau_1,\tau_2)\to \left[\lambda(\tau_1)\lambda(\tau_2)\right]^{2-2q}G^{FF}(\tau_1,\tau_2),
\\
&G^{\psi \bar\psi}(\tau_1,\tau_2)\to \left[\lambda(\tau_1)\lambda(\tau_2)\right]^{2-q}G^{\psi \bar\psi}(\tau_1,\tau_2),\quad G^{\bar\psi \psi}(\tau_1,\tau_2)\to \left[\lambda(\tau_1)\lambda(\tau_2)\right]^{2-q}G^{\bar\psi \psi}(\tau_1,\tau_2).
\end{split}
\label{eqn:scaling}
\end{equation}	 
Unlike the reparametrization symmetry and $U(1)$ affine symmetry, the global part of the local scaling symmetry does not leave the low energy solution \eqref{eqn:conformalAnsatz} invariant. Following \cite{Fu:2016vas} we expect an effective action of the form $J\int (\lambda(\tau)-1)^2 d\tau$, which suppresses the deviation from the value of $\lambda$ determined by the UV.
\end{itemize}

\subsection{The RG flow and supersymmetry breaking}
\label{sec:rgflow}
 As with the bosonic model discussed in \S\ref{sec:BosonicKT} the attainment of the conformal fixed point is predicated upon suitable fine-tuning in the system. The issue again is to due to divergences arising in the bosonic sector which remain despite the presence of supersymmetry. This is another sign that the melonic structure in this class of supertensor theories does not gel with the supersymmetry. Inspired by our bosonic model discussion we will now present the mass counter-terms we include to ensure that the flow starting from the free UV theory lands on the fixed point we picked out from the truncation of the Schwinger-Dyson equations.

At finite temperature, the boson $\phi$ and auxiliary field $F$ acquire IR effective masses given by the self-energies at zero frequency\footnote{ The self-energy of fermion $\psi$ has no zero frequency limit because the frequency is half-integer quantized. The effective mass of the auxiliary field $F$ is defined such that the renormalized action contains the mass term
\begin{equation}
\int^{\beta}_{0} d\tau\, \frac{1}{2} \,\beta^{2}\, (m^{F}_{\rm eff})^2\, (F^{A_q})^2,
\end{equation}
where the explicit $\beta$ is included to preserve the classical dimension of $F$.}
\begin{equation}\label{eqn:effMasses}
\begin{split}
(m^{\phi}_{\rm eff})^2&=-\widetilde\Sigma^{\phi\phi}_c(0)={1\over \widetilde G^{\phi\phi}_c(0)}=\frac{( \beta J) ^{2 \Delta_\phi+1} \Gamma (1- \Delta_\phi )}{\beta^2(b_\phi J^{2\Delta_\phi+1}) \pi ^{2 \Delta_\phi-\frac{1}{2}}\Gamma \left(\frac{1}{2}-\Delta_\phi \right)},
\\
(m^{F}_{\rm eff})^2&=-\beta^{-2}\widetilde\Sigma^{FF}_c(0)={1\over \beta^2\widetilde G^{FF}_c(0)}=\frac{ (\beta J)^{2 \Delta_F-1} \Gamma (1- \Delta_F )}{\beta^2(b_FJ^{2\Delta_F-1})\pi ^{2 \Delta_F-\frac{1}{2}}\Gamma \left(\frac{1}{2}-\Delta_F \right)},\end{split}
\end{equation}
 which go to zero in the zero temperature limit $\beta \to \infty$ while fixing the dimensionless combinations $\beta J$, $b_\phi J^{2\Delta_\phi+1}$, and $b_FJ^{2\Delta_F-1}$.

\begin{figure}[h]
\centering
\subfloat{
\includegraphics[width=.5\textwidth]{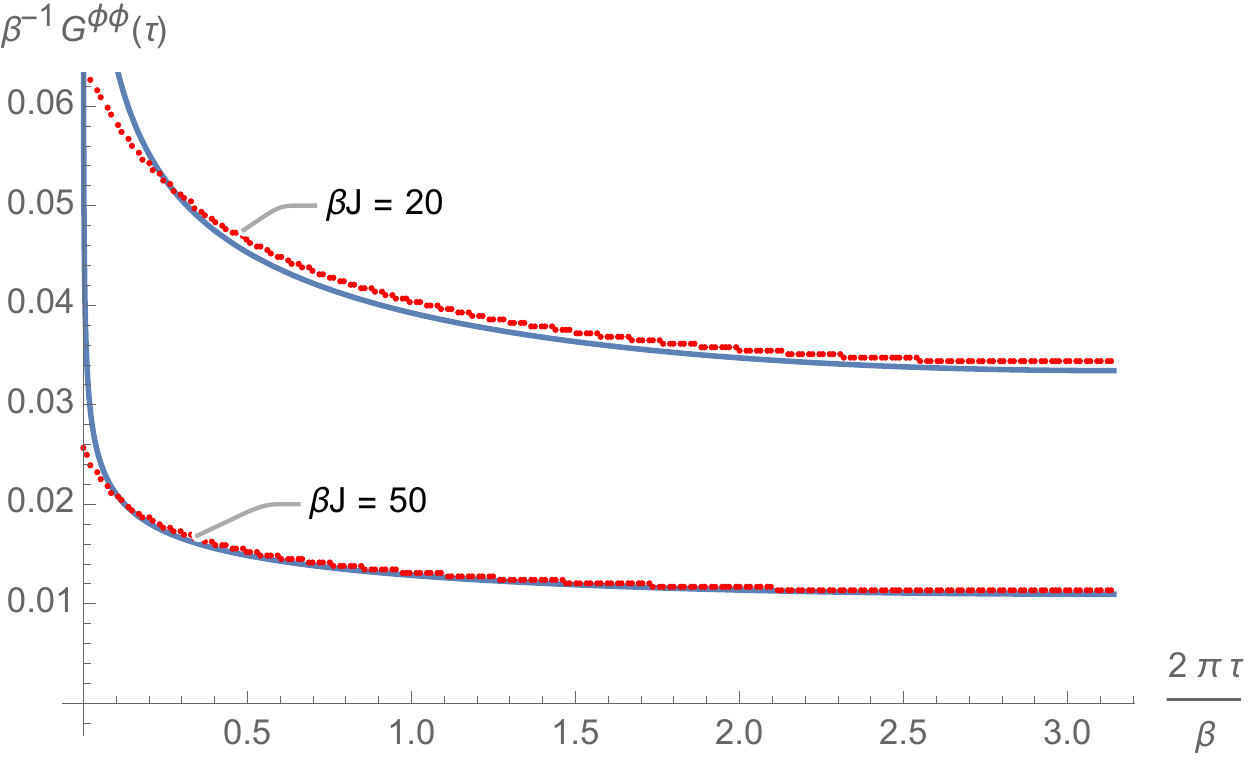}
}
\subfloat{
\includegraphics[width=.5\textwidth]{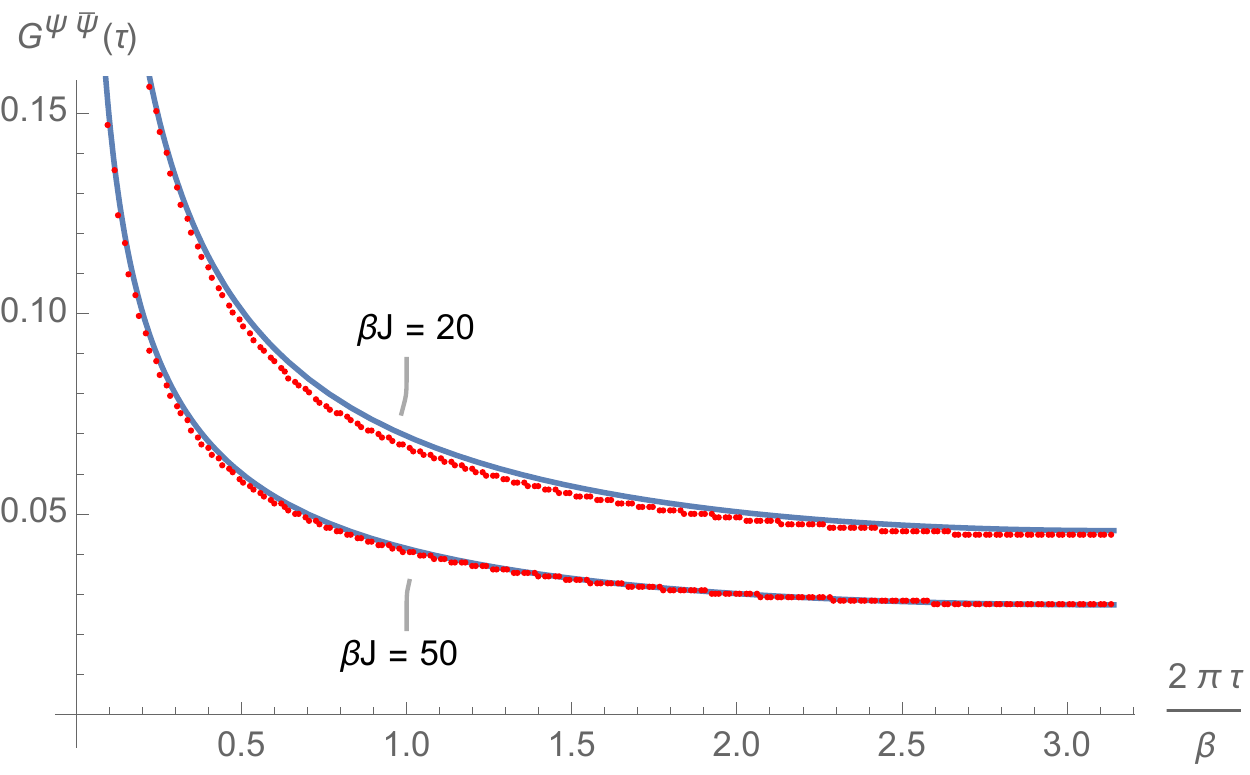}
}
\\
\subfloat{
\includegraphics[width=.5\textwidth]{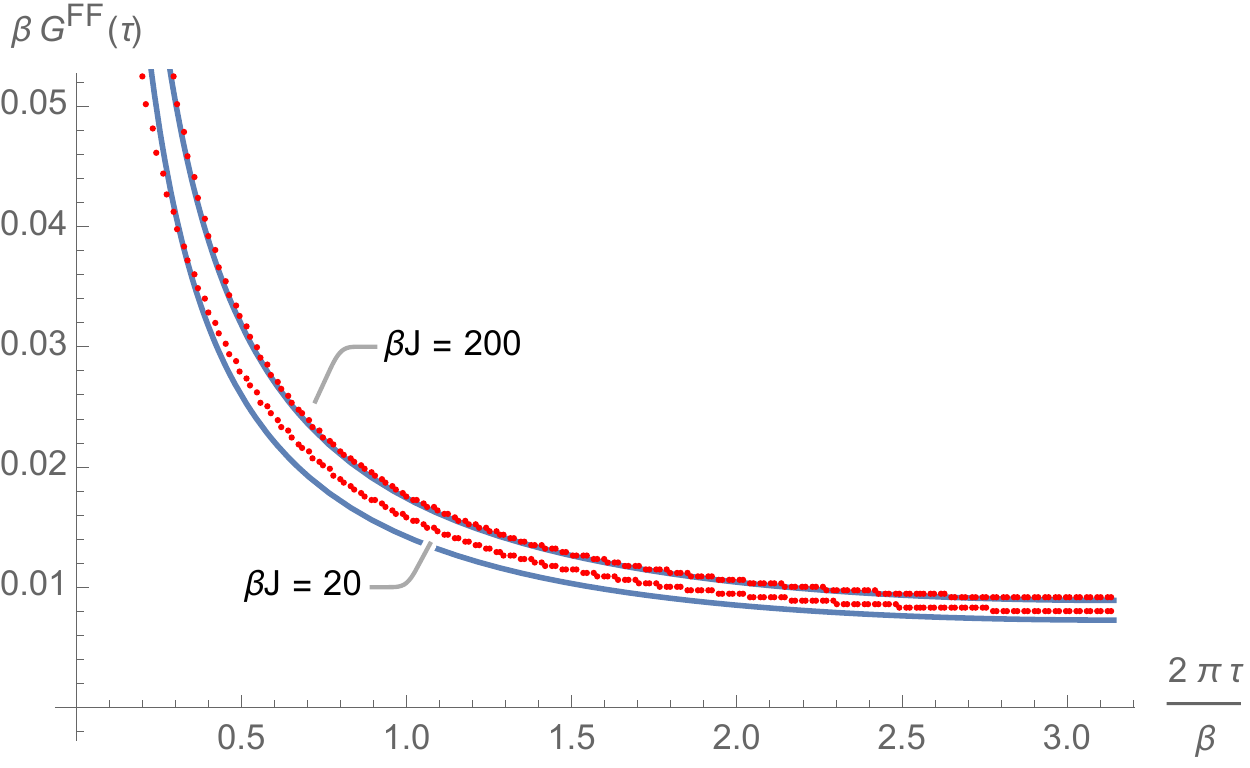}
}
\caption{Comparison of numerical (red) and analytic (blue) solutions of the supersymmetric Schwinger-Dyson equations for $q=6$ and $b_\phi \, J^{2\Delta_\phi+1}=1$ (for different choices of the dimensionless coupling $\beta J$ as indicated). The numerical calculation is still with a discretized temporal grid with 200 points (see footnote \ref{fn:resolution} for comments on increasing the resolution).  For $G^{FF}$ we have specifically chosen a larger value of $\beta J$ to separate out the curves for ease of visualization. 
}
\label{Fig:susyKTG}
\end{figure}

Similar to the bosonic tensor model, for the theory to flow to the conformal fixed points, we need to add bare mass terms to the UV action \eqref{eqn:susyAction}, and fine tune the masses such that the following renormalization conditions are satisfied,
\begin{equation}\label{eqn:renorCondSUSY}
\begin{split}
\lim_{\beta J\to\infty}(\beta J)^{-2\Delta_\phi-1}\left[(m^\phi_{\rm bare})^2-\widetilde\Sigma^{\phi\phi}(0)\right]&={(m^{\phi}_{\rm eff})^2\over(\beta J)^{2\Delta_\phi+1}}=\frac{\Gamma (1- \Delta_\phi )}{\beta^2 (b_\phi J^{2\Delta_\phi+1})\pi ^{2 \Delta_\phi-\frac{1}{2}}\Gamma \left(\frac{1}{2}-\Delta_\phi \right)},
\\
\lim_{\beta J\to\infty}(\beta J)^{-2\Delta_F+1}\left[(m^F_{\rm bare})^2-\beta^{-2}\widetilde\Sigma^{FF}(0)\right]&={(m^{F}_{\rm eff})^2\over(\beta J)^{2\Delta_F-1}}=\frac{\Gamma (1- \Delta_F )}{\beta^2(b_F J^{2\Delta_F-1}) \pi ^{2 \Delta_F-\frac{1}{2}}\Gamma \left(\frac{1}{2}-\Delta_F \right)}.
\end{split}
\end{equation}
The bare mass terms explicitly break supersymmetry. We cannot find a supersymmetry-preserving regulator that flows to the conformal fixed point, which is consistent with the analytic result that supersymmetry is broken for the conformal solution. With this regularization scheme we can numerically solve the full Schwinger-Dyson equations \eqref{eqn:superSDeqn} (with the bare mass term included) all along the flow. The results are plotted in Fig.~\ref{Fig:susyKTG} and we see reasonable convergence in the strong coupling limit to the fixed point solution determined earlier. The coefficients $b_\phi$ and $b_F$ explicitly appear in our renormalization conditions. Hence, the bare masses break the scaling symmetry \eqref{eqn:globalScaling} and determine the values of the coefficients $b_\phi$, $b_\psi$, and $b_F$.

There are a couple of fringe situations that deserve some additional commentary:
\begin{itemize}
\item For $q=4$ the conformal dimension of the auxiliary field $F^{a_1a_2a_3}$ is equal to its classical scaling dimension $\Delta_F=\frac{1}{2} $ from Table~\ref{table:confDims}.\footnote{ For $\Delta_F=\frac{1}{2} $, one may consider a different ansatz
\begin{equation}
G^{FF}_c(\tau)=b'_F \, \delta(\tau).
\end{equation}
However, by the limit representation
$\delta(\tau) = \displaystyle \lim_{\epsilon\to 0}\, \frac{1}{2} \, \epsilon\, \left|\pi\over \beta\sin{\pi\tau\over \beta}\right|^{1-\epsilon}$, this is equivalent to the original ansatz \eqref{eqn:conformalAnsatz} with 
$b_F=\frac{1}{2} \,\epsilon \,b_F'$ and $\Delta_F=\frac{1}{2}\, (1-\epsilon)$.} 
By \eqref{eqn:bsol} and \eqref{eqn:effMasses}, we find that the effective mass $m^F_{\rm eff}$ diverges with the dimensionless coefficient $b_\phi J^{2\Delta_\phi+1}$ keeping fixed.
Hence, the auxiliary field decouples due to the infinitely large mass.
\item There is potentially a  different supersymmetric solution to \eqref{eqn:extraConstraint} given by 
$\Delta_{\phi} = 0$, $\Delta_{\psi} = 1/2$, and $\Delta_{F} = 1$. If this solution is indeed supersymmetric, the coefficients $b_\phi$, $b_\psi$ and $b_F$ must be related by $b_F=(2\Delta_\phi+1)b_\psi=2\Delta_\phi(2\Delta_\phi+1)b_\phi$. Equations \eqref{eqn:bsol} then implies that
\ie
b_\phi^{q}J^{q}&={1\over 4\pi^2(q-1)\Delta_\phi^2}+{\cal O}(\Delta_\phi^{-1}),
\fe
which diverges at $\Delta_\phi=0$. We can expand the boson propagator as
\ie
G_c^{\phi\phi}(\tau) = \frac{b_{\phi}}{ |\tau|^{2\Delta_{\phi}}}=b_\phi -\widetilde b_\phi\log |\tau|+\cdots,
\fe
where $\widetilde b_\phi = 2 b_\phi \Delta_\phi $. The coefficient of the logarithmic term goes to zero as well as the coefficients $b_\psi$ and $b_F$, i.e. $\widetilde b_\phi\sim b_\psi\sim b_F\sim \Delta_\phi^{1-{2\over q}}$. 

\end{itemize}

\paragraph{Comments on supersymmetry breaking:}
As noted earlier, the origins of supersymmetry breaking in our model have to do with the need to regularize the boson self-energy piece. While it may be intuitively hard to grasp why a quantum mechanical system has UV divergences, the origins of the same, of course lie in the fact that the critical dynamics drives the boson dimension too low. One can check that as long as $\Delta_{boson} > \frac{1}{2}$  there is no divergence in the boson self-energy. However, from Table~\ref{table:confDims} we see that this does not pertain in our conformal limit for any choice of $q$. 

One way to think about the supersymmetry breaking is to first ask what are the solutions to the truncated IR Schwinger-Dyson equations. We have a-priori seen that  solutions cannot be found respecting the constraints arising from supersymmetry, which would demand $\Delta_\psi = \Delta_\phi+\frac{1}{2}$ and $\Delta_F = \Delta_\phi +1$ in \S\ref{sec:cfix}. This observation then prompts us to explore regularization schemes that will attain the fixed point solution, without preserving supersymmetry along the RG flow. Put different, our choice of supersymmetry breaking regularization is predicated upon the attainment of non-trivial fixed point in the IR. Had we refrained from doing so the flow would have drifted away and we guess that the result would be similar to the observations made in the context of bosonic models in \cite{Murugan:2017eto,Azeyanagi:2017drg,Azeyanagi:2017mre}. A consequence of this explicit breaking is that we do not expect a goldstino in the low energy spectrum; the analysis of operator spectrum in \S\ref{sec:4pt} will confirm this intuition.

One might wonder whether the supersymmetry breaking phenomenon is peculiar to the melonic dominance. In the context of quiver quantum mechanical theories, the authors of  \cite{Anninos:2016szt} noticed a similar feature.\footnote{ We thank Juan Maldacena for recalling this reference to our attention.} These models are qualitatively similar to the SYK family of theories (with $q=2$, i.e., random Gaussian couplings for fermions) as already noted in their discussion. The low energy Schwinger-Dyson equations in that case admit solutions which preserve supersymmetry as well as those that break it. Arguments were given in favour of the former circumstance being relevant in that context. At a cursory level this is similar to our discussion where a-prioiri there does exist a solution with $\Delta_\phi =0$. As argued above we believe this solution is unphysical since the physical Green's function diverges. Our numerical explorations also support the absence of a supersymmetric low energy fixed point; the Schwinger-Dyson equations do not converge and at best could be suggesting the existence of a trivial gapped phase.

\section{Four-point functions and operator spectrum}
\label{sec:4pt}

We have seen that the strong coupling limit ($\beta \, J \gg 1$)  has emergent conformal invariance with non-trivial anomalous dimensions for the fields as given in Table~\ref{table:confDims}. We now turn to analyzing a part of the low-energy spectrum of the theory, organizing it in terms of conformal dimensions in the IR effective field theory. Our analysis will be based on looking at four-point functions of the elementary fields in the model, following similar analyses in the SYK model literature \cite{Maldacena:2016hyu}. For the fermionic channels, we will derive some new results on fermionic $SL(2,\mathbb{R})$ wavefunctions that we will need for the corresponding four-point functions.

\subsection{Resumming ladder supergraphs}
\label{sec:la}

Let us consider the four-point function of superfields $\Phi^{A_q}\equiv\Phi^{a_1\ldots a_{q-1}}$,
\begin{equation}
\begin{split}
&\frac{1}{N^{q-1}}\sum_{a_i,b_i=1}^N\big<{\cal T}\left(\Phi^{A_q}(X_1)\, \Phi^{A_q}(X_2) \, \Phi^{B_q}(X_3)\, \Phi^{B_q}(X_4)\right)\big>
\\
&\qquad =N^{q-1}{\cal G}(X_1,X_2){\cal G}(X_3,X_4)+{\cal F}(X_1,X_2,X_3,X_4)+{\cal O}(N^{-1}).
\end{split}
\end{equation}
The leading term is a product of free super-propagators and is given by  a disconnected diagram. The sub-leading correction term ${\cal F}$ can be computed by summing over ladder diagrams
\begin{equation}
\begin{split}
{\cal F}(X_1,X_2,X_3,X_4)=\sum_{n=0}^\infty {\cal F}_n(X_1,X_2,X_3,X_4),
\end{split}
\end{equation}
where ${\cal F}_n$ is the contribution from the ladder diagram with $n$ rungs.

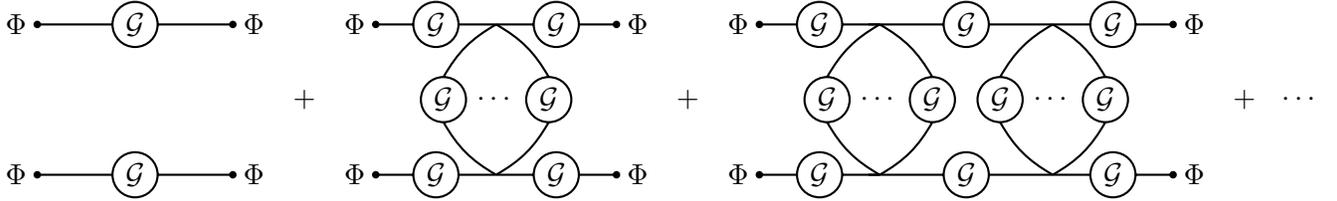
\begin{figure}[H]
\centering
\subfloat{
\begin{tikzpicture}

\draw[thick, shift={(0,1)}] (0,0) to (1,0) (1.3,0) circle (.3cm) (1.3,0) node{${\cal G}$} (1.6,0) to (2.6,0) (0,0) node[circle,fill,inner sep=1pt]{} (0,0) node[left]{$\Phi$} (2.6,0) node[circle,fill,inner sep=1pt]{} (2.6,0) node[right]{$\Phi$};
\draw[thick, shift={(0,-1)}] (0,0) to (1,0) (1.3,0) circle (.3cm) (1.3,0) node{${\cal G}$} (1.6,0) to (2.6,0)
(0,0) node[circle,fill,inner sep=1pt]{} (0,0) node[left]{$\Phi$} (2.6,0) node[circle,fill,inner sep=1pt]{} (2.6,0) node[right]{$\Phi$};

\draw (3.55,0) node{$+$};

\draw[thick, shift={(4.5,1)}] (0,0) to (.5,0) (.8,0) circle (.3cm) (.8,0) node{${\cal G}$} (1.1,0) to (1.6,0) (0,0) node[circle,fill,inner sep=1pt]{} (0,0) node[left]{$\Phi$};
\draw[thick, shift={(4.5,-1)}] (0,0) to (.5,0) (.8,0) circle (.3cm) (.8,0) node{${\cal G}$} (1.1,0) to (1.6,0) (0,0) node[circle,fill,inner sep=1pt]{} (0,0) node[left]{$\Phi$};

\draw[thick,shift={(6.1,0)}] 
(0,-1) to[out=30,in=-120] (.7,-.3)
(.7,0) circle (.3cm)
(.7,0) node{${\cal G}$}
(.7,.3) to[out=120,in=-30] (0,1)
(0,0) node{$\cdots$}
(0,-1) to[out=150,in=-60] (-.7,-.3)
(-.7,0) circle (.3cm)
(-.7,0) node{${\cal G}$}
(-.7,.3) to[out=60,in=210] (0,1);

\draw[thick, shift={(6.1,1)}] (0,0) to (.5,0) (.8,0) circle (.3cm) (.8,0) node{${\cal G}$} (1.1,0) to (1.6,0) 
(1.6,0) node[circle,fill,inner sep=1pt]{} (1.6,0) node[right]{$\Phi$};
\draw[thick, shift={(6.1,-1)}] (0,0) to (.5,0) (.8,0) circle (.3cm) (.8,0) node{${\cal G}$} (1.1,0) to (1.6,0) 
(1.6,0) node[circle,fill,inner sep=1pt]{} (1.6,0) node[right]{$\Phi$};

\draw (8.65,0) node{$+$};

\draw[thick, shift={(9.6,1)}] (0,0) to (.5,0) (.8,0) circle (.3cm) (.8,0) node{${\cal G}$} (1.1,0) to (1.6,0) (0,0) node[circle,fill,inner sep=1pt]{} (0,0) node[left]{$\Phi$};
\draw[thick, shift={(9.6,-1)}] (0,0) to (.5,0) (.8,0) circle (.3cm) (.8,0) node{${\cal G}$} (1.1,0) to (1.6,0) (0,0) node[circle,fill,inner sep=1pt]{} (0,0) node[left]{$\Phi$};

\draw[thick,shift={(11.2,0)}] 
(0,-1) to[out=30,in=-120] (.7,-.3)
(.7,0) circle (.3cm)
(.7,0) node{${\cal G}$}
(.7,.3) to[out=120,in=-30] (0,1)
(0,0) node{$\cdots$}
(0,-1) to[out=150,in=-60] (-.7,-.3)
(-.7,0) circle (.3cm)
(-.7,0) node{${\cal G}$}
(-.7,.3) to[out=60,in=210] (0,1);

\draw[thick, shift={(11.55,1)}] (-.5,0) to (.5,0) (.8,0) circle (.3cm) (.8,0) node{${\cal G}$} (1.1,0) to (1.95,0);
\draw[thick, shift={(11.55,-1)}] (-.5,0) to (.5,0) (.8,0) circle (.3cm) (.8,0) node{${\cal G}$} (1.1,0) to (1.95,0);

\draw[thick,shift={(13.5,0)}] 
(0,-1) to[out=30,in=-120] (.7,-.3)
(.7,0) circle (.3cm)
(.7,0) node{${\cal G}$}
(.7,.3) to[out=120,in=-30] (0,1)
(0,0) node{$\cdots$}
(0,-1) to[out=150,in=-60] (-.7,-.3)
(-.7,0) circle (.3cm)
(-.7,0) node{${\cal G}$}
(-.7,.3) to[out=60,in=210] (0,1);

\draw[thick, shift={(13.5,1)}] (0,0) to (.5,0) (.8,0) circle (.3cm) (.8,0) node{${\cal G}$} (1.1,0) to (1.6,0) 
(1.6,0) node[circle,fill,inner sep=1pt]{} (1.6,0) node[right]{$\Phi$};
\draw[thick, shift={(13.5,-1)}] (0,0) to (.5,0) (.8,0) circle (.3cm) (.8,0) node{${\cal G}$} (1.1,0) to (1.6,0) 
(1.6,0) node[circle,fill,inner sep=1pt]{} (1.6,0) node[right]{$\Phi$};

\draw (16.05,0) node{$+$};
\draw (16.8,0) node{$\cdots$};
\end{tikzpicture}
}
\caption{The leading $1/N$ correction to the four-point function of superfield $\Phi^{A_q}$.}
\label{Fig:ladder}
\end{figure}

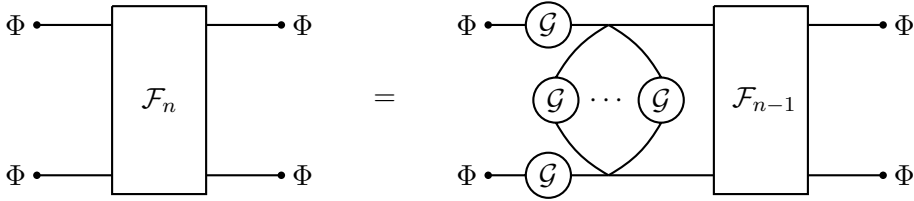
\begin{figure}[H]
\centering
\subfloat{
\begin{tikzpicture}

\draw[thick, shift={(0,1)}] (0,0) to (1,0)
(2.25,0) to (3.25,0)
(0,0) node[circle,fill,inner sep=1pt]{}
(0,0) node[left]{$\Phi$}
(3.25,0) node[circle,fill,inner sep=1pt]{}
(3.25,0) node[right]{$\Phi$};

\draw[thick, shift={(0,-1)}] (0,0) to (1,0)
(2.25,0) to (3.25,0)
(0,0) node[circle,fill,inner sep=1pt]{}
(0,0) node[left]{$\Phi$}
(3.25,0) node[circle,fill,inner sep=1pt]{}
(3.25,0) node[right]{$\Phi$};

\draw[thick, shift={(1,0)}] (0,1.25) to (1.25,1.25) to (1.25,-1.25) to (0,-1.25) to (0,1.25)
(.25,0) node[right]{${\cal F}_n$};

\draw (4.625,0) node{$=$};

\draw[thick, shift={(6,1)}] (0,0) to (.5,0) (.8,0) circle (.3cm) (.8,0) node{${\cal G}$} (1.1,0) to (1.6,0) (0,0) node[circle,fill,inner sep=1pt]{} (0,0) node[left]{$\Phi$};
\draw[thick, shift={(6,-1)}] (0,0) to (.5,0) (.8,0) circle (.3cm) (.8,0) node{${\cal G}$} (1.1,0) to (1.6,0) (0,0) node[circle,fill,inner sep=1pt]{} (0,0) node[left]{$\Phi$};

\draw[thick,shift={(7.6,0)}] 
(0,-1) to[out=30,in=-120] (.7,-.3)
(.7,0) circle (.3cm)
(.7,0) node{${\cal G}$}
(.7,.3) to[out=120,in=-30] (0,1)
(0,0) node{$\cdots$}
(0,-1) to[out=150,in=-60] (-.7,-.3)
(-.7,0) circle (.3cm)
(-.7,0) node{${\cal G}$}
(-.7,.3) to[out=60,in=210] (0,1);

\draw[thick] 
(7.6,-1) to (9,-1)
(7.6,1) to (9,1);

\draw[thick, shift={(9,0)}] (0,1.25) to (1.25,1.25) to (1.25,-1.25) to (0,-1.25) to (0,1.25)
(.1,0) node[right]{${\cal F}_{n-1}$};

\draw[thick,shift={(.25,0)}] 
(10,-1) to (11,-1)
(10,1) to (11,1)
(11,-1) node[circle,fill,inner sep=1pt]{}
(11,-1) node[right]{$\Phi$}
(11,1) node[circle,fill,inner sep=1pt]{}
(11,1) node[right]{$\Phi$};

\end{tikzpicture}
}
\caption{The recurrence relation of the ladder diagram ${\cal F}_n$.}
\label{Fig:ladder}
\end{figure}

The ladder diagrams with $n$ rungs are related to the ladder diagrams with $n-1$ rungs by a recurrence relation
\begin{equation}\label{eqn:superladderRecurrence}
{\cal F}_n(X_1,X_2,X_3,X_4) =\int dX dX'\; {\cal K}(X_1,X_2;X,X')\,{\cal F}_{n-1}(X,X',X_3,X_4),
\end{equation}
where the kernel ${\cal K}(X_1,X_2;X_3,X_4)$ is
\begin{equation}
{\cal K}(X_1,X_2;X_3,X_4) =(q-1)\,J^q\,{\cal G}(X_1,X_3)\, {\cal G}(X_2,X_4) \, {\cal G}(X_3,X_4)^{q-2}.
\end{equation}
Since supersymmetry is broken in the IR, it is more convenient to work with component fields, and expand the four-point function and the kernel as
\begin{equation}
\begin{split}
{\cal F}(X_1,X_2,X_3,X_4)&={\cal F}^{\phi\phi\phi\phi}(\tau_1,\tau_2,\tau_3,\tau_4)+\theta_1 \bar\theta_2\, {\cal F}^{\bar\psi\psi\phi\phi}(\tau_1,\tau_2,\tau_3,\tau_4)+\cdots,
\\
{\cal K}(X_1,X_2,X_3,X_4)&=K^{\phi\phi,\phi\phi}(\tau_1,\tau_2,\tau_3,\tau_4)+\theta_1 \bar\theta_2\, K^{\bar\psi\psi,\phi\phi}(\tau_1,\tau_2,\tau_3,\tau_4)+\cdots.
\end{split}
\end{equation}

The recurrence relation \eqref{eqn:superladderRecurrence} can be written for the components in a compact notation using a variable $\sigma$
to designate the fields, i.e.,  $\sigma \in \{\phi = \bar{\phi}, \; \psi, \,\bar{\psi},\; F = \bar{F}\}$. We have:
\begin{equation}\label{eqn:ladderRecurrence}
{\cal F}^{\sigma_1\sigma_2\sigma_3\sigma_4}_n(\tau_1,\tau_2,\tau_3,\tau_4) 
=\sum_{\sigma,\sigma'}\int d\tau d\tau'\, K^{\sigma_1\sigma_2 ,\bar\sigma \bar\sigma'}(\tau_1,\tau_2;\tau,\tau')
\; {\cal F}^{\sigma\sigma'\sigma_3\sigma_4}_{n-1}(\tau,\tau',\tau_3,\tau_4) \,.
\end{equation}
 Let us denote $({\bf F}^{\sigma_3\sigma_4})^{\sigma_1\sigma_2}={\cal F}^{\sigma_1\sigma_2\sigma_3\sigma_4}$ as a vector and $({\bf K})^{\sigma_1\sigma_2,\sigma_3\sigma_4}=K^{\sigma_1\sigma_2,\sigma_3\sigma_4}$ as a matrix. The recurrence relation \eqref{eqn:ladderRecurrence} can be written in matrix notation as
\begin{equation}
{\bf F}^{\sigma_3\sigma_4}_n(\tau_1,\tau_2,\tau_3,\tau_4)
=\int d\tau d\tau'\; {\bf K}(\tau_1,\tau_2;\tau,\tau')\, {\bf F}^{\sigma_3\sigma_4}_{n-1}(\tau,\tau',\tau_3,\tau_4).
\end{equation}

The sum of all ladder diagrams is a geometric series, which can be resummed and formally written as
\begin{equation}\label{eqn:sumGeo}
{\bf F}^{\sigma_3\sigma_4}=\sum_{n=0}^\infty {\bf K}^n{\bf F}^{\sigma_3\sigma_4}_0 = (1-{\bf K})^{-1}{\bf F}^{\sigma_3\sigma_4}_0.
\end{equation}
Let us consider the conformal limit, and add a subscript $c$ to the four-point functions and kernels. 
Denote the eigenvectors of the kernel ${\bf K}_c$ by ${\bf \Psi}^i_h$, where the dimension $h$ is related to the eigenvalue of the Casimir operator that will be discussed later, and $i$ denotes other quantum numbers. The eigenequation is
\begin{equation}
\int d\tau d\tau'\, {\bf K}_c(\tau_1,\tau_2;\tau,\tau') \,{\bf \Psi}^i_h(\tau,\tau',\tau_3,\tau_4)=k_i(h){\bf \Psi}^i_h(\tau_1,\tau_2,\tau_3,\tau_4),
\end{equation}
where $k_i(h)$ are the eigenvalues. The equation \eqref{eqn:sumGeo} can be rewritten in the basis of the eigenvectors ${\bf \Psi}^i_h$ as
\begin{equation}\label{eqn:4ptfFormal}
{\bf F}_c^{\sigma_3\sigma_4}= \sum_{i,h}{\bf \Psi}^i_h \; \frac{1}{1-k_i(h)} \; 
\frac{\la{\bf \Psi}^i_h,{\bf F}^{\sigma_3\sigma_4}_{c,0}\ra}{ \la{\bf \Psi}^i_h,{\bf \Psi}^i_h\ra}.
\end{equation}

In the following subsections, we discuss various ingredients that appear in the above formula, and make this formal expression explicit. In \S\ref{sec:efns}, we discuss the eigenvectors of the kernel ${\bf K}_c$, which are organized by the conformal eigenfunctions of the IR conformal algebra.  In \S\ref{sec:speccomm}, we compute the eigenvalues $k_i(h)$ of the kernel ${\bf K}_c$, and extract the spectrum of operators that appear in the $\sigma_1\times \sigma_2$ OPE. In \S\ref{sec:4ptf}, we compute the inner products between the tree-level four-point functions ${\cal F}^{\sigma_1\sigma_2\sigma_3\sigma_4}_{c,0}$ and the conformal eigenfunctions, and  give explicit expressions for the four-point functions.

\subsection{Conformal eigenfunctions}
\label{sec:efns}

As in the case of the SYK model studied in \cite{Kitaev:2015aa,Polchinski:2016xgd,Maldacena:2016hyu}, the kernel 
${\bf K}_c(\tau_1,\tau_2;\tau_3,\tau_4)$ commutes with an IR $SL(2,\bR)$ algebra, whose generators $\hat D$, $\hat P$ and $\hat K$ are
\begin{equation}
\hat D=-\tau\partial_\tau-\Delta,\quad \hat P=\partial_\tau,\quad \hat K=\tau^2\partial_\tau+2\tau\Delta.
\end{equation}
This implies that the kernel also commutes with the Casimir operator built from the sum of the $SL(2,\bR)$ generators acting on
 $\tau_1$ and $\tau_2$
\begin{equation}
C_{1+2}=(\hat D_1+\hat D_2)^2-\frac{1}{2} \{\hat K_1+\hat K_2,\hat P_1+\hat P_2\}.
\end{equation}

The $SL(2,\bR)$ invariance of the kernel implies that the four-point function only depends on the conformal invariant cross-ratio $\chi={\tau_{12}\tau_{34}\over \tau_{13}\tau_{24}}$ and the ordering of the points $(\tau_1,\tau_2,\tau_3,\tau_4)$. In particular, after partially fixing to the ordering $\tau_1<\tau_3<\tau_4$ and $\tau_2<\tau_4$, the four-point function takes the form as a function of $\chi$ times a suitable conformal factor.\footnote{ For the other orderings, the four-point function takes the same form as \eqref{eqn:4ptConformalForm} but the function ${\cal F}_{c}^{\sigma_1\sigma_2\sigma_3\sigma_4}(\chi)$ may be different.} We will make a convenient (but somewhat non-traditional) choice for the prefactor and define: 
\begin{equation}\label{eqn:4ptConformalForm}
\begin{split}
{\cal F}_{c}^{\sigma_1\sigma_2\sigma_3\sigma_4}(\tau_1,\tau_2,\tau_3,\tau_4)
&= {\sqrt{b_{\sigma_1}b_{\sigma_2}b_{\sigma_3}b_{\sigma_4}}}{\rm sgn}(\tau_{12})^{|\sigma_1||\sigma_2|} {\rm sgn}(\tau_{34})^{|\sigma_3||\sigma_4|}
\\
&\qquad\times\mathcal{P}^{\sigma_1\sigma_2\sigma_3\sigma_4}(\tau_1, \tau_2, \tau_3, \tau_4) \; 
{\cal F}_{c}^{\sigma_1\sigma_2\sigma_3\sigma_4}(\chi) \,,  \\
\mathcal{P}^{\sigma_1\sigma_2\sigma_3\sigma_4}(\tau_1, \tau_2, \tau_3, \tau_4) &\equiv 
\frac{1}{ |\tau_{12}|^{\Delta_1+\Delta_2} |\tau_{34}|^{\Delta_3+\Delta_4}}\left| \frac{\tau_{23}}{ \tau_{14}}\right|^{\frac{1}{2} (\Delta_{12}-\Delta_{34})}
\; \left| \frac{\tau_{24}}{ \tau_{13}}\right|^{\frac{1}{2} (\Delta_{12}+\Delta_{34})} \,.
\end{split}
\end{equation}
where $\Delta_{ij} = \Delta_i  - \Delta_j$ and $|\sigma|$ is an even (odd) integer if $\sigma$ is a boson (fermion).
The Casimir operator acting on this parametrization of the four-point function reduces to a simple second order differential operator in terms of the cross-ratio:
\begin{equation}
\begin{split}
&C_{1+2}{\cal F}_{c}^{\sigma_1\sigma_2\sigma_3\sigma_4}(\tau_1,\tau_2,\tau_3,\tau_4) = 
	 {\sqrt{b_{\sigma_1}b_{\sigma_2}b_{\sigma_3}b_{\sigma_4}}}{\rm sgn}(\tau_{12})^{|\sigma_1||\sigma_2|} {\rm sgn}(\tau_{34})^{|\sigma_3||\sigma_4|}
	 \\
	& \hspace{50mm}\times\mathcal{P}^{\sigma_1\sigma_2\sigma_3\sigma_4}(\tau_1, \tau_2, \tau_3, \tau_4)\	{\cal C}{\cal F}_{c}^{\sigma_1\sigma_2\sigma_3\sigma_4}(\chi),
\\
&{\cal C} \equiv \chi^2(1-\chi)\partial^2_\chi-\chi^2\partial_\chi+{4\Delta_{12}\Delta_{34}\chi-(\Delta_{12}+\Delta_{34})^2\chi^2\over 4(1-\chi)}.
\end{split}
\end{equation}
We will continue to refer the differential operator $\cC$ as the  Casimir operator. 

It is convenient to expand the four-point function in the basis of the eigenfunctions of the Casimir operator ${\cal C}$.
The eigenfunctions of the Casimir operator are solutions to the hypergeometric equation
\begin{equation}\label{eqn:CasimirEq}
{\cal C}\Psi_h(\chi)=h(h-1)\Psi_h(\chi).
\end{equation}

To pick out the wavefunctions of interest we need to ensure that the operator ${\cal C}$ is Hermitian. This however depends on the choice of norm imposed on the wavefunctions. We will discuss Hermiticity  with respect to four different norms, indexed by a 
pair $m,n=0,1$. The norms are chosen to be:
\begin{equation}\label{eqn:norms}
\begin{split}
&\la f,g \ra_{m,n} = \frac{1}{2} \int^\infty_{-\infty} {d\chi\over \chi^2} \,{\rm sgn}(\chi^m(\chi-1)^n)\,f^*(\chi) g(\chi)\,.
\end{split}
\end{equation}
Most of the discussion in the literature concerns itself with the $\la\cdot,\cdot\ra_{0,0}$ norm, which is the natural inner product we can impose on bosonic wavefunctions. Two of the other norms $\la\cdot,\cdot\ra_{1,1}$ and $\la\cdot,\cdot\ra_{1,0}$ become relevant when we have fermionic intermediate states in the 4-point function.  For each of the norms \eqref{eqn:norms} and for a fixed eigenvalue of the Casimir operator, there are two linearly independent solutions to the Casimir equation \eqref{eqn:CasimirEq}. They are summarized in Appendix \ref{sec:sl2Rwf}.  The dimension $h$ can be continuous $h\in\frac{1}{2} +i\bR^+$, or discrete $h\in\bZ^+$ and $h\in\bZ^++{1\over 2}$. For continuum states, the eigenfunctions have integral representations \eqref{eqn:EFintrep}, \eqref{eqn:BosWave1234}, \eqref{eqn:FermWave1423}, and \eqref{eqn:FermWave1324}.

Let us introduce  the conformal three-point functions:
\begin{equation}\label{eqn:3ptfns}
\begin{split}
    \vev{\sigma_1(\tau_1) \, \sigma_2(\tau_2) {\cal O}_h(\tau_0)}^{m,n,p}
 & =  {{\rm sgn}(\tau_{10})^m\, {\rm sgn}(\tau_{20})^n\,{\rm sgn}(\tau_{12})^p \over |\tau_{10}|^{\Delta_1+h-\Delta_2}\, |\tau_{20}|^{\Delta_2+h-\Delta_1}\, |\tau_{12}|^{\Delta_1+\Delta_2-h}} \,, 
\end{split}
\end{equation}
in terms of which  the integrals \eqref{eqn:EFintrep}, \eqref{eqn:BosWave1234}, \eqref{eqn:FermWave1423}, and \eqref{eqn:FermWave1324} can be rewritten in the shadow representation (after reinstating  the conformal factors), using
\begin{equation}\label{eqn:shadow}
\begin{split}
\Psi^s_h(\tau_1,\tau_2,\tau_3,\tau_4) 
&= 
	\mathcal{P}^{\sigma_1\sigma_2\sigma_3\sigma_4}(\tau_1, \tau_2, \tau_3, \tau_4) \, \Psi^s_h(\chi) \\
&= \int d\tau_0\, \vev{\sigma_1(\tau_1) \, \sigma_2(\tau_2) {\cal O}_h(\tau_0)}^{0,0,0}  \  
	\vev{ \sigma_3(\tau_3) \, \sigma_4(\tau_4) \,{\cal O}_{1-h}(\tau_0) }^{0,0,0}  \,, \\
\Psi^a_h(\tau_1,\tau_2,\tau_3,\tau_4) 
&= 
	\mathcal{P}^{\sigma_1\sigma_2\sigma_3\sigma_4}(\tau_1, \tau_2, \tau_3, \tau_4) \, \Psi^a_h(\chi) \\
&= \int d\tau_0\, \vev{\sigma_1(\tau_1) \, \sigma_2(\tau_2) {\cal O}_h(\tau_0)}^{1,1,1}  \  
	\vev{ \sigma_3(\tau_3) \, \sigma_4(\tau_4)\,{\cal O}_{1-h}(\tau_0) }^{1,1,1}  \,, \\
\Psi^{12}_h(\tau_1,\tau_2,\tau_3,\tau_4) 
&= 
	\mathcal{P}^{\sigma_1\sigma_2\sigma_3\sigma_4}(\tau_1, \tau_2, \tau_3, \tau_4) \, \Psi^{12}_h(\chi) \\
&= \int d\tau_0\, \vev{\sigma_1(\tau_1) \, \sigma_2(\tau_2) {\cal O}_h(\tau_0)}^{1,1,0}  \  
	\vev{ \sigma_3(\tau_3) \, \sigma_4(\tau_4)\,{\cal O}_{1-h}(\tau_0) }^{0,0,0}  \,, \\
\Psi^{34}_h(\tau_1,\tau_2,\tau_3,\tau_4) 
&= 
	\mathcal{P}^{\sigma_1\sigma_2\sigma_3\sigma_4}(\tau_1, \tau_2, \tau_3, \tau_4) \, \Psi^{34}_h(\chi) \\
&= \int d\tau_0\, \vev{\sigma_1(\tau_1) \, \sigma_2(\tau_2) {\cal O}_h(\tau_0)}^{0,0,1}  \  
	\vev{ \sigma_3(\tau_3) \, \sigma_4(\tau_4)\,{\cal O}_{1-h}(\tau_0) }^{1,1,1}  \,, \\
\Psi^{14}_h(\tau_1,\tau_2,\tau_3,\tau_4) 
&= 
	\mathcal{P}^{\sigma_1\sigma_2\sigma_3\sigma_4}(\tau_1, \tau_2, \tau_3, \tau_4) \, \Psi^{14}_h(\chi) \\
&= \int d\tau_0\, \vev{\sigma_1(\tau_1) \, \sigma_2(\tau_2) {\cal O}_h(\tau_0)}^{1,0,1}  \  
	\vev{ \sigma_3(\tau_3) \, \sigma_4(\tau_4)\,{\cal O}_{1-h}(\tau_0) }^{0,1,1}  \,, \\
\Psi^{23}_h(\tau_1,\tau_2,\tau_3,\tau_4) 
&= 
	\mathcal{P}^{\sigma_1\sigma_2\sigma_3\sigma_4}(\tau_1, \tau_2, \tau_3, \tau_4) \, \Psi^{23}_h(\chi) \\
&= \int d\tau_0\, \vev{\sigma_1(\tau_1) \, \sigma_2(\tau_2) {\cal O}_h(\tau_0)}^{0,1,0}  \  
	\vev{ \sigma_3(\tau_3) \, \sigma_4(\tau_4)\,{\cal O}_{1-h}(\tau_0) }^{1,0,0}  \,, \\
	\Psi^{13}_h(\tau_1,\tau_2,\tau_3,\tau_4) 
&= 
	\mathcal{P}^{\sigma_1\sigma_2\sigma_3\sigma_4}(\tau_1, \tau_2, \tau_3, \tau_4) \, \Psi^{13}_h(\chi) \\
&= \int d\tau_0\, \vev{\sigma_1(\tau_1) \, \sigma_2(\tau_2) {\cal O}_h(\tau_0)}^{1,0,0}  \  
	\vev{ \sigma_3(\tau_3) \, \sigma_4(\tau_4)\,{\cal O}_{1-h}(\tau_0) }^{1,0,0}  \,, \\
	\Psi^{24}_h(\tau_1,\tau_2,\tau_3,\tau_4) 
&= 
	\mathcal{P}^{\sigma_1\sigma_2\sigma_3\sigma_4}(\tau_1, \tau_2, \tau_3, \tau_4) \, \Psi^{24}_h(\chi) \\
&= \int d\tau_0\, \vev{\sigma_1(\tau_1) \, \sigma_2(\tau_2) {\cal O}_h(\tau_0)}^{0,1,1}  \  
	\vev{ \sigma_3(\tau_3) \, \sigma_4(\tau_4)\,{\cal O}_{1-h}(\tau_0) }^{0,1,1}  \,. \\
\end{split}
\end{equation}

The eigenvectors of the kernel ${\bf K}_c(\tau_1,\tau_2;\tau,\tau')$ are vectors, whose components are the conformal eigenfunctions $\Psi^s_h(\tau,\tau',\tau_3,\tau_4),\,\Psi^a_h(\tau,\tau',\tau_3,\tau_4),\,\cdots$. We would like to compute the action of the kernel ${\bf K}_c$ on the conformal eigenfunctions. For the continuum states, it suffices to consider the kernel acting on the integrand of the conformal three-point functions that appear in the shadow representation \eqref{eqn:shadow}. The eigenvalues of the discrete states can be obtained by analytic continuing the eigenvalues of the continuum states.

\subsection{Spectrum of operators}
\label{sec:speccomm}
As discussed in \cite{Kitaev:2015aa,Polchinski:2016xgd,Maldacena:2016hyu}, the solutions to the equation $k_i(h)=1$ correspond to the spectrum of operators that appear in the $\sigma_1\times\sigma_2$ OPE. Depending on the statistics and the $U(1)_R$ charges of the component fields $\sigma_1$ and $\sigma_2$, the operators that appear in the OPE can be charged bosons, neutral bosons, or charged fermions. In Table \ref{table:spec}, we summarize the spectrum of the  first few low dimension operators for $q=4, \;6$, and $100$.

\afterpage{\clearpage
\begin{sidewaystable}[H]
\centering
\begin{tabular}{| l | l | l | l}
\hline
${\shadeB q}$ & {\shadeB  charged bosons} & {\shadeB  neutral bosons }
\\
\hline
\multirow{ 2}{*}{$4$}  & \multirow{ 2}{*}{$h_{\rm CB}=1,\,3.46,\,5.53,\,7.56,\,\cdots$} & $h_{\rm NB}^t=1,\,2,\,2.21,\,3.82,\,4.30,\,5.78,\,6.32,\,7.75,\,\cdots$ 
\\
&& $h_{\rm NB}^s=1,\,2.84,\,4.79,\,6.76,\,\cdots$ 
\\
\hline
\multirow{ 2}{*}{$6$} & \multirow{ 2}{*}{$h_{\rm CB}=1,\,3.42,\,5.48,\,7.50,\,\cdots$} & $h_{\rm NB}^t=0.832,\,1,\,1.84,\,2,\,2.91,\,3.68,\,4.18,\,4.91,\,5.64,\,6.20,\,6.91,\,7.63,\,\cdots$ 
\\
&& $h_{\rm NB}^s=1,\,2.73,\,4.66,\,6.63,\,\cdots$
\\
\hline
\multirow{ 2}{*}{$100$} & \multirow{ 2}{*}{$h_{\rm CB}=1,\,3.33,\,5.36,\,7.37,\,\cdots$} & $h_{\rm NB}^t=1,\,1.34,\,2,\,2.78,\,,\,3.44,\,4.01,\,4.77,\,5.42,\,6.01,\,6.77,\,7.41,\,\cdots$  
\\
&& $h_{\rm NB}^s=1,\,2.50,\,4.44,\,6.42,\,\cdots$
\\
\hline
\multicolumn{1}{c}{} 
\\
\end{tabular}
\vspace*{2cm}
\begin{tabular}{| l | l | l | l |}
\hline
${\shadeB q}$ & {\shadeB charged fermions }
\\
\hline
\multirow{ 2}{*}{$4$}   & $h^1_{\rm F}={7\over 6}$
\\
& $h^2_{\rm F}={7\over 6}$
\\
\hline
\multirow{ 2}{*}{$6$}  & $h^1_{\rm F}=1.17,\,1.76,\,1.83,\,2.26,\,2.44,\,3.74,\,3.75,\,4.36,\,\cdots$
\\
 & $h_{\rm F}^2=0.63,\,0.83,\,1.48,\,2.75,\,2.76,\,3.33,\,3.42,\,4.74\,\cdots$
\\
\hline
\multirow{ 2}{*}{$100$} &  $h^1_{\rm F}=0.57,\,1.19,\,1.60,\,1.77,\, 1.81,\,2.30,\, 3.58,\,3.58,\,\cdots$ 
\\
 & $h_{\rm F}^2=0.82,\,1.42,\,2.59,\,3.11,\,3.26,\,4.58,\,4.58,\, 5.17,\,\cdots$ 
\\
\hline
\end{tabular}
\caption{Spectrum of first few low dimension operators for $q=4,6$ and $100$ (rounded off to two decimals). We have only retained solutions that lie in the range $h> \frac{1}{2}$. As we discuss in the main text there are no complex solutions for the bosonic states, but the $h_F^2$ branch of fermions has a pair of complex roots (for $q>4$) which however do not lie on the principal continuous line $h = \frac{1}{2}+i\,s$.}
\label{table:spec}
\end{sidewaystable}
\clearpage}

Let us make a few observations about the spectrum:\footnote{ We are grateful to Igor Klebanov for 
raising  important questions regarding the spectrum, especially the stability of the conformal limit and the decoupling of certain modes (with $h=1$) from the spectrum.}   
\begin{itemize}
\item Among the neutral bosons in the  spectrum, we universally find an operator with $h=2$, corresponding to the emergent conformal (time-reparametrization) symmetry as in \eqref{eqn:reparametrization}  in the IR. Away from the strict IR limit where the kinetic term is relevant, as in the SYK model, it will acquire Schwarzian dynamics. In the dual theory this is the part captured by the JT theory in the emergent near-
AdS$_2$ region.
\item The are additionally two  operators with $h=1$ in the neutral boson channel. One of these corresponds to the conserved $R$-current which operates as a local phase rotation in the IR, cf.,   \eqref{eq:u1affine}. This symmetry appears in the presence of additional charges as noted in earlier discussions \cite{Gross:2016kjj,Fu:2016vas,Davison:2016ngz}. 
The second $h=1$ operator corresponds to the local scaling symmetry \eqref{eqn:scaling} which is additionally present in our model. We comment on these modes below.
\item A slightly more peculiar operator is the $h=1 $ charged boson mode which arises in the $\psi \times \psi$ OPE. A similar operator was found in \cite{Bulycheva:2017ilt}; its existence appears to be accidental and we do not anticipate it being part of a new IR symmetry and will argue below that it decouples from the spectrum.
\item We also expect the theory to have $\mathcal{O}(N^2)$ light-modes corresponding to the time-dependent $O(N)^{q-1}$ rotations 
$\Phi^{A_{q}} \rightarrow \sum_{B_q}[M(t)]^{A_q}_{B_q}\Phi^{B_{q}}$; however, these are not singlets so we do not expect to see them in the OPE for the channels that we consider. 
\end{itemize}

Let us discuss the $h=1$ modes in the theory, which we have three of, with two being neutral and one carrying a $U(1)_R$ charge. To understand their role one can work out the OPE coefficient for this mode along the lines of \cite{Maldacena:2016hyu}.
\begin{itemize}
\item For the charged boson sector we find that the OPE coefficient is proportional to $ \cot(\tfrac{\pi}{2}h)$. As this  vanishes for $h=1$, we infer that the mode in question decouples from the spectrum -- similar observations were made in \cite{Gross:2016kjj}.
\item One of the $h=1$ neutral bosons behaves similarly.  Naively one would like therefore to argue that it too decouples from the spectrum. However, in this case as alluded to above we have a local scaling symmetry \eqref{eqn:scaling} which was related to the fact that we had a one-parameter family of conformal solutions, cf., \eqref{eqn:bsol}. We believe that while this mode decouples in the strict IR it returns to the spectrum once we step back and include the kinetic term. This would be consistent with the interpretation offered in  \cite{Fu:2016vas} for the scaling symmetry to correspond to a redefinition of the supersymmetry generators along with an effective action of the form $J\int (\lambda(\tau)-1)^2 d\tau$.
\item The third neutral $h=1$  mode corresponds to the local phase rotations \eqref{eq:u1affine}. This symmetry is  local only in the strict IR limit for the truncated  low-energy Schwinger-Dyson equations \eqref{eqn:SDconformal}. Away from the conformal limit, it gets broken (as for the $h=2$  reparametrization mode) to a global transformation, leaving behind the corresponding  pseudo-Nambu-Goldstone modes in the spectrum. The affine $U(1)$ $R$-symmetry is broken down to a global phase rotation with soft dynamics.
\end{itemize}

Before moving on to the details on obtaining the spectrum, let us also remark here that we have checked that there are no bosonic composite operators with complex dimensions on a wide domain of the complex $h$ plane. In particular, the presence of such modes along the principal continuous series line of $SL(2,{\mathbb R})$, viz.,
$h =\frac{1}{2} + i s$  would correspond to states that violate the unitarity bound in the IR (or equivalently the Breitenlohner-Freedman bound \cite{Breitenlohner:1982jf} in the AdS$_2$ geometry), and affects the stability of the fixed point. Such complex modes were seen in earlier analysis of bosonic tensor models \cite{Giombi:2017dtl} in various dimensions as well as in the bosonic SYK model in $d=2$ \cite{Murugan:2017eto}. It is reassuring to note that the model is indeed free of such pathologies. 

Furthermore, the bosonic states can be matched directly with composite operators:
\begin{itemize}
\item Composite charged bosons are identified with primary operators of the form $\partial^m \psi \partial^n \psi$
with $m+n >0$ odd. The spectrum of states roughly has dimensions $2k+1 + 2\, \Delta_\psi +\epsilon(q)$  with $k>0$ where $\epsilon(q) \to 0 $ as  $q\to \infty$. 
\item Composite neutral bosons in the singlet channel are primaries of the form $\partial^m  \bar{\psi} \partial^n \psi$ with $m+n >0$ even (as the singlet channel is symmetric under $\tau_1 \leftrightarrow \tau_2$). Their dimensions are $2k + 2\, \Delta_\psi +\epsilon(q)$  with $k>0$ and $\epsilon(q) \to 0 $ as  $q\to \infty$. 
\item Composite neutral bosons in the triplet channels come in three sets: (a) 
$\partial^m  \bar{\psi} \partial^n \psi$,  (b) $\partial^m  \phi \partial^n \phi$, and (c) $\partial^m  F \partial^n F$.
For the first case, the derivatives are antisymmetrically distributed between the two fermions, while in the latter two cases we symmetrize the derivatives. The conformal dimensions approach $2k+1+2\Delta_\psi $, $2k + 2\, \Delta_\phi$ and $2k+2\,\Delta_F$, respectively, with $k>0$, in the large $q$ limit. For $q=4$ the last set involving the auxiliary field is not present due to $F$ decoupling.
\end{itemize}
It is easy to check the presence of states corresponding to every one of these primaries in the spectrum (we only list the leading few in Table \ref{table:spec}).

The story for fermionic excitations in contrast is a bit more confusing.  We have also able to identify many of the fermionic excitations with primaries of the form $\partial^m \phi\, \partial^n \psi$ and $\partial^m F\, \partial^n \psi$, respectively.  For instance the solution with $h=0.57$ is a $\psi F$ composite, while the solution with $h=1.19$  well approximates $\psi \partial \phi$ (it is the one state that converges really well at large $q$).

There are however other solutions which seem to fall outside this set. For instance, we find some states with complex dimensions but these are off the $\frac{1}{2} + i s$ line. The precise locations for different choices of $q$ do not seem to have any particular significance (for instance the lowest such solutions are at  $h = 1.37 \pm 0.37 i$ for $q=6$ and  $h = 1.31 \pm 0.55 \,i$ for $q=100$ (there is no complex solution for $q=4$). We believe these to be benign and not part of the spectrum. While we have not carefully analyzed the decomposition of the 4-point function in the shadow representation to see if these states would contribute, we believe that the contour deformation arguments used for example in \cite{Maldacena:2016hyu} can be used to show that such modes do not correspond to physical states of the low energy theory. Let us also note that the decoupling of the auxiliary field leaves a strong impact on the fermion spectrum -- for $q=4$ we have only two degenerate operators  with real dimension.  Overall the fermion spectrum deserves to be understood better.

\subsubsection{Charged bosons}

We first consider the four-point function ${\cal F}_c^{\psi\psi \bar\psi\bar\psi}(\tau_1,\tau_2,\tau_3,\tau_4)$. The $\psi\times\psi$ OPE contains bosonic operators of $U(1)_R$ charge $2$ and decouples from other sectors. 
The recurrence relation \eqref{eqn:ladderRecurrence} specialized to this case is
\begin{equation}
{\cal F}^{\psi\psi\bar\psi\bar\psi}_{c,n}(\tau_1,\tau_2,\tau_3,\tau_4)=
\int d\tau d\tau' K_c^{\psi\psi,\bar\psi\bar\psi}(\tau_1,\tau_2;\tau,\tau') \; {\cal F}^{\psi\psi\bar\psi\bar\psi}_{c,n-1}(\tau,\tau',\tau_3,\tau_4),
\end{equation}
where the kernel $K_c^{\psi\psi,\bar\psi\bar\psi}$ is
\begin{equation}\label{eqn:KcB}
K_c^{\psi\psi,\bar\psi\bar\psi}(\tau_1,\tau_2;\tau_3,\tau_4)=(q-1)\, J^q \,
G_c^{\psi \bar\psi}(\tau_{13}) \, G_c^{\psi \bar\psi}(\tau_{24})\, G_c^{\phi\phi}(\tau_{34})^{q-2}.
\end{equation}

Due to the fermion statistics, the four-point function ${\cal F}^{\psi\psi\bar\psi\bar\psi}_c$ is odd under exchanging $\tau_1$ and $\tau_2$. 
We consider the eigenfunction of the kernel \eqref{eqn:KcB},
\ie
{{\rm sgn}(\tau_{12})\over |\tau_{10}|^h|\tau_{20}|^h |\tau_{12}|^{2\Delta_\psi-h}}.
\fe
The eigenvalue is computed by
\begin{equation}
\begin{split}
&\int d\tau_3 d\tau_4 \, K^{\psi\psi,\bar\psi\bar\psi}_c(\tau_1,\tau_2;\tau_3,\tau_4)\; 
{{\rm sgn}(\tau_{34})\over |\tau_{30}|^h|\tau_{40}|^h |\tau_{34}|^{2\Delta_\psi-h}}=k_{\rm CB}(h) \; {{\rm sgn}(\tau_{12})\over |\tau_{10}|^h|\tau_{20}|^h |\tau_{12}|^{2\Delta_\psi-h}},
\\
&k_{\rm CB}(h)=(q-1) \, b_\psi^2 b_\phi^{q-2}J^q \, k_2(2\Delta_\psi,2-h-2\Delta_\psi) \, k_1(2\Delta_\psi,1-h),
\end{split}
\end{equation}
where the functions $k_1(A,B,\tau)$ and $k_2(A,B,\tau)$ are given in \eqref{eqn:integrals}. 

The spectrum of the charged bosons, that appears in the $\psi\times\psi$ OPE, is then given by the solutions  $h= h_{\rm CB}$ to the equation
\begin{equation}
k_{\rm CB}(h)=1.
\end{equation}
The first few solutions to this equation are summarized in Table \ref{table:spec}. As noted earlier, there is a peculiar marginal mode in this sector whose origin is mysterious. 

\subsubsection{Neutral bosons}

We next turn to  the four-point functions ${\cal F}_c^{\sigma_1\sigma_2\sigma_3\sigma_4}$ with $(\sigma_1,\sigma_2)\in\{(\phi,\phi),(F,F),(\psi,\bar\psi),(\bar\psi,\psi)\}$. The $(\sigma_3,\sigma_4)$ should also belong to the set $\{(\phi,\phi),(F,F),(\psi,\bar\psi),(\bar\psi,\psi)\}$, but 
the precise nature of the  $(\sigma_3,\sigma_4)$  operators will not materially affect the discussion below. 
The $\sigma_1\times\sigma_2$ OPE now contains bosonic operators of zero $U(1)_R$ charge. The recurrence relation \eqref{eqn:ladderRecurrence} specialized to this case gives
\begin{align}
&{\cal F}^{\phi\phi\sigma_3\sigma_4}_{c,n}(\tau_1,\tau_2,\tau_3,\tau_4)=\int d\tau d\tau' \bigg[K_c^{\phi\phi,\phi\phi}(\tau_1,\tau_2;\tau,\tau'){\cal F}_{c,n-1}^{\phi\phi\sigma_3\sigma_4}(\tau,\tau',\tau_3,\tau_4)
\nonumber \\
&\hspace{47mm}+K_c^{\phi\phi, FF}(\tau_1,\tau_2;\tau,\tau'){\cal F}^{FF\sigma_3\sigma_4}_{c,n-1}(\tau,\tau',\tau_3,\tau_4)
\nonumber \\
&\hspace{47mm}+K_c^{\phi\phi,\bar\psi\psi}(\tau_1,\tau_2;\tau,\tau')\left({\cal F}_{c,n-1}^{\psi\bar\psi\sigma_3\sigma_4}(\tau,\tau',\tau_3,\tau_4)+{\cal F}_{c,n-1}^{\bar\psi\psi\sigma_3\sigma_4}(\tau,\tau',\tau_3,\tau_4)\right)\bigg],
\nonumber \\
&{\cal F}_{c,n}^{\psi\bar\psi\sigma_3\sigma_4}(\tau_1,\tau_2,\tau_3,\tau_4)=\int d\tau d\tau' \bigg[K_c^{\psi\bar\psi,\phi\phi}(\tau_1,\tau_2;\tau,\tau'){\cal F}_{n-1}^{\phi\phi\sigma_3\sigma_4}(\tau,\tau',\tau_3,\tau_4)
\nonumber \\
&\hspace{47mm}+K_c^{\psi\bar\psi,\bar\psi\psi}(\tau_1,\tau_2;\tau,\tau'){\cal F}_{c,n-1}^{\psi\bar\psi\sigma_3\sigma_4}(\tau,\tau',\tau_3,\tau_4)\bigg],
\nonumber \\
&{\cal F}_{c,n}^{\bar\psi\psi\sigma_3\sigma_4}(\tau_1,\tau_2,\tau_3,\tau_4)=\int d\tau d\tau' \bigg[K_c^{\psi\bar\psi,\phi\phi}(\tau_1,\tau_2;\tau,\tau'){\cal F}_{c,n-1}^{\phi\phi\sigma_3\sigma_4}(\tau,\tau',\tau_3,\tau_4)
\nonumber \\
&\hspace{47mm}+K_c^{\bar\psi\psi,\psi\bar\psi}(\tau_1,\tau_2;\tau,\tau'){\cal F}_{c,n-1}^{\bar\psi\psi\sigma_3\sigma_4}(\tau,\tau',\tau_3,\tau_4)\bigg],
\nonumber \\
&{\cal F}_n^{FF\sigma_3\sigma_4}(\tau_1,\tau_2,\tau_3,\tau_4)=\int d\tau d\tau'K_c^{FF,\phi\phi}(\tau_1,\tau_2;\tau,\tau'){\cal F}_{c,n-1}^{\phi\phi\sigma_3\sigma_4}(\tau,\tau',\tau_3,\tau_4) \,.
\end{align}
The primary kernels relevant to our computation and appearing  in the above are  
\begin{equation}
\begin{split}
K_c^{\psi\bar\psi,\bar\psi\psi}&=-(q-1)J^q \, G_c^{\psi\bar\psi}(\tau_{13})\, G_c^{\psi\bar\psi}(\tau_{24})\, G_c^{\phi\bar\phi}(\tau_{34})^{q-2},
\\
K_c^{\phi\phi,\bar\psi\psi}&=(q-1)(q-2)J^q\, G_c^{\phi\phi}(\tau_{13})\, G_c^{\phi\phi}(\tau_{24})\, G_c^{\psi\bar\psi}(\tau_{34})\, G_c^{\phi\phi}(\tau_{34})^{q-3},
\\
K_c^{\psi\bar\psi,\phi\phi}&=-(q-1)(q-2)J^q \, G_c^{\psi\bar\psi}(\tau_{13})\, G_c^{\psi\bar\psi}(\tau_{24})\, G_c^{\psi\bar\psi}(\tau_{34})\, G_c^{\phi\phi}(\tau_{34})^{q-3},
\\
K_c^{\phi\phi, FF}&=(q-1)J^q \, G_c^{\phi\phi}(\tau_{13})\, G_c^{\phi\phi}(\tau_{24})\, G_c^{\phi\phi}(\tau_{34})^{q-2},
\\
K_c^{FF,\phi\phi}&=(q-1)J^q \, G_c^{FF}(\tau_{13})\, G_c^{FF}(\tau_{24})\, G_c^{\phi\phi}(\tau_{34})^{q-2},
\\
K_c^{\phi\phi,\phi\phi}&=(q-1)(q-2)(q-3)J^q \, G_c^{\phi\phi}(\tau_{13})\, G_c^{\phi\phi}(\tau_{24})\, G_c^{\psi\bar\psi}(\tau_{34})^2\, G_c^{\phi\phi}(\tau_{34})^{q-4}
\\
&\quad+(q-1)(q-2)J^q \, G_c^{\phi\phi}(\tau_{13})\, G_c^{\phi\phi}(\tau_{24})\, G_c^{FF}(\tau_{34})\, G_c^{\phi\phi}(\tau_{34})^{q-3} \,.
\end{split}
\end{equation}
The remaining kernels are determined by the relations
\begin{equation}\label{eqn:psikernelrelations}
\begin{split}
&K_c^{\phi\phi,\psi\bar\psi}= K_c^{\phi\phi,\bar\psi\psi},\quad K_c^{\bar\psi\psi,\phi\phi}= K_c^{\psi\bar\psi,\phi\phi},\quad  K_c^{\bar\psi\psi,\psi\bar\psi}=K_c^{\psi\bar\psi,\bar\psi\psi}.
\end{split}
\end{equation}

Using the relations \eqref{eqn:psikernelrelations}, it is convenient to organize the four-point functions as a triplet and a singlet
\begin{equation}\label{eqn:tripletAndsinglet}
\begin{pmatrix}
{\cal F}_c^{\phi\phi\sigma_3\sigma_4}
\\
{\cal F}_c^{\psi\bar\psi\sigma_3\sigma_4}+{\cal F}_c^{\bar\psi\psi\sigma_3\sigma_4}
\\
{\cal F}_c^{FF\sigma_3\sigma_4}
\end{pmatrix},\quad {\cal F}_c^{\psi\bar\psi\sigma_3\sigma_4}-{\cal F}_c^{\bar\psi\psi\sigma_3\sigma_4}.
\end{equation}
The kernels are organized as a 3$\times$3 matrix
\begin{equation}\label{eqn:3x3kernel}
\begin{pmatrix}
K_c^{\phi\phi,\phi\phi}&K_c^{\phi\phi,\bar\psi\psi}&K_c^{\phi\phi, FF}
\\
2K_c^{\psi\bar\psi,\phi\phi}&K_c^{\psi\bar\psi,\bar\psi\psi} & 0
\\
K_c^{FF,\phi\phi} &0&0
\end{pmatrix},
\end{equation}
which acts on the triplet, and the kernel $K_c^{\psi\bar\psi,\bar\psi\psi}$ acts on the singlet. 

\paragraph{Triplet four-point function}

Let us first focus on the $3\times 3$ matrix \eqref{eqn:3x3kernel}.  The first and third components of the triplet are even under the exchange $\tau_1\leftrightarrow \tau_2$, while the second component is odd under this exchange. Hence, we consider the vector
\begin{equation}\label{eqn:tripletVec}
\begin{pmatrix}
{b_\phi\over |\tau_{10}|^h|\tau_{20}|^h|\tau_{12}|^{2\Delta_\phi-h}}
\\
{b_\psi {\rm sgn}(\tau_{12})\over |\tau_{10}|^h|\tau_{20}|^h|\tau_{12}|^{2\Delta_\psi-h}}
\\
{b_F\over |\tau_{10}|^h|\tau_{20}|^h|\tau_{12}|^{2\Delta_F-h}}
\end{pmatrix}.
\end{equation}
The $3\times 3$ kernel matrix \eqref{eqn:3x3kernel} acts on the vector \eqref{eqn:tripletVec} as the matrix
\begin{equation} \label{eqn:nbSDE}
{\bf k}^t_{\rm NB}(h)\equiv\begin{pmatrix}
k_t^{\phi\phi,\phi\phi}(h)&k_t^{\phi\phi,\bar\psi\psi}(h)&k_t^{\phi\phi, FF}(h)
\\
2k_t^{\psi\bar\psi,\phi\phi}(h)&k_t^{\psi\bar\psi,\bar\psi\psi}(h) & 0
\\
k_t^{FF,\phi\phi}(h) &0&0
\end{pmatrix},
\end{equation}
in which 
\begin{equation}
\begin{split}
k_t^{\psi\bar\psi,\bar\psi\psi}(h)&=-(q-1)b_\psi^2 b_\phi^{q-2}J^q\, k_2(2\Delta_\psi,2-h-2\Delta_\psi) \, k_1(2\Delta_\psi,1-h),
\\
k_t^{\phi\phi,\bar\psi\psi}(h)&=(q-1)(q-2)b_\psi^2 b_\phi^{q-2}J^q k_0(2\Delta_\phi,2-h-2\Delta_\phi)\, 
k_0(2\Delta_\phi,1-h),
\\
k_t^{\psi\bar\psi,\phi\phi}(h)&=-(q-1)(q-2)b_\psi^2 b_\phi^{q-2}J^q\, k_2(2\Delta_\psi,2-h-2\Delta_\psi)\, k_1(2\Delta_\psi,1-h),
\\
k_t^{\phi\phi, FF}(h)&=(q-1)b_F b_\phi^{q-1}J^q\, 
k_0(2\Delta_\phi,2-h-2\Delta_\phi)\, 
k_0(2\Delta_\phi,1-h),
\\
k_t^{FF,\phi\phi }(h)&=(q-1)b_F b_\phi^{q-1}J^q\, 
k_0(2\Delta_F,2-h-2\Delta_F)\, 
k_0(2\Delta_F,1-h),
\\
k_{t}^{\phi\phi,\phi\phi}(h)&=(q-1)(q-2)\left[(q-3)b_\psi^2 b_\phi^{q-2}+b_F b_\phi^{q-1}\right]J^q\, 
k_0(2\Delta_\phi,2-h-2\Delta_\phi)\, 
k_0(2\Delta_\phi,1-h).
\end{split}
\end{equation}

We would like to solve for the dimensions $h = h^t_{\rm NB}$ such that any of the eigenvalues of ${\bf k}^t_{\rm NB}(h)$ equals to unity. This is equivalent to the equation $\det({\bf k}^t_{\rm NB}(h)-1)=0$. The first few solutions are listed in Table \ref{table:spec}.
Amongst them we note the presence of two light degrees of freedom corresponding to the local $U(1)$ symmetry and the emergent conformal symmetry with  $h=1$ and $h=2$, respectively.

\paragraph{Singlet four-point function}

Next, we consider the kernel $K^{\psi\bar\psi,\bar\psi\psi}_c$ that acts on the singlet in \eqref{eqn:tripletAndsinglet}.  The singlet is symmetric under the exchange $\tau_1\leftrightarrow\tau_2$.  Hence, we consider the symmetric eigenfunction
\begin{equation}
{b_\psi \over |\tau_{10}|^h|\tau_{20}|^h|\tau_{12}|^{2\Delta_\psi-h}}.
\end{equation}
The eigenvalue is
\begin{equation}
k_{\rm NB}^{s}(h) =-(q-1)b_\psi^2 b_\phi^{q-2} \, k_1(2\Delta_\psi,2-h-2\Delta_\psi) \, k_2(2\Delta_\psi,1-h).
\end{equation}
The first  few solutions $h = h^s_{\rm NB}$ to the  equation $k^s_{\rm NB}(h)=1$  are  listed in Table \ref{table:spec}. As noted earlier we have a single light mode with $h=1$ in this sector corresponding to a local scaling symmetry.

\subsubsection{Charged fermions}

Finally, we consider the four-point functions ${\cal F}_c^{\sigma_1\sigma_2\sigma_3\sigma_4}$ with $(\sigma_1,\sigma_2)=\{(\psi,\phi),(\phi,\psi),(\psi,F),(F,\psi)\}$. The $(\sigma_3,\sigma_4)$ should belong to the set $\{(\bar\psi,\phi),(\phi,\bar\psi),(\bar\psi,F),(F,\bar\psi)\}$, but the precise operators will be immaterial for what follows. The $\sigma_1\times\sigma_2$ OPE contains fermionic operators of $U(1)_R$ charge 
$1$.  The recurrence relation \eqref{eqn:ladderRecurrence} now gives
\begin{equation}
\begin{split}
&{\cal F}_{c,n}^{\psi\phi\sigma_3\sigma_4}(\tau_1,\tau_2,\tau_3,\tau_4)=\int d\tau d\tau' 
\bigg[K^{\psi\phi, \phi\bar \psi}(\tau_1,\tau_2;\tau,\tau') \, {\cal F}_{c,n-1}^{\phi\psi\sigma_3\sigma_4}(\tau',\tau,\tau_3,\tau_4)
\\
&\hspace{47mm}+K^{\psi\phi,\bar \psi F}(\tau_1,\tau_2;\tau,\tau') \, {\cal F}_{c,n-1}^{\psi F\sigma_3\sigma_4}(\tau,\tau',\tau_3,\tau_4)\bigg],
\\
&{\cal F}_n^{\phi\psi\sigma_3\sigma_4}(\tau_1,\tau_2,\tau_3,\tau_4)=\int d\tau d\tau' \bigg[K^{\phi\psi, \bar \psi\phi}(\tau_1,\tau_2;\tau,\tau' )\, {\cal F}_{n-1}^{\psi\phi\sigma_3\sigma_4}(\tau,\tau',\tau_3,\tau_4)
\\
&\hspace{47mm}+K^{\phi\psi, F\bar \psi}(\tau_1,\tau_2;\tau,\tau') \, {\cal F}_{n-1}^{ F\psi\sigma_3\sigma_4}(\tau,\tau',\tau_3,\tau_4)\bigg],
\\
&{\cal F}_{c,n}^{\psi F\sigma_3\sigma_4}(\tau_1,\tau_2,\tau_3,\tau_4)=\int d\tau d\tau' \, 
K^{\psi F,\bar\psi \phi}(\tau_1,\tau_2;\tau,\tau')\, {\cal F}_{c,n-1}^{\psi\phi\sigma_3\sigma_4}(\tau,\tau',\tau_3,\tau_4),
\\
&{\cal F}_n^{ F\psi\sigma_3\sigma_4}(\tau_1,\tau_2,\tau_3,\tau_4)=\int d\tau d\tau' K^{ F\psi, \phi\bar\psi}(\tau_1,\tau_2;\tau,\tau') \, {\cal F}_{n-1}^{\phi\psi\sigma_3\sigma_4}(\tau,\tau',\tau_3,\tau_4),
\end{split}
\end{equation}
where the kernels are given explicitly as
\begin{equation}
\begin{split}
&K_c^{\psi\phi ,\phi \bar\psi}(\tau_1,\tau_2;\tau_3,\tau_4)=(q-1)(q-2)J^q \, G_c^{\psi\bar{\psi}}(\tau_{13})\, G_c^{\phi\phi}(\tau_{24})\, G_c^{\psi\bar\psi}(\tau_{34})\, G_c^{\phi\phi}(\tau_{34})^{q-3},
\\
&K_c^{\phi \psi, \bar\psi\phi}(\tau_1,\tau_2;\tau_3,\tau_4)=-(q-1)(q-2)J^q \,  G_c^{\phi\phi}(\tau_{13}) \, 
G_c^{\psi\bar{\psi}}(\tau_{24}) \, G_c^{\psi\bar\psi}(\tau_{34}) \, G_c^{\phi\phi}(\tau_{34})^{q-3},
\\
&K_c^{\psi \phi,\bar \psi F}(\tau_1,\tau_2;\tau_3,\tau_4)=(q-1)J^q \, G_c^{\psi \bar\psi}(\tau_{13})\, G_c^{\phi\phi}(\tau_{24})\, G_c^{\phi\phi}(\tau_{34})^{q-2},
\\
&K_c^{\psi F,\bar \psi\phi}(\tau_1,\tau_2;\tau_3,\tau_4)=(q-1)J^q \, G_c^{\psi\bar\psi}(\tau_{13})\, G_c^{FF}(\tau_{24})\, G_c^{\phi\phi}(\tau_{34})^{q-2}.
\\
&K_c^{ \phi\psi,F\bar \psi }(\tau_1,\tau_2;\tau_3,\tau_4)=(q-1)J^q \, G_c^{\phi\phi}(\tau_{13}) \, G_c^{\psi \bar\psi}(\tau_{24}) \, G_c^{\phi\phi}(\tau_{34})^{q-2},
\\
&K_c^{F\psi ,\phi\bar \psi}(\tau_1,\tau_2;\tau_3,\tau_4)=(q-1)J^q \, G_c^{FF}(\tau_{13}) \, G_c^{\psi\bar\psi}(\tau_{24})\, G_c^{\phi\phi}(\tau_{34})^{q-2}.
\end{split}
\end{equation}
The kernels form a $4\times 4$ matrix
\begin{equation}\label{eqn:2x2MK}
\begin{split}
\begin{pmatrix}
0 & K^{\psi\phi, \phi \bar\psi}&K^{\psi \phi,\bar \psi F} & 0
\\
K^{\phi\psi,  \bar\psi\phi} & 0 & 0 &K^{\phi\psi , F\bar \psi }
\\
K^{\psi F,\bar \psi\phi}&0 &0 &0
\\
0&K^{F \psi,\phi\bar \psi}&0&0
\end{pmatrix}.
\end{split}
\end{equation}
We consider the vectors
\begin{equation}
\begin{split}
\begin{pmatrix}
{\sqrt{b_\psi b_\phi}{\rm sgn}(\tau_{10})\over |\tau_{10}|^{\Delta_\psi+h-\Delta_\phi}|\tau_{20}|^{\Delta_\phi+h-\Delta_\psi}|\tau_{12}|^{\Delta_\phi+\Delta_\psi-h}}
\\
{\sqrt{b_\psi b_\phi}{\rm sgn}(\tau_{20})\over |\tau_{10}|^{\Delta_\phi+h-\Delta_\psi}|\tau_{20}|^{\Delta_\psi+h-\Delta_\phi}|\tau_{12}|^{\Delta_\phi+\Delta_\psi-h}}
\\
{\sqrt{b_\psi b_F}{\rm sgn}(\tau_{20}){\rm sgn}(\tau_{12})\over |\tau_{10}|^{\Delta_\psi+h-\Delta_F}|\tau_{20}|^{\Delta_F+h-\Delta_\psi}|\tau_{12}|^{\Delta_\psi+\Delta_F-h}}
\\
{\sqrt{b_\psi b_F}{\rm sgn}(\tau_{10}){\rm sgn}(\tau_{12})\over |\tau_{10}|^{\Delta_F+h-\Delta_\psi}|\tau_{20}|^{\Delta_\psi+h-\Delta_F}|\tau_{12}|^{\Delta_\psi+\Delta_F-h}}
\end{pmatrix},
\quad
\begin{pmatrix}
{\sqrt{b_\psi b_\phi}{\rm sgn}(\tau_{20}){\rm sgn}(\tau_{12})\over |\tau_{10}|^{\Delta_\psi+h-\Delta_\phi}|\tau_{20}|^{\Delta_\phi+h-\Delta_\psi}|\tau_{12}|^{\Delta_\phi+\Delta_\psi-h}}
\\
{\sqrt{b_\psi b_\phi}{\rm sgn}(\tau_{10}){\rm sgn}(\tau_{12})\over |\tau_{10}|^{\Delta_\phi+h-\Delta_\psi}|\tau_{20}|^{\Delta_\psi+h-\Delta_\phi}|\tau_{12}|^{\Delta_\phi+\Delta_\psi-h}}
\\
{\sqrt{b_\psi b_F}{\rm sgn}(\tau_{10})\over |\tau_{10}|^{\Delta_\psi+h-\Delta_F}|\tau_{20}|^{\Delta_F+h-\Delta_\psi}|\tau_{12}|^{\Delta_\psi+\Delta_F-h}}
\\
{\sqrt{b_\psi b_F}{\rm sgn}(\tau_{20})\over |\tau_{10}|^{\Delta_F+h-\Delta_\psi}|\tau_{20}|^{\Delta_\psi+h-\Delta_F}|\tau_{12}|^{\Delta_\psi+\Delta_F-h}}
\end{pmatrix}.
\end{split}
\end{equation}
The kernel matrix \eqref{eqn:2x2MK} acts on the vector as the matrices
\begin{equation}
\begin{split}\label{eqn:cfSDeq}
&{\bf k}^1_{\rm F}(h)\equiv\begin{pmatrix}
0 & k_1^{\psi\phi, \phi \bar\psi}(h)&k_1^{\psi \phi,\bar \psi F}(h) & 0
\\
k_1^{\phi\psi,  \bar\psi\phi}(h) & 0 & 0 &k_1^{\phi\psi , F\bar \psi }(h)
\\
k_1^{\psi F,\bar \psi\phi}(h)&0 &0 &0
\\
0&k_1^{F \psi,\phi\bar \psi}(h)&0&0
\end{pmatrix},
\\
&{\bf k}^2_{\rm F}(h)\equiv\begin{pmatrix}
0 & k_2^{\psi\phi, \phi \bar\psi}(h)&k_2^{\psi \phi,\bar \psi F}(h) & 0
\\
k_2^{\phi\psi,  \bar\psi\phi}(h) & 0 & 0 &k_2^{\phi\psi , F\bar \psi }(h)
\\
k_2^{\psi F,\bar \psi\phi}(h)&0 &0 &0
\\
0&k_2^{F \psi,\phi\bar \psi}(h)&0&0
\end{pmatrix},
\end{split}
\end{equation}
where the components are
\begin{equation}
\begin{split}
k_1^{\psi\phi, \phi \bar\psi}(h)&=-(q-1)(q-2) b_\psi^2 b_\phi^{q-2}J^q\, k_2(2\Delta_\psi,2-h-\Delta_\phi-\Delta_\psi)\, 
k_0(2\Delta_\phi,1-h-\Delta_\phi+\Delta_\psi),
\\
k_1^{\phi\psi,  \bar\psi\phi}(h)&=-(q-1)(q-2) b_\psi^2 b_\phi^{q-2}J^q\,k_1(2-h-\Delta_\phi-\Delta_\psi,2\Delta_\phi)\, k_2(2\Delta_\psi,1-h+\Delta_\phi-\Delta_\psi),
\\
k_1^{\psi \phi,\bar \psi F}(h)&=-(q-1) \sqrt{b_\psi^2 b_F b_\phi^{2q-3}}J^q\, k_2(2\Delta_\psi,2-h-\Delta_\phi-\Delta_\psi) \, 
k_0(2\Delta_\phi,1-h-\Delta_\phi+\Delta_\psi),
\\
k_1^{\psi F,\bar\psi \phi }(h)&=(q-1) \sqrt{b_\psi^2 b_F b_\phi^{2q-3}}J^q\, k_1(2\Delta_\psi,2-h-\Delta_F-\Delta_\psi)\, k_1(1-h+\Delta_\phi-\Delta_\psi,2\Delta_F),
\\
k_1^{\phi\psi,  F \bar\psi}(h)&=(q-1)  \sqrt{b_\psi^2 b_F b_\phi^{2q-3}}J^q\,k_1(2-h-\Delta_\phi-\Delta_\psi,2\Delta_\phi)\,k_2(2\Delta_\psi,1-h+\Delta_\phi-\Delta_\psi),
\\
k_1^{F\psi,  \phi \bar\psi}(h)&=-(q-1)  \sqrt{b_\psi^2 b_F b_\phi^{2q-3}}J^q\, k_0(2-h-\Delta_F-\Delta_\psi,2\Delta_F)\, k_1(2\Delta_\psi,1-h-\Delta_\phi+\Delta_\psi),
\\
k_2^{\psi\phi, \phi \bar\psi}(h)&=(q-1)(q-2) b_\psi^2 b_\phi^{q-2}J^q\, k_1(2\Delta_\psi,2-h-\Delta_\phi-\Delta_\psi)\, 
k_1(1-h-\Delta_\phi+\Delta_\psi,2\Delta_\phi),
\\
k_2^{\phi\psi,  \bar\psi\phi}(h)&=(q-1)(q-2) b_\psi^2 b_\phi^{q-2}J^q\,k_0(2-h-\Delta_\phi-\Delta_\psi,2\Delta_\phi)\, k_1(2\Delta_\psi,1-h+\Delta_\phi-\Delta_\psi),
\\
k_2^{\psi \phi,\bar \psi F}(h)&=(q-1) \sqrt{b_\psi^2 b_F b_\phi^{2q-3}}J^q\, k_1(2\Delta_\psi,2-h-\Delta_\phi-\Delta_\psi) \, 
k_1(1-h-\Delta_\phi+\Delta_\psi,2\Delta_\phi),
\\
k_2^{\psi F,\bar\psi \phi }(h)&=-(q-1) \sqrt{b_\psi^2 b_F b_\phi^{2q-3}}J^q\, k_2(2\Delta_\psi,2-h-\Delta_F-\Delta_\psi)\, k_0(2\Delta_F,1-h+\Delta_\phi-\Delta_\psi),
\\
k_2^{\phi\psi,  F \bar\psi}(h)&=-(q-1)  \sqrt{b_\psi^2 b_F b_\phi^{2q-3}}J^q\,k_0(2-h-\Delta_\phi-\Delta_\psi,2\Delta_\phi)\,k_1(2\Delta_\psi,1-h+\Delta_\phi-\Delta_\psi),
\\
k_2^{F\psi,  \phi \bar\psi}(h)&=(q-1)  \sqrt{b_\psi^2 b_F b_\phi^{2q-3}}J^q\, k_1(2-h-\Delta_F-\Delta_\psi,2\Delta_F)\, k_2(2\Delta_\psi,1-h-\Delta_\phi+\Delta_\psi),
\end{split}
\end{equation}
We denote the solutions to the equation $\det({\bf k}^1_{\rm F}(h)-1)=0$ and $\det({\bf k}^2_{\rm F}(h)-1)=0$ by $h=h^1_{\rm F}$ and $h=h^2_{\rm F}$, respectively. The first few solutions are listed in Table \ref{table:spec}.

\subsection{Four-point functions}
\label{sec:4ptf}

In this section, we collect all the ingredients and write down explicit formulae of the four-point functions. First, the nontrivial tree-level four-point functions are
\ie
{\cal F}_{c,0}^{\phi\phi\phi\phi}(\chi)&=|\chi|^{2\Delta_\phi}+\left|{\chi\over \chi-1}\right|^{2\Delta_\phi},\hspace{10mm}
{\cal F}_{c,0}^{FFFF}(\chi)=|\chi|^{2\Delta_F}+\left|{\chi\over \chi-1}\right|^{2\Delta_F},
\\
{\cal F}_{c,0}^{\psi\bar\psi\psi\bar\psi}(\chi)&=-{\rm sgn}\left(\chi\over\chi-1\right)\left|{\chi\over \chi-1}\right|^{2\Delta_\psi},\quad{\cal F}_{c,0}^{\bar\psi\psi\psi\bar\psi}(\chi)=-{\rm sgn}(\chi)|\chi|^{2\Delta_\psi},
\\
{\cal F}_{c,0}^{\psi\psi\bar\psi\bar\psi}(\chi)&=-{\rm sgn}(\chi)|\chi|^{2\Delta_\psi}-{\rm sgn}\left(\chi\over\chi-1\right)\left|{\chi\over \chi-1}\right|^{2\Delta_\psi},
\\
{\cal F}_{c,0}^{\phi\psi \phi\bar\psi}(\chi)&=-|\chi|^{\Delta_\phi+\Delta_\psi},\quad {\cal F}_{c,0}^{\psi\phi \phi\bar\psi}(\chi)=-\left|\chi\over \chi-1\right|^{\Delta_\phi+\Delta_\psi},
\\
{\cal F}_{c,0}^{\psi F \bar\psi F}(\chi)&=-|\chi|^{\Delta_F+\Delta_\psi},\quad {\cal F}_{c,0}^{ F\psi \bar\psi F}(\chi)=-{\rm sgn}(1-\chi)\left|\chi\over \chi-1\right|^{\Delta_F+\Delta_\psi}.
\fe
The inner products of them and the conformal eigenfunctions are
\ie
\la\Psi^s_h,{\cal F}^{\psi\psi\bar\psi\bar\psi}_{c,0}\ra_{0,0}
&=-{1\over 2}k_2(2\Delta_\psi,2-h-2\Delta_\psi)k_1(2\Delta_\psi,1-h),
\\
\la\Psi^s_h,{\cal F}^{\phi\phi\phi\phi}_{c,0}\ra_{0,0}&={1\over 2}k_0(2\Delta_\phi,2-h-2\Delta_\phi)k_0(2\Delta_\phi,1-h),
\\
\la\Psi^s_h,{\cal F}^{FFFF}_{c,0}\ra_{0,0}&=\la\Psi^s_h,{\cal F}^{\phi\phi\phi\phi}_{c,0}\ra_{0,0}\big|_{\Delta_\phi\to \Delta_F},
\\
\la\Psi^s_h,{\cal F}^{\psi\bar\psi\psi\bar\psi}_{c,0}+{\cal F}^{\bar\psi\psi\psi\bar\psi}_{c,0}\ra_{0,0}
&=\la\Psi^s_h,{\cal F}^{\psi\psi\bar\psi\bar\psi}_{c,0}\ra_{0,0},
\\
\la\Psi^a_h,{\cal F}^{\psi\bar\psi\psi\bar\psi}_{c,0}-{\cal F}^{\bar\psi\psi\psi\bar\psi}_{c,0}\ra_{0,0}
&={1\over 2}k_1(2\Delta_\psi,2-h-2\Delta_\psi)k_2(2\Delta_\psi,1-h),
\\
\la\Psi^{23}_h,{\cal F}^{\phi\psi\phi\bar\psi}_{c,0}\ra_{1,0}
&=-{1\over 2}k_1(2-h-\Delta_\phi-\Delta_\psi,2\Delta_\phi)\, k_2(2\Delta_\psi,1-h+\Delta_\phi-\Delta_\psi),
\\
\la\Psi^{13}_h,{\cal F}^{\psi\phi\phi\bar\psi}_{c,0}\ra_{1,1}&=\la\Psi^{23}_h,{\cal F}^{\phi\psi\phi\bar\psi}_{c,0}\ra_{1,0},
\\
\la\Psi^{23}_h,{\cal F}^{\psi F \bar\psi F}_0\ra_{1,0}&=-{1\over 2}k_1(2-h-\Delta_\psi-\Delta_F,2\Delta_\psi)k_2(2\Delta_F,1-h+\Delta_\psi-\Delta_F),
\\
\la\Psi^{13}_h,{\cal F}^{ F \psi\bar\psi F}_0\ra_{1,1}&={1\over 2}k_2(2-h-\Delta_F-\Delta_\psi,2\Delta_F)k_0(1-\Delta_F-\Delta_\psi+h,2\Delta_\psi).
\fe
Let us define the linear functionals
\ie
&{\cal I}_s\,:\,f\mapsto\int_0^\infty ds\,{4h-2\over \pi\tan \pi h}f(h)\Big|_{h={1\over 2}+is}+\sum_{h\in 2\bZ^+}{4h-2\over \pi^2}f(h),
\\
&{\cal I}_a\,:\,f\mapsto\int_0^\infty ds\,{4h-2\over \pi\tan \pi h}f(h)\Big|_{h={1\over 2}+is}+\sum_{h\in 2\bZ^+-1}{4h-2\over \pi^2}f(h),
\\
&{\cal I}_{\rm F}\,:\,f\mapsto\int_{-\infty}^\infty ds\,{2-4h\over \pi \cot\pi h} f(h)\Big|_{h={1\over 2}+is}+\sum_{h\in \bZ^++{1\over 2}}{4h-2\over \pi^2}f(h).
\fe
The four-point functions are written explicitly as
\ie\label{eqn:4ptfExp}
{\cal F}_c^{\psi\psi\bar\psi\bar\psi}(\chi)&={\cal I}_s{\la\Psi^s,{\cal F}^{\psi\psi\bar\psi\bar\psi}_{c,0}\ra_{0,0}\over 1-k_{\rm CB}}\Psi^s(\chi),
\\
\begin{pmatrix}
{\cal F}_c^{\phi\phi\sigma_3\sigma_4}(\chi)
\\
{\cal F}_c^{\psi\bar\psi\sigma_3\sigma_4}(\chi)+{\cal F}_c^{\bar\psi\psi\sigma_3\sigma_4}(\chi)
\\
{\cal F}_c^{FF\sigma_3\sigma_4}(\chi)
\end{pmatrix}
&={\cal I}_s\left[1-{\bf k}^t_{\rm NB}\right]^{-1}\begin{pmatrix}
\la\Psi^s,{\cal F}_{c,0}^{\phi\phi\sigma_3\sigma_4}\ra_{0,0}\Psi^s(\chi)
\\
\la\Psi^s,{\cal F}_{c,0}^{\psi\bar\psi\sigma_3\sigma_4}+{\cal F}_{c,0}^{\bar\psi\psi\sigma_3\sigma_4}\ra_{0,0}\Psi^s(\chi)
\\
\la\Psi^s,{\cal F}_{c,0}^{FF\sigma_3\sigma_4}\ra_{0,0}\Psi^s(\chi)
\end{pmatrix},
\\
{\cal F}_c^{\psi\bar\psi\psi\bar\psi}(\chi)-{\cal F}_c^{\bar\psi\psi\psi\bar\psi}(\chi)&={\cal I}_a{\la\Psi^a,{\cal F}^{\psi\bar\psi\psi\bar\psi}_{c,0}-{\cal F}^{\bar\psi\psi\psi\bar\psi}_{c,0}\ra_{0,0}\over 1-k_{\rm NB}^s}\Psi^a(\chi),
\\
\begin{pmatrix}
{\cal F}_c^{\psi\phi\phi \bar\psi}(\chi)
\\
{\cal F}_c^{\phi \psi\phi \bar\psi}(\chi)
\\
{\cal F}_c^{\psi F\phi \bar\psi}(\chi)
\\
{\cal F}_c^{F \psi \phi \bar\psi}(\chi)
\end{pmatrix}
&={\cal I}_{\rm F}\left[1-{\bf k}^1_{\rm F}\right]^{-1}\begin{pmatrix}
\la\Psi^{13},{\cal F}_c^{\psi\phi\phi \bar\psi}\ra_{1,1}\Psi^{13}(\chi)
\\
\la\Psi^{23},{\cal F}_c^{\phi\psi\phi \bar\psi}\ra_{1,0}\Psi^{23}(\chi)
\\
0
\\
0
\end{pmatrix},
\\
\begin{pmatrix}
{\cal F}_c^{\psi\phi\bar\psi F}(\chi)
\\
{\cal F}_c^{\phi \psi\bar\psi F}(\chi)
\\
{\cal F}_c^{\psi F\bar\psi F}(\chi)
\\
{\cal F}_c^{F \psi \bar\psi F}(\chi)
\end{pmatrix}
&={\cal I}_{\rm F}\left[1-{\bf k}^2_{\rm F}\right]^{-1}\begin{pmatrix}
0
\\
0
\\
\la\Psi^{13},{\cal F}_c^{F \psi \bar\psi F}\ra_{1,1}\Psi^{13}(\chi)
\\
\la\Psi^{23},{\cal F}_c^{\psi F\bar\psi F}\ra_{1,0}\Psi^{23}(\chi)
\end{pmatrix},
\fe
where the matrices ${\bf k}^t_{\rm NB}(h)$, ${\bf k}^1_{\rm F}(h)$, ${\bf k}^2_{\rm F}(h)$ and the functions $k_{\rm CB}(h)$, $k_{\rm NB}^s(h)$ are given explicitly in the previous subsection.

On the second and third equations of \eqref{eqn:4ptfExp}, the $h=2$ and $h=1$ terms in the sum over discrete states diverge, because $\det({\bf k}^t_{\rm NB}(2)-1)=0$ and $k^s_{\rm NB}(1)=1$. They correspond to the soft modes associated to the emergent time-reparametrization symmetry and the local $U(1)$ $R$-symmetry. The proper treatment of the contribution from the soft modes to the four-point functions requires moving slightly away from the conformal limit \cite{Maldacena:2016hyu,Bulycheva:2017uqj}.

\section{Discussion}
\label{sec:discuss}

The primary thrust of our analysis was to examine the interplay between melonic dominance in a class of supersymmetric quantum mechanical models  with dynamical bosons and supersymmetry. 
Somewhat curiously we find that these theories do not exhibit any particular simplification with the inclusion of supersymmetry and in fact non-trivial low energy vacua are non-supersymmetric. One might somewhat facilely characterize the situation as melonic supertensors not wanting to be supermelonic. 
Modulo this peculiarity, we find that they behave for all intents and purposes like the melonic tensor models analyzed in the literature. More specifically, there is a non-trivial conformal fixed point with a spectrum of singlet operators that can be computed. The low energy dynamics has an emergent time-reparametrization symmetry and an affine $U(1)$ $R$-symmetry, in addition to a peculiar local scaling symmetry. The latter symmetry has also been noticed in other supersymmetric constructions \cite{Fu:2016vas}.  

The origins of supersymmetry breaking in our system are in the regularization scheme we employ to attain the low energy conformal fixed point. In this sense the IR theory has explicitly broken supersymmetry and therefore no associate goldstino modes in the spectrum. Supersymmetry restoration occurs only in the deep UV where the kinetic term dominates over the interaction term. We did note that there exists a formal solution to the Schwinger-Dyson equations with spectrum appearing to preserve supersymmetry. Upon closer examination we find that the Green's function actually diverges in this limit, leading us to discard this solution. The situation we encounter here is analogous to earlier observations made in quiver quantum mechanical models \cite{Anninos:2016szt} as noted at the end of \S\ref{sec:melons}, where also one finds supersymmetric and non-supersymmetric low-energy vacua. In that context, however, the authors argue  the supersymmetry preserving vacuum  to be the appropriate one, in contrast to our discussion, where this seems to be untenable. 

Along with establishing the existence of a non-supersymmetric fixed point, we have also computed the spectrum of composite operators in the theory in the singlet sector. The spectrum is free of any pathologies (all bosonic composite operators have real conformal dimension) and shows  the low energy fixed point to be stable. We do find some curious features involving fermion composite operators -- there are some solutions to the eigenvalue equation with complex dimensions, but these we believe are not part of the spectrum as they do not propagate in the intermediate channels. In the process of computing the spectrum, we have also derived explicitly the expressions for the four-point functions of the fundamental tensor fields of our model. This information suffices for instance to read off the chaos correlator as in \cite{Maldacena:2016hyu} and note that the leading contribution comes from the reparametrization mode as expected. This observation further lends support to the argument of \cite{Choudhury:2017tax} who noted that the out-of-time-order four-point function that captures the growth of chaos in the system continues to be exponential and saturates the chaos bound, despite the presence of $\mathcal{O}(N^2)$ light non-singlet states. 

One can also engineer disordered SYK models where we encounter similar behaviour. For instance, we can take a $\mathcal{N}=2$ $N$-component  real vector superfield $\Phi^i$ and construct a SYK action with random couplings, viz., 
\begin{equation}
S=\int d\tau d\theta d\bar \theta\left( \frac{1}{2} D_\theta \Phi^iD_{\bar \theta} \Phi^i+j_{i_1\ldots i_q} \Phi^{i_1}\cdots \Phi^{i_q}\right),
\end{equation}
where $q$ must be an even integer for the action to be bosonic. The couplings $j_{i_1\ldots i_q}$ are independent Gaussian random variables with mean zero and variance $\la j_{i_1\ldots i_q}^2\ra={1\over q}J^q N^{1-q}$. By a similar argument as in \cite{Witten:2016iux,Klebanov:2016xxf}, one can show that the leading large $N$ limit of this theory is dominated by the same set of melon diagrams as in the ${\cal N}=2$ tensor model introduced in \S\ref{sec:SUSYKT}. This suffices to infer the existence of a supersymmetry breaking vacuum.  

One can also attempt to relate the construction of the ${\cal N}=2$ SYK model studied in \cite{Fu:2016vas} to our analysis. Consider a Fermi superfield $\Upsilon$ and its complex conjugate $\overline \Upsilon$ which satisfy the  conditions
\begin{equation}
D_{\bar\theta} \Upsilon=0 \,, \qquad D_\theta \overline \Upsilon =0 \,.
\end{equation}
The  Fermi superfield $\Upsilon$ can be expanded in terms of component fields as
\begin{equation}
\Upsilon=\psi+\theta F+ \theta\bar\theta\partial_\tau \psi,
\end{equation}
where $\psi$ is a complex fermion and $F$ is a bosonic auxiliary field. We could take a model of $N$ Fermi superfields $\Upsilon^i$ having an action
\begin{equation}
S=\int d\tau d\bar \theta\, \overline\Upsilon^iD_{ \theta} \Upsilon^i+i^{q-1\over 2}\int d\tau\left[\int d\theta\, j_{i_1\ldots i_q} \Upsilon^{i_1}\cdots \Upsilon^{i_q}+\int d\bar\theta\, j^*_{i_1\ldots i_q} \overline\Upsilon^{i_1}\cdots \overline\Upsilon^{i_q}\right] .
\end{equation}
The couplings $j_{i_1\ldots i_q}$ are independent complex Gaussian random variables with mean zero and variance $\la j_{i_1\ldots i_q}j^*_{i_1\ldots i_q}\ra={1\over q}J N^{1-q}$. 
In this situation  $q$ must be an odd integer for the action to be bosonic, and it is therefore unclear how to promote this  to a melonic tensor model.

We have primarily analyzed models with two supercharges, so one might wonder if the situation can be improved, vis-a-vis supersymmetry preservation, by working with a different number of supercharges. While our analysis has not been exhaustive, we find that extended supersymmetry fails to help (a preliminary analysis is reported in Appendix~\ref{sec:N=1a4KT}). The trouble here is that a superfield interaction term which one naively one expects to be melonic, results in derivative couplings. In addition we do not anticipate the bosonic sector of the theory to behave any better than in the $\mathcal{N}=2$ case. More importantly, all extended multiplets will generically contain dynamical bosons which, as we have seen, is problematic. This suggests a general lesson that melonic dominance is intrinsically at tension with supersymmetry. One might wonder if this is further suggestive of such theories not naturally being embeddable into string theory.

Another natural question is whether the melonic tensors can be used to construct novel fixed points in higher dimensions.\footnote{ We thank Igor Klebanov and Shiraz Minwalla for interesting discussions on this issue.} Analysis of bosonic models in \cite{Giombi:2017dtl} reveals some intricate interplay, and potentially suggests the existence of a fixed point in the neighbourhood of $d=3$ dimensions at large $N$. Analysis of the $q=4$, $\mathcal{N}=2$ model uplifted to $d=3$ similarly reveals a weakly coupled large $N$ fixed point in the $\epsilon$-expansion. In attempting  to gauge the large global symmetry of these tensor models, one might wonder if in $d=3$, a suitable Chern-Simons tensor model would lead to a new class of conformal field theories. It is easy to see that the Chern-Simons couplings will lead to interactions that are non-melonic (for instance, the so-called pillow vertices arise after integrating out the gauge field or auxiliary fields). Taming these appears to drive one towards the weak-coupling limit of the Chern-Simons gauging, suggesting the absence of a non-trivial fixed point. We hope to report further on these constructions in the near future. 

Finally, let us note an interesting corollary of our analysis which could potentially have bearing  in more familiar contexts of the AdS/CFT correspondence.\footnote{ We thank Juan Maldacena for emphasizing this point to us.}
The fact that we have a theory with two supercharges with a supersymmetry broken vacuum could have implications for  counting  black hole entropy for $\frac{1}{16}$ BPS black holes in 
AdS$_5 \times {\bf S}^5$. The current status quo for these black holes is that they are supersymmetric solutions of Type IIB supergravity with $\mathcal{O}(N^2)$ entropy. But field theory analysis reveals both the index \cite{Kinney:2005ej} and explicit enumeration of states (preserving 2 supercharges) at small $N$ \cite{Chang:2013fba} to have far fewer states falling short of the black hole entropy. The analogy to draw here would be the potential for supersymmetry breaking effects due in the $1/N$ expansion (either perturbatively beyond leading order or non-perturbatively)  could make the supergravity solutions fail to be supersymmetric in the full quantum theory. Whether this is really the case, remains to be explored, but the class of models discussed here and in \cite{Anninos:2016szt} leave open this intriguing possibility.

\acknowledgments

It is a pleasure to thank  Tarek Anous, Frederik Denef, Tudor Dimofte, Michael Geracie, Igor Klebanov, R.~Loganayagam, Juan Maldacena,  Shiraz Minwalla,  David Ramirez, Steve Shenker, Douglas Stanford, Grigory Tarnopolsky for useful discussions and correspondence. We would like to especially thank Igor Klebanov, Juan Maldacena, and Douglas Stanford for  feedback on a draft of the paper.
CC and MR are supported  by U.S.\ Department of Energy grant DE-SC0009999 and by funds from the University of California. 
MR would like to thank  ICTS-TIFR, Bengaluru and the Galileo Galilei Institute, Florence for hospitality during the course of the workshops ``20 years of AdS/CFT and beyond'' and ``Entanglement in Quantum Systems'' held during the concluding stages of this work.

\newpage

\appendix

\section{Tensor models with various supercharges}
\label{sec:N=1a4KT}

We undertake a quick examination of  tensor models with different amounts of supersymmetry to demonstrate that 
the model considered in the main text was the ideal starting point. In particular, we will show that with $\mathcal{N}=1$ supersymmetry we do not get a reasonably quantum theory with melonic couplings. Likewise increasing the supersymmetry to $\mathcal{N}=4$ fails to help for we end up with  non-linear interactions that prevent the solvability of the large $N$ theory. 

First, consider an $\mathcal{N}=1$ supersymmetric model with fermionic superfield (along the lines of \cite{Fu:2016vas})
\begin{equation}
\Psi^{A_{q}}(t,\theta) = \psi^{A_q}(t)+\theta \, b^{A_{q}}(t) \,,
\end{equation}
that transforms in the $(q-1)$-fundamental representation of $O(N)^{q-1}$ with $q \geq 4$ even. As we want melonic dominance we should ensure that the index contraction follows the all-body coupling described in the text. It is easy to see that the only way to do this is to have a superpotential term $W(\Psi^{A_q}) = [\Psi^{q}]$. Integrating this over superspace will give us the desired action, which including the kinetic term takes the form:
\begin{equation}\label{eqn:N=1Action}
\begin{split}
		S &= \int dt\,d\theta\bigg(-\frac{1}{2}\Psi^{A_q} D\Psi^{A_q} + g\frac{1}{q}[\Psi^{q}]\bigg) 
	\\	&= \int dt\,\bigg(\frac{i}{2}\psi^{A_{q}}\partial_{t}\psi^{A_{q}}-\frac{1}{2}b^{A_{q}}b^{A_{q}}+g\frac{1}{q}\sum_{\mathrm{perms.}\;\sigma}
	\,\mathrm{sgn}(\sigma)\,[b\psi^{q-1}]\bigg)
\end{split}
\end{equation}
where $D = \partial_{\theta}+i\theta\partial_{t}$ is the superderivative and the sum runs over all permutations $\sigma \in S_{q}$ of $b$. However, since $q$ is even, this gives a potential with an odd number of fermions, which does not lead to a sensible theory.

Alternatively, we can consider higher supersymmetry, for instance, $\mathcal{N}=4$ supersymmetry.  Focusing for simplicity on $q=4$ we  have the bosonic superfield
\begin{equation} 
\label{eqn:N=4Superfield}
\begin{split}
\Phi^{abc}  &= 
	\phi^{abc} + \theta^{\alpha} \, \bar{\psi}_{\alpha}^{abc}
	 - \bar{\theta}_{\alpha} \, \big(\psi^{abc}\big)^{\alpha}
	+\theta^{\alpha}\bar{\theta}_{\beta}\, (B^{abc})_{\alpha}^{\beta}\\
&
	+\frac{i}{4}(\theta\theta)\bar{\theta}_{\alpha}\partial_{t}\big(\bar{\psi}^{abc}\big)^{\alpha}
		-\frac{i}{4}(\bar{\theta}\bar{\theta})\theta^{\alpha}\partial_{t}\psi_{\alpha}^{abc}+\frac{1}{16}(\theta\theta)(\bar{\theta}\bar{\theta})\partial_{t}^2\phi^{abc},
\end{split}
\end{equation}
where $\phi^{abc}$ is a bosonic field, $(B^{abc})_{\alpha}^{\beta} = (\sigma_{i})_{\alpha}^{\beta}B_{i}^{abc}$ are three auxiliary bosonic fields, and $(\psi^{abc})^{\alpha},(\bar{\psi}^{abc})^{\alpha}$ are four fermionic fields. Naively, this model seems very interesting since the expansion of the superfield with tetrahedral contractions,
 $\Phi^{a_{1}b_{1}c_{1}}\Phi^{a_{1}b_{2}c_{2}}\Phi^{a_{2}b_{1}c_{2}}\Phi^{a_{2}b_{2}c_{1}}$ in terms of component fields includes interactions of the form $\psi^{a_{1}b_{1}c_{1}}\bar{\psi}^{a_{1}b_{2}c_{2}}\psi^{a_{2}b_{1}c_{2}}\bar{\psi}^{a_{2}b_{2}c_{1}}$, which is analogous to the original interaction of the fermionic tensor model  \cite{Klebanov:2016xxf}.  However, interactions of the form: 
 $\phi^{a_{1}b_{1}c_{1}}\phi^{a_{1}b_{2}c_{2}}\phi^{a_{2}b_{1}c_{2}}\partial_{t}^2\phi^{a_{2}b_{2}c_{1}}$ and 
 $(\psi^{a_{1}b_{1}c_{1}})^{\alpha}(\partial_{t}\bar{\psi}_{\alpha}^{a_{1}b_{2}c_{2}})\phi^{a_{2}b_{1}c_{2}}\phi^{a_{2}b_{2}c_{1}}$, which are present lead to pathologies.

Attempts to write down models using non-linear $\sigma$-model intuition (cf., \cite{Donets:1999jx}) fails owing to having to engineer melonic index contraction of the tensors. As such it is not clear how to proceed to write down models with higher amounts of supersymmetry that lead to solvable Schwinger-Dyson equations. Based on these arguments it should be transparent that this problem is only exacerbated for higher supersymmetry.

\section{$SL(2,\bR)$ invariant wavefunctions}
\label{sec:sl2Rwf}

The solutions to the Casimir equation \eqref{eqn:CasimirEq} take the general form
\begin{equation}\label{eqn:solToCasimir}
\begin{split}
\Psi_h(\chi) &= (1-\chi)^{\frac{1}{2} (\Delta_{12}-\Delta_{34})}\bigg[ A(h) \, \chi^h \,  {}_2F_1(h+\Delta_{12},h-\Delta_{34};2h;\chi)
\\
&\hspace{33mm} + B(h)\, \chi^{1-h} \,  {}_2F_1(1-h+\Delta_{1_2},1-h-\Delta_{34};2-2h;\chi)\bigg],
\end{split}
\end{equation}
where $ A(h)$ and $ B(h)$ are integration constants. Demanding that the Casimir operator $\cC$ has real eigenvalues, the dimension $h$ can take the value in $h\in \bR$ or $h\in \frac{1}{2} +i\bR$. Due to the obvious symmetry $h\to 1-h$ of the Casimir equation \eqref{eqn:CasimirEq}, we can restrict the possible values of dimension $h$ to be $h\ge \frac{1}{2} $ or $h\in \frac{1}{2} +i\bR^+$. 

The eigenfunction $\Psi_h(\chi)$ is not analytic at $\chi=0,1$ and $\infty$, which correspond to the points $\tau_2=\tau_1,\tau_3$ and $\tau_4$, respectively.\footnote{ Recall that we chose a time ordering $\tau_1 < \tau_3 < \tau_4$ and $\tau_2 < \tau_4$ which has three possible coincidence limits as listed.} Consider the three regions $\chi<0$, $0<\chi<1$, and $1<\chi$. The constants $A(h)$ and $B(h)$ in different regions are in general different.  
A set of matching conditions, that relates the $A(h)$ and $B(h)$ in different regions, can be derived from the Casimir equation at $\chi=0,1,\infty$ and the hermiticity condition of the Casimir operator.\footnote{ We thank Douglas Stanford for a useful discussion on the matching conditions.} On functions $f(\chi)$ and  $g(\chi)$, the hermiticity condition reads
\begin{equation}\label{eqn:hermOnEigenfunctions}
0=\la {\cal C} f,g \ra_{m,n} -\la  f,{\cal C} g \ra_{m,n} = \frac{1}{2} \int^\infty_{-\infty}d\chi\,{\rm sgn}(\chi^m(1-\chi)^n)\,\partial_\chi
\left[(1-\chi)(f^*\partial_\chi g-g\partial_\chi f^* )\right],
\end{equation}
where the integrand is a total derivative. The integral has ``boundaries" at $\chi=0^\pm,1^\pm,\pm\infty$. We need to ensure that the boundary terms all cancel. We will first analyze the three different limits and then assemble the eigenfunctions used in the main text.  For technical reasons, we will assume $|\Delta_{12}|,|\Delta_{34}|<{1\over 2}$. We leave the analysis for general $\Delta_{12},\Delta_{34}$ to future work.

\subsection{Matching conditions}
We examine the Casimir equation in the neighborhood of the boundaries of the three domains discussed above. As with any Schr\"odinger equation we will see that the matching conditions will relate the expansion coefficients across domains, and potentially could give  a quantization condition for the eigenvalue $h$.

\subsubsection{$\chi=1$}

In the limit $\chi\to 1$, the Casimir equation \eqref{eqn:CasimirEq} reduces to
\begin{equation}
\partial_\chi\big[(1-\chi)\partial_\chi\Psi_h(\chi)\big]-{\delta^2\over 4(1-\chi)}\Psi_h(\chi)=0,
\end{equation}
where $\delta=\Delta_{12}-\Delta_{34}$. The field redefinition, 
\begin{equation}
\Psi_h(\chi)=|1-\chi|^{-\frac{1}{2} |\delta|}\phi_h(\chi) \,, 
\end{equation}
results in  $\phi_h(\chi)$ satisfying the following equation near $\chi=0$:
\begin{equation}\label{eqn:casmirphi}
\partial_\chi\big[(1-\chi)\partial_\chi\phi_h(\chi)+|\delta| \phi_h(\chi)\big]=0.
\end{equation}
The solutions are simply:
\begin{equation}
\begin{split}
\delta \neq 0: \qquad 
\phi_h(\chi) & =
\begin{cases}
	{\mathfrak a}_{1^-}(h)(1-\chi)^{|\delta|}+{\mathfrak b}_{1^-}(h)&{\rm for}\quad\chi\to1^-,
	\\
	 {\mathfrak a}_{1^+}(h)(\chi-1)^{|\delta|}+ {\mathfrak b}_{1^+}(h)&{\rm for}\quad\chi\to1^+.
\end{cases} \\
\delta = 0: \qquad 
\phi_h(\chi) & =
	\begin{cases}
	{\mathfrak a}_{1^-}(h)+{\mathfrak b}_{1^-}(h)\log(1-\chi)&{\rm for}\quad\chi\to1^-,
	\\
 	{\mathfrak a}_{1^+}(h)+ {\mathfrak b}_{1^+}(h)\log(\chi-1)&{\rm for}\quad\chi\to1^+.
\end{cases}
\end{split}
\end{equation}

To obtain the matching between the coefficients $\mathfrak{a}_{1\pm}$ and $\mathfrak{b}_{1\pm}$ we can first integrate the 
Casimir equation \eqref{eqn:casmirphi} from $\chi=1-\epsilon$ to $\chi=1+\epsilon$ for $\epsilon>0$. In the limit $\epsilon\to 0$, we find
\begin{equation}
{\mathfrak b}(h)\equiv {\mathfrak b}_{1^-}(h)= {\mathfrak b}_{1^+}(h).
\end{equation}
The hermiticity condition \eqref{eqn:hermOnEigenfunctions} for $f=\Psi_h$ and $g=\widetilde \Psi_{\widetilde h}$ implies
\begin{equation}
(1-\chi)(\Psi_h^*\partial_\chi \widetilde\Psi_{\widetilde h}-\widetilde\Psi_{\widetilde h}\partial_\chi \Psi_h^* )\Big|_{\chi\to 1^+}=\pm(1-\chi)(\Psi_h^*\partial_\chi \widetilde \Psi_{\widetilde h}-\widetilde \Psi_{\widetilde h}\partial_\chi \Psi_h^* )\Big|_{\chi\to 1^-},
\end{equation}
which further constrains
\begin{equation}
 {\mathfrak a}_{1^-}(h)= \pm{\mathfrak a}_{1^+}(h)+ c {\mathfrak b}(h),
\end{equation}
where $c$ is a real number, the $+$ sign is for the norms $\la\cdot,\cdot\ra_{0,0}$ and $\la\cdot,\cdot\ra_{1,0}$, and the $-$ sign is for the norms $\la\cdot,\cdot\ra_{0,1}$ and $\la\cdot,\cdot\ra_{1,1}$ defined in \eqref{eqn:norms}. The zero-rung four-point function ${\cal F}^{\sigma_1\sigma_2\sigma_3\sigma_4}_0$ has no discontinuity at $\tau_2=\tau_3$, so we would use the conformal eigenfunctions with $c=0$. We will refer to the matching condition with the $+$ sign as the ``standard matching condition", and with the $-$ sign as the ``twisted matching condition".

\subsubsection{$\chi\to\pm\infty$}

The analysis in the limit $\chi\to\pm\infty$ is parallel to the above. First, note that the Casimir equation reduces to
\begin{equation}
\chi^2\partial^2_\chi\Psi_h(\chi)+\chi\partial_\chi\Psi_h(\chi)-{\widetilde\delta^2\over 4}\Psi_h(\chi)=0,
\end{equation}
where $\widetilde \delta=\Delta_{12}+\Delta_{34}$. The redefinition 
\begin{equation}
\Psi_h(\chi)=|\chi|^{1+{|\widetilde \delta|\over 2}}\varphi_h(\chi) \,,
\end{equation}
results in a simple equation for the function $\phi_h(\chi)$ as $\chi\to\pm\infty$, viz.,  
\begin{equation}\label{eqn:casmirtildephi}
\partial_\chi\left[\chi^2 \partial_\chi \varphi_h(\chi)+(1+|\widetilde \delta|)\chi\varphi_h(\chi)\right]=0.
\end{equation}
The solutions to \eqref{eqn:casmirtildephi} are easily determined to be 
\begin{equation}
\begin{split}
\widetilde \delta\neq 0: \qquad 
\varphi_h(\chi)&=
\begin{cases}
	{\mathscr A}_{+\infty}(h)\chi^{-1-|\widetilde\delta|}+{\mathscr B}_{+\infty}(h)\chi^{-1}&{\rm for}\quad \chi\to+\infty,
	\\
	{\mathscr A}_{-\infty}(h)\chi^{-1-|\widetilde\delta|}+{\mathscr B}_{-\infty}(h)\chi^{-1}&{\rm for}\quad \chi\to -\infty,
\end{cases}
\\
\widetilde \delta = 0 : \qquad
\varphi_h(\chi) &=
	\begin{cases}{\mathscr A}_{+\infty}(h)\chi^{-1}+{\mathscr B}_{+\infty}(h)\chi^{-1}\log(\chi)&{\rm for}\quad \chi\to+\infty,
	\\
	{\mathscr A}_{-\infty}(h)\chi^{-1}+{\mathscr B}_{-\infty}(h)\chi^{-1}\log(-\chi)&{\rm for}\quad \chi\to -\infty.
\end{cases}	
\end{split}
\end{equation}

We integrate the Casimir equation \eqref{eqn:casmirtildephi} along the region $\chi \in (-\infty,-\Lambda]\cup[\Lambda,+\infty)$ for $\Lambda>0$. In the limit $\Lambda\to +\infty$, we obtain the condition
\begin{equation}
{\mathscr B}(h)\equiv{\mathscr B}_{-\infty}(h)={\mathscr B}_{+\infty}(h).
\end{equation}
The hermiticity condition \eqref{eqn:hermOnEigenfunctions} implies
\begin{equation}
\chi(\Psi_h^*\partial_\chi \Psi_{h'}-\Psi_{h'}\partial_\chi \Psi_h^* )\Big|_{\chi\to +\infty}=\pm\chi(\Psi_h^*\partial_\chi \Psi_{h'}-\Psi_{h'}\partial_\chi \Psi_h^* )\Big|_{\chi\to -\infty},
\end{equation}
which further constrains
\begin{equation}
 {\mathscr A}_{-\infty}(h)=\pm {\mathscr A}_{+\infty}(h)+ c \, {\mathscr B}(h),
\end{equation}
where $c$ is a real number, the $+$ sign is for the norms $\la\cdot,\cdot\ra_{0,0}$ and $\la\cdot,\cdot\ra_{1,1}$, and the $-$ sign is for the norms $\la\cdot,\cdot\ra_{0,1}$ and $\la\cdot,\cdot\ra_{1,0}$ defined in \eqref{eqn:norms}. The zero-rung four-point function ${\cal F}^{\sigma_1\sigma_2\sigma_3\sigma_4}_0$ has no discontinuity at $\tau_2=\tau_4$, so we would use the conformal eigenfunctions with $c=0$. We will refer to the matching condition with the $+$ sign as the ``standard matching condition", and with the $-$ sign as the ``twisted matching condition".

\subsubsection{$\chi=0$}

The Casimir equation at $\chi=0$ does not give any useful condition. In the $\chi\to 0$ limit, the solution to the Casimir equation takes the form as
\begin{equation}\label{eqn:Phiat0}
\Psi_h(\chi)=\begin{cases}
{ A}_{0^+}(h) \chi^{h}+{ B}_{0^+}(h)\chi^{1-h}&{\rm for}\quad\chi\to0^+,
\\
{ A}_{0^-}(h)(-\chi)^{h}+ { B}_{0^-}(h)(-\chi)^{1-h}&{\rm for}\quad\chi\to0^-.
\end{cases}
\end{equation}
The hermiticity condition \eqref{eqn:hermOnEigenfunctions} implies 
\begin{equation}\label{eqn:hermat0}
(\Psi_h^*\partial_\chi \Psi_{h'}-\Psi_{h'}\partial_\chi \Psi_h^* )\Big|_{\chi\to 0^+}=\pm(\Psi_h^*\partial_\chi \Psi_{h'}-\Psi_{h'}\partial_\chi \Psi_h^* )\Big|_{\chi\to 0^-}.
\end{equation}
Plugging \eqref{eqn:Phiat0} into the above equation gives
\begin{equation}
B_{0^+}(h)=0=B_{0^-}(h)\quad{\rm when}\quad h>\frac{1}{2} \,,
\end{equation}
which will end up picking out $h \in {\mathbb Z}^+$ or $h \in {\mathbb Z}^++{1\over 2}$.  The condition \eqref{eqn:hermat0} does not give any constraints when $h\in\frac{1}{2} +i\bR^+$. 

\subsection{Solutions}
\label{sec:eigenSol}

In this subsection, we use the matching conditions discussed in the previous subsection to determine the bases of conformal eigenfunctions with respect to the four different norms \eqref{eqn:norms}.

\subsubsection{Bosonic wavefunctions} 

\paragraph{$\la\cdot,\cdot\ra_{0,0}$ norm:}
Let us start with the region $0<\chi<1$. General solutions to the Casimir equation can be written as linear combinations of the following two conformal eigenfunctions as in \eqref{eqn:solToCasimir}. To wit,
\begin{equation}
\begin{split}
&\Psi_h^s(\chi)=\frac{1}{2} (1-\chi)^{\frac{1}{2} (\Delta_{12}-\Delta_{34})}\Big[{\Gamma(h-\Delta_{34})\Gamma(h+\Delta_{34})\over \Gamma(2h)}(1+\cos\pi\Delta_{34}\sec\pi h)\chi^h{}_2F_1(h+\Delta_{12},h-\Delta_{34};2h;\chi)
\\
&\quad+{\Gamma(1-h-\Delta_{12})\Gamma(1-h+\Delta_{12})\over \Gamma(2-2h)}(1-\cos\pi\Delta_{12}\sec\pi h)\chi^{1-h}{}_2F_1(1-h+\Delta_{12},1-h-\Delta_{34};2-2h;\chi)\Big],
\\
&\Psi_h^a(\chi)=\frac{1}{2} (1-\chi)^{\frac{1}{2} (\Delta_{12}-\Delta_{34})}\Big[{\Gamma(h-\Delta_{34})\Gamma(h+\Delta_{34})\over \Gamma(2h)}(-1+\cos\pi\Delta_{34}\sec\pi h)\chi^h{}_2F_1(h+\Delta_{12},h-\Delta_{34};2h;\chi)
\\
&\quad+{\Gamma(1-h-\Delta_{12})\Gamma(1-h+\Delta_{12})\over \Gamma(2-2h)}(-1-\cos\pi\Delta_{12}\sec\pi h)\chi^{1-h}{}_2F_1(1-h+\Delta_{12},1-h-\Delta_{34};2-2h;\chi)\Big].
\end{split}
\end{equation}
Applying the standard matching condition at $\chi=1$, we obtain the conformal eigenfunctions $\Psi^s_h$ and $\Psi^a_h$ in the region $\chi>1$,
\begin{equation}
\begin{split}
\Psi_h^s(\chi)&=-\frac{\pi  (\chi -1)^{\frac{1}{2} \left(\Delta _{12}-\Delta _{34}\right)}\chi ^h \csc
   \left(\frac{\pi}{2}   \left(\Delta _{12}-\Delta _{34}\right)\right)  \sin
   \left(\frac{\pi}{2}   \left(h-\Delta _{12}\right)\right) \csc \left(\frac{\pi}{2}  
   \left(h-\Delta _{34}\right)\right) \Gamma \left(1-h+\Delta _{12}\right)}{2\Gamma
   \left(1+\Delta _{12}-\Delta _{34}\right) \Gamma \left(1-h+\Delta _{34}\right)}
\\
&\quad \times{}_2F_1\left(h+\Delta _{12},h-\Delta _{34};1+\Delta _{12}-\Delta _{34};1-\chi
   \right)
\\
&+\frac{\pi  (\chi -1)^{\frac{1}{2} \left(\Delta _{34}-\Delta _{12}\right)}\chi ^{1-h} \csc
   \left(\frac{\pi}{2}   \left(\Delta _{12}-\Delta _{34}\right)\right)  \cos
   \left(\frac{\pi}{2}   \left(h+\Delta _{34}\right)\right) \sec \left(\frac{\pi}{2}  
   \left(h+\Delta _{12}\right)\right) \Gamma \left(h+\Delta _{34}\right)}{2\Gamma \left(1-\Delta _{12}+\Delta _{34}\right) \Gamma \left(h+\Delta
   _{12}\right)}
\\
&\quad\times{}_2F_1\left(1-h-\Delta _{12},1-h+\Delta _{34};1-\Delta _{12}+\Delta _{34};1-\chi
   \right),
\\
\Psi_h^a(\chi)&=\frac{\pi  (\chi -1)^{\frac{1}{2} \left(\Delta _{12}-\Delta _{34}\right)}\chi ^h \csc
   \left(\frac{\pi}{2}   \left(\Delta _{12}-\Delta _{34}\right)\right)  \cos
   \left(\frac{\pi}{2}   \left(h-\Delta _{12}\right)\right) \sec \left(\frac{\pi}{2}  
   \left(h-\Delta _{34}\right)\right) \Gamma \left(1-h+\Delta _{12}\right)}{2\Gamma
   \left(1+\Delta _{12}-\Delta _{34}\right) \Gamma \left(1-h+\Delta _{34}\right)}
\\
&\quad \times{}_2F_1\left(h+\Delta _{12},h-\Delta _{34};1+\Delta _{12}-\Delta _{34};1-\chi
   \right)
\\
&-\frac{\pi  (\chi -1)^{\frac{1}{2} \left(\Delta _{34}-\Delta _{12}\right)}\chi ^{1-h} \csc
   \left(\frac{\pi}{2}   \left(\Delta _{12}-\Delta _{34}\right)\right)  \sin
   \left(\frac{\pi}{2}   \left(h+\Delta _{34}\right)\right) \csc \left(\frac{\pi}{2}  
   \left(h+\Delta _{12}\right)\right) \Gamma \left(h+\Delta _{34}\right)}{2\Gamma \left(1-\Delta _{12}+\Delta _{34}\right) \Gamma \left(h+\Delta
   _{12}\right)}
\\
&\quad\times{}_2F_1\left(1-h-\Delta _{12},1-h+\Delta _{34};1-\Delta _{12}+\Delta _{34};1-\chi
   \right).
 \end{split}
\end{equation}
Applying the standard matching condition at $\chi\to\pm\infty$, we obtain the conformal eigenfunctions $\Psi^s_h$ and $\Psi^a_h$ in the region $\chi<0$,
\begin{equation}
\begin{split}
&\Psi_h^s(\chi)=\frac{1}{2} (1-\chi)^{\frac{1}{2} (\Delta_{12}-\Delta_{34})}\Big[{\Gamma(h-\Delta_{34})\Gamma(h+\Delta_{34})\over \Gamma(2h)}(1+\cos\pi\Delta_{34}\sec\pi h)(-\chi)^h{}_2F_1(h+\Delta_{12},h-\Delta_{34};2h;\chi)
\\
&\quad+{\Gamma(1-h-\Delta_{12})\Gamma(1-h+\Delta_{12})\over \Gamma(2-2h)}(1-\cos\pi\Delta_{12}\sec\pi h)(-\chi)^{1-h}{}_2F_1(1-h+\Delta_{12},1-h-\Delta_{34};2-2h;\chi)\Big],
\\
&\Psi_h^a(\chi)=\frac{1}{2} (1-\chi)^{\frac{1}{2} (\Delta_{12}-\Delta_{34})}\Big[{\Gamma(h-\Delta_{34})\Gamma(h+\Delta_{34})\over \Gamma(2h)}(1-\cos\pi\Delta_{34}\sec\pi h)(-\chi)^h{}_2F_1(h+\Delta_{12},h-\Delta_{34};2h;\chi)
\\
&\quad+{\Gamma(1-h-\Delta_{12})\Gamma(1-h+\Delta_{12})\over \Gamma(2-2h)}(1+\cos\pi\Delta_{12}\sec\pi h)(-\chi)^{1-h}{}_2F_1(1-h+\Delta_{12},1-h-\Delta_{34};2-2h;\chi)\Big].
\end{split}
\end{equation}
The conformal eigenfunctions $\Psi^s_h$ and $\Psi^a_h$ satisfy the equations
\begin{equation}
\begin{split}
&\Psi^s_{1-h}(\chi)={(\cos\pi\Delta_{12}+\cos\pi h)\Gamma(h-\Delta_{12})\Gamma(h+\Delta_{12})\over (\cos\pi\Delta_{34}+\cos\pi h)\Gamma(h-\Delta_{34})\Gamma(h+\Delta_{34})}\Psi^s_{h}(\chi),
\\
&\Psi^a_{1-h}(\chi)={(\cos\pi\Delta_{12}-\cos\pi h)\Gamma(h-\Delta_{12})\Gamma(h+\Delta_{12})\over (\cos\pi\Delta_{34}-\cos\pi h)\Gamma(h-\Delta_{34})\Gamma(h+\Delta_{34})}\Psi^a_{h}(\chi).
\end{split}
\end{equation}
Hence, we can restrict the possible values of dimension $h$ to be $h\ge \frac{1}{2} $ or $h\in \frac{1}{2} +i\bR^+$. When  $h>\frac{1}{2} $, the matching condition at $\chi=0$ constrains the dimension to be $h\in \bZ^+$. We will refer to the conformal eigenfunctions with dimension $h\in \bZ^+$ as discrete states, and the conformal eigenfunctions with dimension $h\in\frac{1}{2} +i\bR^+$ as continuum states. 

For the continuum states, the conformal eigenfunctions $\Psi^s_h$ and $\Psi^a_h$ have integral representations as
\begin{equation}\label{eqn:EFintrep}
\begin{split}
&\Psi^s_h(\chi)=\frac{1}{2} \int^\infty_{-\infty} dy{|\chi|^{h}\over |y|^{\Delta_{12}+h}|\chi-y|^{h-\Delta_{12}}|1-y|^{\Delta_{34}+(1-h)}\left|1-\chi\right|^{\frac{1}{2} (\Delta_{12}-\Delta_{34})}},
\\
&\Psi^a_h(\chi)=-\frac{1}{2} \int^\infty_{-\infty} dy{|\chi|^{h}{\rm sgn}(y){\rm sgn}(\chi-y){\rm sgn}(1-y){\rm sgn}(\chi)\over |y|^{\Delta_{12}+h}|\chi-y|^{h-\Delta_{12}}|1-y|^{\Delta_{34}+(1-h)}\left|1-\chi\right|^{\frac{1}{2} (\Delta_{12}-\Delta_{34})}}.
\end{split}
\end{equation}
The inner products of the continuum states are
\ie
\la\Psi^s_{h},\Psi^s_{h'}\ra_{0,0}&=N_s(h)\times2\pi i\,\delta(h-h'),
\\
\la\Psi^a_{h},\Psi^a_{h'}\ra_{0,0}&=N_a(h)\times2\pi i\,\delta(h-h'),
\\
\la\Psi^s_{h},\Psi^a_{h'}\ra_{0,0}&=0,
\fe
where the functions $N_s(h)$ and $N_a(h)$ are
\ie
N_s(h)=&{\tan\pi h\over 2(2h-1)\pi}(\cos\pi\Delta_{12}-\cos \pi h)(\cos\pi\Delta_{34}+\cos \pi h)
\\
&\times\Gamma(1-h+\Delta_{12})\Gamma(1-h-\Delta_{12})\Gamma(h+\Delta_{34})\Gamma(h-\Delta_{34}),
\\
N_a(h)=&{\tan\pi h\over 2(2h-1)\pi}(\cos\pi\Delta_{12}+\cos \pi h)(\cos\pi\Delta_{34}-\cos \pi h)
\\
&\times\Gamma(1-h+\Delta_{12})\Gamma(1-h-\Delta_{12})\Gamma(h+\Delta_{34})\Gamma(h-\Delta_{34}).
\fe

For the discrete states, the conformal eigenfunctions $\Psi^s_h$ and $\Psi^a_h$ are proportional to each other
\ie
&\tan\left(\pi \Delta_{34}\over 2\right)\Psi^s_h(\chi)=-\tan\left(\pi \Delta_{12}\over 2\right)\Psi^a_h(\chi)\quad{\rm for}\quad h\in 2\bZ^+,
\\
&\tan\left(\pi \Delta_{12}\over 2\right)\Psi^s_h(\chi)=-\tan\left(\pi \Delta_{34}\over 2\right)\Psi^a_h(\chi)\quad{\rm for}\quad h\in 2\bZ^+-1.
\fe
When $\Delta_{12}=0$ and $\Delta_{34}\neq 0$, $\Psi^a_h$ is non-normalizable when $h\in 2\bZ^+$, and $\Psi^s_h$ is non-normalizable when $h\in 2\bZ^+-1$. When $\Delta_{12}\neq 0$ and $\Delta_{34}=0$, $\Psi^a_h$ is zero when $h\in 2\bZ^+$, and $\Psi^s_h$ is zero when $h\in 2\bZ^+-1$. The inner products of the discrete states are given by
\ie
\la\Psi^s_{h},\Psi^s_{h'}\ra_{0,0} =  {dN_s(h)\over dh}\,\delta_{h,h'}.
\fe

\paragraph{$\la\cdot,\cdot\ra_{0,1}$ norm:} Let us start with the region $0<\chi<1$. General solutions to the Casimir equation can be written as linear combinations of the following two conformal eigenfunctions as in \eqref{eqn:solToCasimir}. To wit,
\ie
&\Psi_h^{12}(\chi)={1\over 2}(1-\chi)^{{1\over 2}(\Delta_{12}-\Delta_{34})}\Big[{\Gamma(h-\Delta_{34})\Gamma(h+\Delta_{34})\over \Gamma(2h)}(1+\cos\pi\Delta_{34}\sec\pi h)\chi^h{}_2F_1(h+\Delta_{12},h-\Delta_{34},2h;\chi)
\\
&\quad+{\Gamma(1-h-\Delta_{12})\Gamma(1-h+\Delta_{12})\over \Gamma(2-2h)}(-1-\cos\pi\Delta_{12}\sec\pi h)\chi^{1-h}{}_2F_1(1-h+\Delta_{12},1-h-\Delta_{34},2-2h;\chi)\Big],
\\
&\Psi_h^{34}(\chi)={1\over 2}(1-\chi)^{{1\over 2}(\Delta_{12}-\Delta_{34})}\Big[{\Gamma(h-\Delta_{34})\Gamma(h+\Delta_{34})\over \Gamma(2h)}(-1+\cos\pi\Delta_{34}\sec\pi h)\chi^h{}_2F_1(h+\Delta_{12},h-\Delta_{34},2h;\chi)
\\
&\quad+{\Gamma(1-h-\Delta_{12})\Gamma(1-h+\Delta_{12})\over \Gamma(2-2h)}(1-\cos\pi\Delta_{12}\sec\pi h)\chi^{1-h}{}_2F_1(1-h+\Delta_{12},1-h-\Delta_{34},2-2h;\chi)\Big].
\fe
Applying the twisted matching condition at $\chi=1$, we obtain the conformal eigenfunctions $\Psi^{12}_h$ and $\Psi^{34}_h$ in the region $\chi>1$,\footnote{ Without loss of generality, we have assumed $\Delta_{34}\le \Delta_{12}$.}
\ie
\Psi_h^{12}(\chi)&=-\frac{\pi  (\chi -1)^{\frac{1}{2} \left(\Delta _{12}-\Delta _{34}\right)}\chi ^h \sec
   \left(\frac{\pi}{2}   \left(\Delta _{12}-\Delta _{34}\right)\right)  \cos
   \left(\frac{\pi}{2}   \left(h-\Delta _{12}\right)\right) \csc \left(\frac{\pi}{2}  
   \left(h-\Delta _{34}\right)\right) \Gamma \left(1-h+\Delta _{12}\right)}{\Gamma
   \left(1+\Delta _{12}-\Delta _{34}\right) \Gamma \left(1-h+\Delta _{34}\right)}
\\
&\quad \times{}_2F_1\left(h+\Delta _{12},h-\Delta _{34};1+\Delta _{12}-\Delta _{34};1-\chi
   \right)
\\
&+\frac{\pi  (\chi -1)^{\frac{1}{2} \left(\Delta _{34}-\Delta _{12}\right)}\chi ^{1-h} \sec
   \left(\frac{\pi}{2}   \left(\Delta _{12}-\Delta _{34}\right)\right)  \cos
   \left(\frac{\pi}{2}   \left(h+\Delta _{34}\right)\right) \csc \left(\frac{\pi}{2}  
   \left(h+\Delta _{12}\right)\right) \Gamma \left(h+\Delta _{34}\right)}{\Gamma \left(1-\Delta _{12}+\Delta _{34}\right) \Gamma \left(h+\Delta
   _{12}\right)}
\\
&\quad\times{}_2F_1\left(1-h-\Delta _{12},1-h+\Delta _{34};1-\Delta _{12}+\Delta _{34};1-\chi
   \right),
\\
\Psi_h^{34}(\chi)&=-\frac{\pi  (\chi -1)^{\frac{1}{2} \left(\Delta _{12}-\Delta _{34}\right)}\chi ^h \sec
   \left(\frac{\pi}{2}   \left(\Delta _{12}-\Delta _{34}\right)\right)  \sin
   \left(\frac{\pi}{2}   \left(h-\Delta _{12}\right)\right) \sec \left(\frac{\pi}{2}  
   \left(h-\Delta _{34}\right)\right) \Gamma \left(1-h+\Delta _{12}\right)}{\Gamma
   \left(1+\Delta _{12}-\Delta _{34}\right) \Gamma \left(1-h+\Delta _{34}\right)}
\\
&\quad \times{}_2F_1\left(h+\Delta _{12},h-\Delta _{34};1+\Delta _{12}-\Delta _{34};1-\chi
   \right)
\\
&+\frac{\pi  (\chi -1)^{\frac{1}{2} \left(\Delta _{34}-\Delta _{12}\right)}\chi ^{1-h} \sec
   \left(\frac{\pi}{2}   \left(\Delta _{12}-\Delta _{34}\right)\right)  \sin
   \left(\frac{\pi}{2}   \left(h+\Delta _{34}\right)\right) \sec \left(\frac{\pi}{2}  
   \left(h+\Delta _{12}\right)\right) \Gamma \left(h+\Delta _{34}\right)}{\Gamma \left(1-\Delta _{12}+\Delta _{34}\right) \Gamma \left(h+\Delta
   _{12}\right)}
\\
&\quad\times{}_2F_1\left(1-h-\Delta _{12},1-h+\Delta _{34};1-\Delta _{12}+\Delta _{34};1-\chi
   \right),
\fe
Applying the twisted matching condition at $\chi\to\pm\infty$, we obtain the conformal eigenfunctions $\Psi^{12}_h$ and $\Psi^{34}_h$ in the region $\chi<0$,\footnote{ Without loss of generality, we have assumed $\Delta_{12}\le -\Delta_{34}$.}
\ie
&\Psi_h^{12}(\chi)={1\over 2}(1-\chi)^{{1\over 2}(\Delta_{12}-\Delta_{34})}\Big[{\Gamma(h-\Delta_{34})\Gamma(h+\Delta_{34})\over \Gamma(2h)}(1+\cos\pi\Delta_{34}\sec\pi h)(-\chi)^h{}_2F_1(h+\Delta_{12},h-\Delta_{34},2h;\chi)
\\
&\quad+{\Gamma(1-h-\Delta_{12})\Gamma(1-h+\Delta_{12})\over \Gamma(2-2h)}(-1-\cos\pi\Delta_{12}\sec\pi h)(-\chi)^{1-h}{}_2F_1(1-h+\Delta_{12},1-h-\Delta_{34},2-2h;\chi)\Big],
\\
&\Psi_h^{34}(\chi)={1\over 2}(1-\chi)^{{1\over 2}(\Delta_{12}-\Delta_{34})}\Big[{\Gamma(h-\Delta_{34})\Gamma(h+\Delta_{34})\over \Gamma(2h)}(1-\cos\pi\Delta_{34}\sec\pi h)(-\chi)^h{}_2F_1(h+\Delta_{12},h-\Delta_{34},2h;\chi)
\\
&\quad+{\Gamma(1-h-\Delta_{12})\Gamma(1-h+\Delta_{12})\over \Gamma(2-2h)}(-1+\cos\pi\Delta_{12}\sec\pi h)(-\chi)^{1-h}{}_2F_1(1-h+\Delta_{12},1-h-\Delta_{34},2-2h;\chi)\Big],
\fe
The conformal eigenfunctions $\Psi^{12}_h$ and $\Psi^{34}_h$ satisfy the equations
\begin{equation}
\begin{split}
&\Psi^{12}_{1-h}(\chi)={(\cos\pi\Delta_{12}-\cos\pi h)\Gamma(h-\Delta_{12})\Gamma(h+\Delta_{12})\over (\cos\pi\Delta_{34}+\cos\pi h)\Gamma(h-\Delta_{34})\Gamma(h+\Delta_{34})}\Psi^{12}_{h}(\chi),
\\
&\Psi^{34}_{1-h}(\chi)={(\cos\pi\Delta_{12}+\cos\pi h)\Gamma(h-\Delta_{12})\Gamma(h+\Delta_{12})\over (\cos\pi\Delta_{34}-\cos\pi h)\Gamma(h-\Delta_{34})\Gamma(h+\Delta_{34})}\Psi^{34}_{h}(\chi).
\end{split}
\end{equation}
Hence, we can restrict the possible values of dimension $h$ to be $h\ge \frac{1}{2} $ or $h\in \frac{1}{2} +i\bR^+$. When  $h>\frac{1}{2} $, the matching condition at $\chi=0$ constrains the dimension to be $h\in \bZ^+$. We will refer to the conformal eigenfunctions with dimension $h\in \bZ^+$ as discrete states, and the conformal eigenfunctions with dimension $h\in\frac{1}{2} +i\bR^+$ as continuum states. 

For the continuum states, the conformal eigenfunctions $\Psi^{12}_h$ and $\Psi^{34}_h$ have integral representations as
\ie\label{eqn:BosWave1234}
&\Phi^{12}_h(\chi)=-{1\over 2}\int^\infty_{-\infty} dy{|\chi|^{h}{\rm sgn}(y){\rm sgn}(\chi-y)\over |y|^{\Delta_{12}+h}|\chi-y|^{h-\Delta_{12}}|1-y|^{\Delta_{34}+(1-h)}\left|1-\chi\right|^{{1\over 2}(\Delta_{12}-\Delta_{34})}},
\\
&\Phi^{34}_h(\chi)={1\over 2}\int^\infty_{-\infty} dy{|\chi|^{h}{\rm sgn}(\chi){\rm sgn}(1-y)\over |y|^{\Delta_{12}+h}|\chi-y|^{h-\Delta_{12}}|1-y|^{\Delta_{34}+(1-h)}\left|1-\chi\right|^{{1\over 2}(\Delta_{12}-\Delta_{34})}}.
\fe
The inner products of the continuum states are
\ie
\la\Psi^{12}_{h},\Psi^{12}_{h'}\ra_{0,1}&=N_{12}(h)\times2\pi i\,\delta(h-h'),
\\
\la\Psi^{34}_{h},\Psi^{34}_{h'}\ra_{0,1}&=N_{34}(h)\times2\pi i\,\delta(h-h'),
\\
\la\Psi^{12}_{h},\Psi^{34}_{h'}\ra_{0,1}&=0,
\fe
where the functions $N_s(h)$ and $N_a(h)$ are
\ie
N_{12}(h)=&-{\tan\pi h\over 2(2h-1)\pi}(\cos\pi\Delta_{12}+\cos \pi h)(\cos\pi\Delta_{34}+\cos \pi h)
\\
&\times\Gamma(1-h+\Delta_{12})\Gamma(1-h-\Delta_{12})\Gamma(h+\Delta_{34})\Gamma(h-\Delta_{34}),
\\
N_{34}(h)=&-{\tan\pi h\over 2(2h-1)\pi}(\cos\pi\Delta_{12}-\cos \pi h)(\cos\pi\Delta_{34}-\cos \pi h)
\\
&\times\Gamma(1-h+\Delta_{12})\Gamma(1-h-\Delta_{12})\Gamma(h+\Delta_{34})\Gamma(h-\Delta_{34}).
\fe
For the discrete states, the conformal eigenfunctions $\Psi^{12}_h$ and $\Psi^{34}_h$ are proportional to each other
\ie
&\Psi^{12}_h(\chi)=\cot\left(\pi \Delta_{12}\over 2\right)\cot\left(\pi \Delta_{34}\over 2\right)\Psi^{34}_h(\chi)\quad{\rm for}\quad h\in 2\bZ^+,
\\
&\Psi^{12}_h(\chi)=\tan\left(\pi \Delta_{12}\over 2\right)\tan\left(\pi \Delta_{34}\over 2\right)\Psi^{34}_h(\chi)\quad{\rm for}\quad h\in 2\bZ^+-1.
\fe
Their inner products are given by
\ie
\la\Psi^{12}_{h},\Psi^{12}_{h'}\ra_{0,1}&={dN_{12}(h)\over dh}\,\delta_{h,h'}.
\fe

\subsubsection{Fermionic wavefunctions} 

\paragraph{$\la\cdot,\cdot\ra_{1,0}$ norm:} Let us start with the region $0<\chi<1$. General solutions to the Casimir equation can be written as linear combinations of the following two conformal eigenfunctions as in \eqref{eqn:solToCasimir}. To wit,
\ie
&\Psi_h^{14}(\chi)={1\over 2}(1-\chi)^{{1\over 2}(\Delta_{12}-\Delta_{34})}\Big[{\Gamma(h-\Delta_{34})\Gamma(h+\Delta_{34})\over \Gamma(2h)}(-1-\sin\pi\Delta_{34}\csc\pi h)\chi^h{}_2F_1(h+\Delta_{12},h-\Delta_{34},2h;\chi)
\\
&\quad+{\Gamma(1-h-\Delta_{12})\Gamma(1-h+\Delta_{12})\over \Gamma(2-2h)}(-1+\sin\pi\Delta_{12}\csc\pi h)\chi^{1-h}{}_2F_1(1-h+\Delta_{12},1-h-\Delta_{34},2-2h;\chi)\Big],
\\
&\Psi_h^{23}(\chi)={1\over 2}(1-\chi)^{{1\over 2}(\Delta_{12}-\Delta_{34})}\Big[{\Gamma(h-\Delta_{34})\Gamma(h+\Delta_{34})\over \Gamma(2h)}(1-\sin\pi\Delta_{34}\csc\pi h)\chi^h{}_2F_1(h+\Delta_{12},h-\Delta_{34},2h;\chi)
\\
&\quad+{\Gamma(1-h-\Delta_{12})\Gamma(1-h+\Delta_{12})\over \Gamma(2-2h)}(1+\sin\pi\Delta_{12}\csc\pi h)\chi^{1-h}{}_2F_1(1-h+\Delta_{12},1-h-\Delta_{34},2-2h;\chi)\Big],
\fe
Applying the standard matching condition at $\chi=1$, we obtain the conformal eigenfunctions $\Psi^{14}_h$ and $\Psi^{23}_h$ in the region $\chi>1$,
\ie
\Psi_h^{14}(\chi)&=\frac{\pi  (\chi -1)^{\frac{1}{2} \left(\Delta _{12}-\Delta _{34}\right)}\chi ^h \csc
   \left(\frac{\pi}{2}   \left(\Delta _{12}-\Delta _{34}\right)\right)  \sin
   \left(\frac{\pi}{2}   \left(h-\Delta _{12}\right)\right) \csc \left(\frac{\pi}{2}  
   \left(h-\Delta _{34}\right)\right) \Gamma \left(1-h+\Delta _{12}\right)}{\Gamma
   \left(1+\Delta _{12}-\Delta _{34}\right) \Gamma \left(1-h+\Delta _{34}\right)}
\\
&\quad \times{}_2F_1\left(h+\Delta _{12},h-\Delta _{34};1+\Delta _{12}-\Delta _{34};1-\chi
   \right)
\\
&-\frac{\pi  (\chi -1)^{\frac{1}{2} \left(\Delta _{34}-\Delta _{12}\right)}\chi ^{1-h} \csc
   \left(\frac{\pi}{2}   \left(\Delta _{12}-\Delta _{34}\right)\right)  \sin
   \left(\frac{\pi}{2}   \left(h+\Delta _{34}\right)\right) \csc \left(\frac{\pi}{2}  
   \left(h+\Delta _{12}\right)\right) \Gamma \left(h+\Delta _{34}\right)}{\Gamma \left(1-\Delta _{12}+\Delta _{34}\right) \Gamma \left(h+\Delta
   _{12}\right)}
\\
&\quad\times{}_2F_1\left(1-h-\Delta _{12},1-h+\Delta _{34};1-\Delta _{12}+\Delta _{34};1-\chi
   \right),
\\
\Psi_h^{23}(\chi)&=-\frac{\pi  (\chi -1)^{\frac{1}{2} \left(\Delta _{12}-\Delta _{34}\right)}\chi ^h \csc
   \left(\frac{\pi}{2}   \left(\Delta _{12}-\Delta _{34}\right)\right)  \cos
   \left(\frac{\pi}{2}   \left(h-\Delta _{12}\right)\right) \sec \left(\frac{\pi}{2}  
   \left(h-\Delta _{34}\right)\right) \Gamma \left(1-h+\Delta _{12}\right)}{\Gamma
   \left(1+\Delta _{12}-\Delta _{34}\right) \Gamma \left(1-h+\Delta _{34}\right)}
\\
&\quad \times{}_2F_1\left(h+\Delta _{12},h-\Delta _{34};1+\Delta _{12}-\Delta _{34};1-\chi
   \right)
\\
&+\frac{\pi  (\chi -1)^{\frac{1}{2} \left(\Delta _{34}-\Delta _{12}\right)}\chi ^{1-h} \csc
   \left(\frac{\pi}{2}   \left(\Delta _{12}-\Delta _{34}\right)\right)  \cos
   \left(\frac{\pi}{2}   \left(h+\Delta _{34}\right)\right) \sec \left(\frac{\pi}{2}  
   \left(h+\Delta _{12}\right)\right) \Gamma \left(h+\Delta _{34}\right)}{\Gamma \left(1-\Delta _{12}+\Delta _{34}\right) \Gamma \left(h+\Delta
   _{12}\right)}
\\
&\quad\times{}_2F_1\left(1-h-\Delta _{12},1-h+\Delta _{34};1-\Delta _{12}+\Delta _{34};1-\chi
   \right).
\fe
Applying the twisted matching condition at $\chi\to\pm\infty$, we obtain the conformal eigenfunctions $\Psi^{14}_h$ and $\Psi^{23}_h$ in the region $\chi<0$,\footnote{ Without loss of generality, we have assumed $\Delta_{12}\le -\Delta_{34}$.}
\ie
&\Psi_h^{14}(\chi)={1\over 2}(1-\chi)^{{1\over 2}(\Delta_{12}-\Delta_{34})}\Big[{\Gamma(h-\Delta_{34})\Gamma(h+\Delta_{34})\over \Gamma(2h)}(1+\sin\pi\Delta_{34}\csc\pi h)(-\chi)^h{}_2F_1(h+\Delta_{12},h-\Delta_{34},2h;\chi)
\\
&\quad+{\Gamma(1-h-\Delta_{12})\Gamma(1-h+\Delta_{12})\over \Gamma(2-2h)}(-1+\sin\pi\Delta_{12}\csc\pi h)(-\chi)^{1-h}{}_2F_1(1-h+\Delta_{12},1-h-\Delta_{34},2-2h;\chi)\Big],
\\
&\Psi_h^{23}(\chi)={1\over 2}(1-\chi)^{{1\over 2}(\Delta_{12}-\Delta_{34})}\Big[{\Gamma(h-\Delta_{34})\Gamma(h+\Delta_{34})\over \Gamma(2h)}(1-\sin\pi\Delta_{34}\csc\pi h)(-\chi)^h{}_2F_1(h+\Delta_{12},h-\Delta_{34},2h;\chi)
\\
&\quad+{\Gamma(1-h-\Delta_{12})\Gamma(1-h+\Delta_{12})\over \Gamma(2-2h)}(-1-\sin\pi\Delta_{12}\csc\pi h)(-\chi)^{1-h}{}_2F_1(1-h+\Delta_{12},1-h-\Delta_{34},2-2h;\chi)\Big].
\fe
The conformal eigenfunctions $\Psi^{14}_h$ and $\Psi^{23}_h$ are related by the equation
\begin{equation}
\begin{split}
&\Psi^{14}_{1-h}(\chi)=-{ (\sin\pi\Delta_{34}+\sin\pi h)\Gamma(1-h-\Delta_{34})\Gamma(1-h+\Delta_{34})\over(\sin\pi\Delta_{12}+\sin\pi h)\Gamma(1-h-\Delta_{12})\Gamma(1-h+\Delta_{12})}\Psi^{23}_{h}(\chi).
\end{split}
\end{equation}
Hence, we only need to consider the conformal eigenfunction $\Psi^{23}_h$ with the range of dimension $h\in  \bR $ or $h\in \frac{1}{2} +i\bR$.  When  $h\in  \bR $, the matching condition at $\chi=0$ constrains the dimension to be $h\in \bZ+{1\over 2}$. We will refer to the conformal eigenfunctions with dimension $h\in \bZ+{1\over 2}$ as discrete states, and the conformal eigenfunctions with dimension $h\in\frac{1}{2} +i\bR$ as continuum states.

For the continuum states, the conformal eigenfunctions $\Psi^{14}_h$ and $\Psi^{23}_h$ have integral representations as
\ie\label{eqn:FermWave1423}
&\Psi^{14}_h(\chi)=-{1\over 2}\int^\infty_{-\infty} dy{|\chi|^{h}{\rm sgn}(\chi){\rm sgn}(y)\over |y|^{\Delta_{12}+h}|\chi-y|^{h-\Delta_{12}}|1-y|^{\Delta_{34}+(1-h)}\left|1-\chi\right|^{{1\over 2}(\Delta_{12}-\Delta_{34})}},
\\
&\Psi^{23}_h(\chi)={1\over 2}\int^\infty_{-\infty} dy{|\chi|^{h}{\rm sgn}(\chi-y){\rm sgn}(1-y)\over |y|^{\Delta_{12}+h}|\chi-y|^{h-\Delta_{12}}|1-y|^{\Delta_{34}+(1-h)}\left|1-\chi\right|^{{1\over 2}(\Delta_{12}-\Delta_{34})}}.
\fe
The inner product of the continuum states is
\ie
\la\Psi^{23}_h,\Psi^{23}_{h'}\ra_{0,1}=N_{23}(h)\times 2\pi i\,\delta(h-h'),
\fe
where the function $N_{23}(h)$ is
\ie
N_{23}(h)=&{\cot\pi h\over 2(2h-1)\pi}(\sin \pi \Delta_{12} +\sin\pi h)( \sin\pi\Delta_{34} - \sin \pi h)
\\
&\times\Gamma(1-h+\Delta_{12})\Gamma(1-h-\Delta_{12})\Gamma(h+\Delta_{34})\Gamma(h-\Delta_{34}).
\fe

For discrete states, the conformal eigenfunction $\Psi^{23}_h$ satisfies
\ie
\Psi^{23}_h(\chi)={\Gamma(1-h+\Delta_{12})\Gamma(h+\Delta_{34})\over \Gamma(h+\Delta_{12})\Gamma(1-h+\Delta_{34})}\Psi^{23}_{1-h}(\chi).
\fe
Hence, we can further restrict the range of the dimension as $h\in \bZ^++{1\over 2}$. The inner products of the discrete states are
\ie
\la\Psi^{23}_{h},\Psi^{23}_{h'}\ra_{0,1} =  {dN_{23}(h)\over dh}\,\delta_{h,h'}.
\fe

\paragraph{$\la\cdot,\cdot\ra_{1,1}$ norm:} Let us start with the region $0<\chi<1$. General solutions to the Casimir equation can be written as linear combinations of the following two conformal eigenfunctions as in \eqref{eqn:solToCasimir}. To wit,
\ie
&\Psi_h^{13}(\chi)={1\over 2}(1-\chi)^{{1\over 2}(\Delta_{12}-\Delta_{34})}\Big[{\Gamma(h-\Delta_{34})\Gamma(h+\Delta_{34})\over \Gamma(2h)}(1-\sin\pi\Delta_{34}\csc\pi h)\chi^h{}_2F_1(h+\Delta_{12},h-\Delta_{34},2h;\chi)
\\
&\quad+{\Gamma(1-h-\Delta_{12})\Gamma(1-h+\Delta_{12})\over \Gamma(2-2h)}(-1+\sin\pi\Delta_{12}\csc\pi h)\chi^{1-h}{}_2F_1(1-h+\Delta_{12},1-h-\Delta_{34},2-2h;\chi)\Big],
\\
&\Psi_h^{24}(\chi)={1\over 2}(1-\chi)^{{1\over 2}(\Delta_{12}-\Delta_{34})}\Big[{\Gamma(h-\Delta_{34})\Gamma(h+\Delta_{34})\over \Gamma(2h)}(-1-\sin\pi\Delta_{34}\csc\pi h)\chi^h{}_2F_1(h+\Delta_{12},h-\Delta_{34},2h;\chi)
\\
&\quad+{\Gamma(1-h-\Delta_{12})\Gamma(1-h+\Delta_{12})\over \Gamma(2-2h)}(1+\sin\pi\Delta_{12}\csc\pi h)\chi^{1-h}{}_2F_1(1-h+\Delta_{12},1-h-\Delta_{34},2-2h;\chi)\Big].
\fe
Applying the twisted matching condition at $\chi=1$, we obtain the conformal eigenfunctions $\Psi^{13}_h$ and $\Psi^{24}_h$ in the region $\chi>1$,\footnote{ Without loss of generality, we have assumed $\Delta_{34}\le \Delta_{12}$.}
\ie
\Psi_h^{13}(\chi)&=\frac{\pi  (\chi -1)^{\frac{1}{2} \left(\Delta _{12}-\Delta _{34}\right)}\chi ^h \sec
   \left(\frac{\pi}{2}   \left(\Delta _{12}-\Delta _{34}\right)\right)  \sin
   \left(\frac{\pi}{2}   \left(h-\Delta _{12}\right)\right) \sec \left(\frac{\pi}{2}  
   \left(h-\Delta _{34}\right)\right) \Gamma \left(1-h+\Delta _{12}\right)}{\Gamma
   \left(1+\Delta _{12}-\Delta _{34}\right) \Gamma \left(1-h+\Delta _{34}\right)}
\\
&\quad \times{}_2F_1\left(h+\Delta _{12},h-\Delta _{34};1+\Delta _{12}-\Delta _{34};1-\chi
   \right)
\\
&+\frac{\pi  (\chi -1)^{\frac{1}{2} \left(\Delta _{34}-\Delta _{12}\right)}\chi ^{1-h} \sec
   \left(\frac{\pi}{2}   \left(\Delta _{12}-\Delta _{34}\right)\right)  \cos
   \left(\frac{\pi}{2}   \left(h+\Delta _{34}\right)\right) \csc \left(\frac{\pi}{2}  
   \left(h+\Delta _{12}\right)\right) \Gamma \left(h+\Delta _{34}\right)}{\Gamma \left(1-\Delta _{12}+\Delta _{34}\right) \Gamma \left(h+\Delta
   _{12}\right)}
\\
&\quad\times{}_2F_1\left(1-h-\Delta _{12},1-h+\Delta _{34};1-\Delta _{12}+\Delta _{34};1-\chi
   \right),
\\
\Psi_h^{24}(\chi)&=\frac{\pi  (\chi -1)^{\frac{1}{2} \left(\Delta _{12}-\Delta _{34}\right)}\chi ^h \sec
   \left(\frac{\pi}{2}   \left(\Delta _{12}-\Delta _{34}\right)\right)  \cos
   \left(\frac{\pi}{2}   \left(h-\Delta _{12}\right)\right) \csc \left(\frac{\pi}{2}  
   \left(h-\Delta _{34}\right)\right) \Gamma \left(1-h+\Delta _{12}\right)}{\Gamma
   \left(1+\Delta _{12}-\Delta _{34}\right) \Gamma \left(1-h+\Delta _{34}\right)}
\\
&\quad \times{}_2F_1\left(h+\Delta _{12},h-\Delta _{34};1+\Delta _{12}-\Delta _{34};1-\chi
   \right)
\\
&+\frac{\pi  (\chi -1)^{\frac{1}{2} \left(\Delta _{34}-\Delta _{12}\right)}\chi ^{1-h} \sec
   \left(\frac{\pi}{2}   \left(\Delta _{12}-\Delta _{34}\right)\right)  \sin
   \left(\frac{\pi}{2}   \left(h+\Delta _{34}\right)\right) \sec \left(\frac{\pi}{2}  
   \left(h+\Delta _{12}\right)\right) \Gamma \left(h+\Delta _{34}\right)}{\Gamma \left(1-\Delta _{12}+\Delta _{34}\right) \Gamma \left(h+\Delta
   _{12}\right)}
\\
&\quad\times{}_2F_1\left(1-h-\Delta _{12},1-h+\Delta _{34};1-\Delta _{12}+\Delta _{34};1-\chi
   \right).
\fe
Applying the standard matching condition at $\chi\to\pm\infty$, we obtain the conformal eigenfunctions $\Psi^{13}_h$ and $\Psi^{24}_h$ in the region $\chi<0$,
\ie
&\Psi_h^{13}(\chi)={1\over 2}(1-\chi)^{{1\over 2}(\Delta_{12}-\Delta_{34})}\Big[{\Gamma(h-\Delta_{34})\Gamma(h+\Delta_{34})\over \Gamma(2h)}(1-\sin\pi\Delta_{34}\csc\pi h)(-\chi)^h{}_2F_1(h+\Delta_{12},h-\Delta_{34},2h;\chi)
\\
&\quad+{\Gamma(1-h-\Delta_{12})\Gamma(1-h+\Delta_{12})\over \Gamma(2-2h)}(1-\sin\pi\Delta_{12}\csc\pi h)(-\chi)^{1-h}{}_2F_1(1-h+\Delta_{12},1-h-\Delta_{34},2-2h;\chi)\Big],
\\
&\Psi_h^{24}(\chi)={1\over 2}(1-\chi)^{{1\over 2}(\Delta_{12}-\Delta_{34})}\Big[{\Gamma(h-\Delta_{34})\Gamma(h+\Delta_{34})\over \Gamma(2h)}(1+\sin\pi\Delta_{34}\csc\pi h)(-\chi)^h{}_2F_1(h+\Delta_{12},h-\Delta_{34},2h;\chi)
\\
&\quad+{\Gamma(1-h-\Delta_{12})\Gamma(1-h+\Delta_{12})\over \Gamma(2-2h)}(1+\sin\pi\Delta_{12}\csc\pi h)(-\chi)^{1-h}{}_2F_1(1-h+\Delta_{12},1-h-\Delta_{34},2-2h;\chi)\Big].
\fe
The conformal eigenfunctions $\Psi^{13}_h$ and $\Psi^{24}_h$ are related by the equation
\begin{equation}
\begin{split}
&\Psi^{24}_{1-h}(\chi)=-{(\sin\pi\Delta_{12}+\sin\pi h)\Gamma(h-\Delta_{12})\Gamma(h+\Delta_{12})\over (\sin\pi\Delta_{34}-\sin\pi h)\Gamma(h-\Delta_{34})\Gamma(h+\Delta_{34})}\Psi^{13}_{h}(\chi).
\end{split}
\end{equation}
Hence, we only need to consider the conformal eigenfunction $\Psi^{13}_h$ with the range of dimension $h\in  \bR $ or $h\in \frac{1}{2} +i\bR$.  When  $h\in  \bR $, the matching condition at $\chi=0$ constrains the dimension to be $h\in \bZ+{1\over 2}$. We will refer to the conformal eigenfunctions with dimension $h\in \bZ+{1\over 2}$ as discrete states, and the conformal eigenfunctions with dimension $h\in\frac{1}{2} +i\bR$ as continuum states.

For the continuum states, the conformal eigenfunctions $\Psi^{13}_h$ and $\Psi^{24}_h$ have integral representations as
\ie\label{eqn:FermWave1324}
&\Psi^{13}_h(\chi)=-{1\over 2}\int^\infty_{-\infty} dy{|\chi|^{h}{\rm sgn}(y){\rm sgn}(1-y)\over |y|^{\Delta_{12}+h}|\chi-y|^{h-\Delta_{12}}|1-y|^{\Delta_{34}+(1-h)}\left|1-\chi\right|^{{1\over 2}(\Delta_{12}-\Delta_{34})}},
\\
&\Psi^{24}_h(\chi)={1\over 2}\int^\infty_{-\infty} dy{|\chi|^{h}{\rm sgn}(\chi){\rm sgn}(\chi-y)\over |y|^{\Delta_{12}+h}|\chi-y|^{h-\Delta_{12}}|1-y|^{\Delta_{34}+(1-h)}\left|1-\chi\right|^{{1\over 2}(\Delta_{12}-\Delta_{34})}}.
\fe
The inner product of the continuum states is
\ie
\la\Psi^{13}_h,\Psi^{13}_{h'}\ra_{1,1}=N_{13}(h)\times 2\pi i\,\delta(h-h'),
\fe
where the function $N_{13}(h)$ is
\ie
N_{13}(h)=&-{\cot\pi h\over 2(2h-1)\pi}(\sin \pi \Delta_{12} -\sin\pi h)( \sin\pi\Delta_{34} - \sin \pi h)
\\
&\times\Gamma(1-h+\Delta_{12})\Gamma(1-h-\Delta_{12})\Gamma(h+\Delta_{34})\Gamma(h-\Delta_{34}).
\fe

For discrete states, the conformal eigenfunction $\Psi^{13}_h$ satisfies
\ie
\Psi^{13}_h(\chi)={\Gamma(1-h+\Delta_{12})\Gamma(h+\Delta_{34})\over \Gamma(h+\Delta_{12})\Gamma(1-h+\Delta_{34})}\Psi^{13}_{1-h}(\chi).
\fe
Hence, we can further restrict the range of the dimension as $h\in \bZ^++{1\over 2}$. The inner products of the discrete states are
\ie
\la\Psi^{13}_{h},\Psi^{13}_{h'}\ra_{1,1} =  {dN_{13}(h)\over dh}\,\delta_{h,h'}.
\fe
%

\section{Useful integrals}
\label{sec:useint}

In this appendix, we list some useful integrals.

\begin{itemize}
\item First consider some basic Fourier transforms that enter into the zero-temperature computations:
\begin{equation}
\int^\infty_{-\infty}d\tau\, e^{i\omega\tau}{1\over |\tau|^{2\Delta}}=2\sin (\pi  \Delta ) \Gamma (1-2 \Delta ) |\omega |^{2 \Delta -1}.
\label{eqn:inta1}
\end{equation}
which converges for $\Delta \in (0,\frac{1}{2})$.
\item We also need the integral for the zero-frequency mode at finite temperature
\begin{equation}
\int^{\beta\over 2}_{-{\beta\over 2}}\left|\pi\over \beta\sin{\pi\tau\over \beta}\right|^{2\Delta}d\tau=\frac{\pi ^{2 \Delta -\frac{1}{2}} \beta ^{1-2 \Delta } \Gamma \left(\frac{1}{2}-\Delta
   \right)}{\Gamma (1-\Delta )}.
\label{eqn:inta2}   
\end{equation}
which  converges in the domain $\Delta < \frac{1}{2}$.

\item Let us also define a class of integrals that enter into our computation for the four-point function:
\begin{equation}\label{eqn:integrals}
\begin{split}
&\int^\infty_{-\infty} d\tau' {1\over |\tau'|^{A} |\tau'-\tau|^{B}}={k_0(A,B)\over \left| \tau \right| ^{A+B-1}},
\\
&\int^\infty_{-\infty} d\tau' {{\rm sign}(\tau')\over |\tau'|^{A}|\tau'-\tau|^{B}}={k_1(A,B){\rm sgn}(\tau)\over  \left| \tau \right| ^{A+B-1}},
\\
&\int^\infty_{-\infty} d\tau' {{\rm sign}(\tau'){\rm sign}(\tau'-\tau)\over |\tau'|^{A} |\tau'-\tau|^{B}}={k_2(A,B)\over \left| \tau \right| ^{A+B-1}},
\\
&\int^\infty_{-\infty} d\tau_0{1\over |\tau_{10}|^A|\tau_{20}|^B |\tau_{30}|^{2-A-B}}={k_0(A,B)\over |\tau_{12}|^{A+B-1}|\tau_{13}|^{1-B}|\tau_{23}|^{1-A}},
\\
&\int^\infty_{-\infty} d\tau_0{{\rm sgn}(\tau_{10}){\rm sgn}(\tau_{20})\over |\tau_{10}|^A|\tau_{20}|^B |\tau_{30}|^{2-A-B}}={k_2(A,B){\rm sgn}(\tau_{13}){\rm sgn}(\tau_{23})\over |\tau_{12}|^{A+B-1}|\tau_{13}|^{1-B}|\tau_{23}|^{1-A}},
\end{split}
\end{equation}
where the functions $k_0(A,B)$, $k_1(A,B)$, $k_2(A,B)$ are explicitly given by
\begin{equation}
\begin{split}
&k_0(A,B)={1\over \pi}\Gamma (1-A) \Gamma (1-B)\Gamma
   (A+B-1)  \left[\sin (\pi  A)+\sin (\pi  B)-\sin (\pi  (A+B))\right] ,
\\
&k_1(A,B)={1\over \pi}  \Gamma (1-A) \Gamma (1-B)\Gamma (A+B-1) \left[\sin (\pi  A)-\sin (\pi  B)-\sin (\pi  (A+B))\right],
\\
&k_2(A,B)={1\over \pi}\Gamma (1-A) \Gamma (1-B)\Gamma
   (A+B-1)  \left[\sin (\pi  A)+\sin (\pi  B)+\sin (\pi  (A+B))\right] .
\end{split}
\end{equation}
They satisfy the relations
\begin{equation}
\begin{split}
k_0(A,B)=k_0(A,2-A-B)=k_0(B,2-A-B),
\\
k_2(A,B)=k_1(A,2-A-B)=k_1(B,2-A-B).
\end{split}
\end{equation}

\item 

Other useful integrals are
\begin{equation}
\begin{split}
&\int d\tau d\tau'{{\rm sgn}(\tau_1-\tau){\rm sgn}(\tau_2-\tau'){\rm sgn}(\tau-\tau')\over |\tau_1-\tau|^{A}|\tau_2-\tau'|^{B}|\tau-\tau'|^{C}}
={{\rm sgn}(\tau_{12})\over |\tau_{12}|^{A+B+C-2}}k_2(A,C)k_1(B,A+C-1),
\\
&\int d\tau d\tau'{1\over |\tau_1-\tau|^{A}|\tau_2-\tau'|^{B}|\tau-\tau'|^{C}}
={1\over |\tau_{12}|^{A+B+C-2}}k_0(A,C)k_0(B,A+C-1),
\\
&\int d\tau d\tau'{{\rm sgn}(\tau_1-\tau){\rm sgn}(\tau-\tau')\over |\tau_1-\tau|^{A}|\tau_2-\tau'|^{B}|\tau-\tau'|^{C}}
=-{1\over |\tau_{12}|^{A+B+C-2}}k_2(A,C)k_0(B,A+C-1),
\\
&\int d\tau d\tau'{{\rm sgn}(\tau_1-\tau)\over |\tau_1-\tau|^{A}|\tau_2-\tau'|^{B}|\tau-\tau'|^{C}}
=-{1\over |\tau_{12}|^{A+B+C-2}}k_1(A,C)k_1(A+C-1,B),
\\
&\int d\tau d\tau'{{\rm sgn}(\tau_1-\tau){\rm sgn}(\tau_2-\tau')\over |\tau_1-\tau|^{A}|\tau_2-\tau'|^{B}|\tau-\tau'|^{C}}
={1\over |\tau_{12}|^{A+B+C-2}}k_1(A,C)k_2(B,A+C-1),
\\
&\int d\tau d\tau'{{\rm sgn}(\tau-\tau')\over |\tau_1-\tau|^{A}|\tau_2-\tau'|^{B}|\tau-\tau'|^{C}}
={{\rm sgn}(\tau_{12})\over |\tau_{12}|^{A+B+C-2}}k_1(C,A)k_1(A+C-1,B).
\end{split}
\end{equation}

\end{itemize}

\clearpage
\include{susymelons-biblio}


\end{document}

%% file: susymelons-macros.tex
\definecolor{rust}{rgb}{0.8,0.2,0.2}

 \def\shadeB{\cellcolor{blue!5}}

\def\bR {\mathbb{R}}
\def\bZ {\mathbb{Z}}

\def\be{\begin{equation}}
\def\ee{\end{equation}}
\def\bea{\begin{eqnarray}}
\def\eea{\end{eqnarray}}

\makeatletter\@addtoreset{equation}{section}\makeatother

\newcommand{\vev}[1]{{\left< {#1} \right>}}

\newcommand{\ket}[1]{{\left| {#1} \right>}}

\newcommand{\cC}{{\mathcal C}}

\def\be{\begin{equation}}
\def\ee{\end{equation}}
\def\bea{\begin{eqnarray}}
\def\eea{\end{eqnarray}}
\def\ie{\begin{equation}\begin{aligned}}
\def\fe{\end{aligned}\end{equation}}

\newcommand{\Tr}{{\rm Tr\,}}
\newcommand{\la}{\langle}
\newcommand{\ra}{\rangle}

%% file: susymelons-biblio.tex
\providecommand{\href}[2]{#2}\begingroup\raggedright\endgroup